\documentclass[]{interact}

\usepackage{epstopdf}% To incorporate .eps illustrations using PDFLaTeX, etc.
\usepackage[caption=false]{subfig}% Support for small, `sub' figures and tables

\usepackage[numbers,sort&compress]{natbib}% Citation support using natbib.sty
\bibpunct[, ]{[}{]}{,}{n}{,}{,}% Citation support using natbib.sty
\renewcommand\bibfont{\fontsize{10}{12}\selectfont}% Bibliography support using natbib.sty
\makeatletter% @ becomes a letter
\def\NAT@def@citea{\def\@citea{\NAT@separator}}% Suppress spaces between citations using natbib.sty
\makeatother% @ becomes a symbol again

\theoremstyle{plain}% Theorem-like structures provided by amsthm.sty
\newtheorem{theorem}{Theorem}[section]

\newtheorem{corollary}[theorem]{Corollary}

\newtheorem{hypothesis}[theorem]{Hypothesis}

\theoremstyle{definition}
\newtheorem{definition}[theorem]{Definition}

\theoremstyle{remark}
\newtheorem{remark}{Remark}

\usepackage[ruled,linesnumbered]{algorithm2e}
\SetKwComment{Comment}{// }{}
\RestyleAlgo{ruled}
\begin{document}

\articletype{ARTICLE TEMPLATE}% Specify the article type or omit as appropriate

\title{Fractional and tempered fractional models for Reynolds-averaged Navier-Stokes equations}

\author{
\name{Pavan Pranjivan Mehta\textsuperscript{a}\thanks{CONTACT Pavan P. Mehta Email: pavan\_pranjivan\_mehta@alumni.brown.edu}}
\affil{\textsuperscript{a} SISSA, International School of Advanced Studies, Trieste, Italy \\ Via Bonomea, 265, 34136 Trieste TS, Italy}
}

\maketitle

\begin{abstract}

Turbulence is a non-local phenomenon and has multiple-scales. Non-locality can be addressed either implicitly or explicitly. Implicitly, by subsequent resolution of all spatio-temporal scales. However, if directly solved for the temporal or spatially averaged fields, a closure problem arises on account of missing information between two points. To solve the closure problem in Reynolds-averaged Navier-Stokes equations (RANS), an eddy-viscosity hypotheses has been a popular modelling choice, where it follows either a linear or non-linear stress-strain relationship. Here, a non-constant diffusivity is introduced. Such a non-constant diffusivity is also characteristic of non-Fickian diffusion equation addressing anomalous diffusion process. An alternative approach, is a fractional derivative based diffusion equations. Thus, in the paper, we formulate a fractional stress-strain relationship using variable-order Caputo fractional derivative. This provides new opportunities for future modelling effort. 

We pedagogically study of our model construction, starting from one-sided model and followed by two-sided model applied to channel, couette and pipe flow. Non-locality at a point is the amalgamation of all the effects, thus we find the two-sided model is physically consistent. Further, our construction can also addresses viscous effects, which is a local process. Thus, our fractional model addresses the amalgamation of local and non-local process. We also show  its validity at infinite Reynolds number limit.  An expression for the fractional order is also found, thereby solving the closure problem for the considered cases.

This study is further extended to tempered fractional calculus, where tempering ensures finite jump lengths, this is an important remark for unbounded flows. Within the context of this paper, we limit ourselves to wall bounded flow. Two tempered definitions are introduced with a smooth and sharp cutoff, by the exponential term and Heaviside function, respectively and we also define the horizon of non-local interactions. We further study the equivalence between the two definitions, as truncating the domain has computational advantages.  

For the above investigations, we carefully designed algorithms, notably, the pointwise version of fractional physics-informed neural network to find the fractional order as an inverse problem. 

\end{abstract}


\begin{tabular}{ |p{0.15\textwidth} p{0.79\textwidth}| }
\hline
\multicolumn{2}{|c|}{\textbf{Notations}} \\
$(.)^+$ & non-dimensional form using friction velocity ($U_\tau$) and viscosity ($\nu$) \\
$\overline{(.)}$ & indicates time-averaged field \\
${ d^\alpha \over d x^\alpha}$ & fractional derivative with an arbitrary order $\alpha \in \mathbb{R}_+$ \\ 
${}_a I^p_x f(x)$ & Riemann-Liouville fractional integral with the lower terminal as $a$ and upper terminal as $x$ of the integral,  $p$ is the order \\
$\Gamma(.)$ & Euler gamma function \\
${}_a^{RL} D^p_x f(x)$ & p-th order left-sided Riemann-Liouville fractional derivative, with the lower terminal as $a$ and upper terminal as $x$ of the integral \\
${}_x^{RL} D^p_b f(x)$ & p-th order right-sided Riemann-Liouville fractional derivative, with the lower terminal as $x$ and upper terminal as $b$ of the integral \\
${}_a^{C} D^p_x f(x)$ & p-th order left-sided Caputo fractional derivative, with the lower terminal as $a$ and upper terminal as $x$ of the integral \\
${}_x^{C} D^p_b f(x)$ & p-th order right-sided Caputo fractional derivative, with the lower terminal as $x$ and upper terminal as $b$ of the integral \\
${}_a^{RZ} D^p_b f(x)$ & p-th order Rierz fractional derivative \\
$^M D_{x}^{\alpha(x)} (\overline{U})$ & This notation is used to refer both one-sided and two-sided fractional model of Reynolds averaged Navier-stokes equations \\ 
${_{[a, b]}} ^T D_{x}^{\alpha(x)} ( \overline{U})$ & Two-sided fractional model of Reynolds averaged Navier-stokes equations. Its a linear combination of left (${}_a^{C} D^{\alpha(x)}_x U$) and right (${}_x^{C} D^{\alpha(x)}_b U$) Caputo fractional derivatives. Note the terminals of integral and the fractional order ($\alpha \in (0,1]$) is a spatially varying.\\
$_{a} ^C D_{x} ^{(\alpha,  {\lambda})} ~~ \overline{U}$ & $\alpha$-th order left-sided tempered fractional derivative, with tempering parameter, $\lambda$; with the lower terminal as $a$ and upper terminal as $x$ of the integral. This definition is formulated by introducing an exponential term in left-sided Caputo fractional derivative \\
$_{x} ^C D_{b} ^{(\alpha,  {\lambda})} ~~ \overline{U}$ & $\alpha$-th order right-sided tempered fractional derivative, with tempering parameter, $\lambda$; with the lower terminal as $x$ and upper terminal as $b$ of the integral. This definition is formulated by introducing an exponential term in right-sided Caputo fractional derivative \\
$_{a} ^C D_{x} ^{(\alpha,  {\delta})} ~~ \overline{U}$ & $\alpha$-th order left-sided truncated fractional derivative, with truncating parameter, $\delta$; with the lower terminal as $a$ and upper terminal as $x$ of the integral. This definition is formulated by introducing an Heaviside function in left-sided Caputo fractional derivative \\
$_{x} ^C D_{b} ^{(\alpha,  {\delta})} ~~ \overline{U}$ & $\alpha$-th order right-sided truncated fractional derivative, with truncating parameter, $\delta$; with the lower terminal as $x$ and upper terminal as $b$ of the integral. This definition is formulated by introducing an Heaviside function in right-sided Caputo fractional derivative \\
${_{[a, b]}} ^T D_{x}^{(\alpha(x),   {\lambda})} \overline{U}$ & Two-sided tempered fractional model of Reynolds averaged Navier-stokes equations. Its a linear combination of left (${}_a^{C} D^{\alpha(x), \lambda}_x U$) and right (${}_x^{C} D^{(\alpha(x), \lambda)}_b U$) tempered fractional derivatives. Note the terminals of integral and  $\alpha \in (0,1]$ is a spatially varying function \\
${_{[a, b]}} ^T D_{x}^{(\alpha(x),   {\delta})} \overline{U}$ & Two-sided truncated fractional model of Reynolds averaged Navier-stokes equations. Its a linear combination of left (${}_a^{C} D^{(\alpha(x), \delta}_x U$) and right (${}_x^{C} D^{(\alpha(x), \delta)}_b U$) truncated fractional derivatives. Note the terminals of integral and $\alpha \in (0,1]$ is a spatially varying function \\
\hline
\end{tabular}

\tableofcontents

\section{Introduction} \label{sec:intro}

Brownian motion is intricately linked with diffusion process. Consider the random walk (\ref{eq:1.1}), where $X_n$ are independent and identical distributed random variable, then, $S_n$ gives the subsequent location of particle after $n$ jumps.

\begin{equation} \label{eq:1.1}
    S_n = X_1 + X_2 + \dots X_n
\end{equation}

Invoking the central limit theorem, the normalised sum ($n^{-1/2} S_n$) converges (in distribution) to a normal distribution as $n \rightarrow \infty$. It can be formally shown, that this normal distribution is a solution of classical diffusion equation (\ref{eq:1.2}) (refer Ch. 1 \cite{Mark_sFC_book}). This implies the microscopic picture of the diffusion equation, is particles (or may be atoms) are performing Brownian motion. This line of thought was used by Einstein in \cite{Einstein} to prove the existence of molecules. It can be trivially extended for the diffusion equation with drift too (refer Ch. 1 \cite{Mark_sFC_book}).

\begin{equation} \label{eq:1.2}
    {d p (x, t) \over d t} = D { d^2 p (x, t) \over d x^2}
\end{equation}

Now, consider, $X_n$ sampled from a heavy tailed distribution, such as Pareto distribution ($ P[X > x] = C x^{-\alpha} $). We now invoke the extended central limit theorem, where the normalised sum ($n^{-1/\alpha} S_n$) converges (in distribution) to a wider class of $\alpha$-stable distribution as $n \rightarrow \infty$. A special case is the normal distribution for $\alpha = 1/2$. Similar to the previous case, it can formally shown that, this now solves a fractional diffusion equation (\ref{eq:1.3}) (refer Ch. 1 \cite{Mark_sFC_book}). Thus fractional diffusion equation address wider class of diffusion process, termed as anomalous diffusion. Fractional Calculus is a natural generalisation of classical notion of derivatives to arbitrary orders. The notation ${ d^\alpha / d x^\alpha}$ denotes a fractional derivative of order $ \alpha \in (1, 2]$, it will be formally introduced in  section \ref{sec:FC}. 

\begin{equation} \label{eq:1.3}
    {d p (x, t) \over d t} = D { d^\alpha p (x, t) \over d x^\alpha} ~~; \alpha \in (0, 1]
\end{equation}

Anomalous diffusion processes are a further generalisation of diffusion process, with two distinct cases, besides Brownian motion associated with classical diffusion equation,
\begin{itemize}
    \item For a sub-diffusion process, the rate of spreading is slower than classical diffusion (Brownian motion). The microscopic picture of the particle is associated with random waiting times. A time-fractional diffusion equation naturally emerges for such process. 
    \item For a super-diffusion process, the rate of spreading is higher (or faster) than Brownian motion.  The microscopic picture is associated with particle performing random jumps (also known as L\'evy flights). A space-fractional diffusion equation naturally emerges for such process. 
\end{itemize}

Fractional derivatives are not the only candidate addressing anomalous diffusion processes, a non-Fickian diffusion equation also address the anomalous diffusion processes. For a non-Fickian case, the diffusivity is no longer a constant, but could be a function of space, then we write, a non-Fickian  diffusion equation as (\ref{eq:1.4}). Note that, the classical diffusion equation (also known as Fick's law of diffusion) is a special case, where the diffusivity ($D$) is a constant.

\begin{equation} \label{eq:1.4}
    {d p (x, t) \over d t} =   {d \over d x} \left (  D (x) { d p (x, t) \over d x} \right )
\end{equation}

Such a non-Fickian diffusion equation for turbulent flows was first constructed by Richardson  \cite{richardson1926atmospheric}, and the diffusivity was experimentally found by a balloon experiment. He used the distance - neighbor graph to discover a power-law behaviour of diffusivity for atmospheric diffusion process. Further, the eddy-viscosity hypothesis solving the closure problem for Reynolds-averaged Navier-Stokes (RANS) equations (\ref{eq:1.5}) \cite{reynolds1895iv} (derived in Appendix \ref{sec:rans}) is of the form of a non-Fickian diffusion equation, similar to one constructed by Richardson  \cite{richardson1926atmospheric}. RANS were derived by time-averaging the Navier-Stokes equations in \cite{reynolds1895iv}, here $\overline{U_i}$ and $\overline{P}$ are mean velocity and pressure. As a result of not resolving all the scales and directly solving for the mean velocity and pressure, a closure problem arises, where there are more unknowns ($\overline{u_{i}u_{j}} $)  than the number of equations.

\begin{equation}\label{eq:1.5} 
 \frac{\partial \overline{U_{i}}} {\partial t} + \overline{U_{j}} \frac{\partial \overline{U_{i}}} {\partial x_{j}} = -\frac{1 } {\rho} \frac{\partial \overline{P}} {\partial x_{i}} + \frac{\partial} {\partial x_{j}} \left( {\nu \frac{\partial \overline{U_{i}}} {\partial x_{j}}} - \overline{u_{i}u_{j}} \right); ~ i, j = {1,2,3}.
\end{equation}

To illustrate equivalence between non-Fickian and eddy-viscosity hypothesis. Consider the linear eddy-viscosity hypothesis for solving the RANS closure problems, here the Reynolds stresses ($\overline{u_{i}u_{j}} $) are linearly proportional to the strain rate as given in (\ref{eq:1.6}). Here, $\nu$ is the molecular viscosity and $ \nu_{t}$ is the eddy viscosity. The use of eddy viscosity (also known as artificial or turbulent viscosity) was proposed in \cite{vonneumann1950method}.  

\begin{equation} \label{eq:1.6}
    \overline{u_i u_j} ~=~ 2/3 k \delta_{ij} ~-~ \nu_t \left ( \frac{\partial \overline{U_{i}}} {\partial x_{j}} + \frac{\partial \overline{U_{j}}} {\partial x_{i}}   \right) ~; ~ i,j = {1,2,3}.
\end{equation}

Upon substituting (\ref{eq:1.6}) in RANS equations (\ref{eq:1.5}) and simplifying, the diffusion terms are given as (\ref{eq:1.7}),

\begin{equation}  \label{eq:1.7}
    \frac{\partial} {\partial x_{j}} \left( { (\nu + \nu_t) \frac{\partial \overline{U_{i}}} {\partial x_{j}}} \right) + \frac{\partial} {\partial x_{j}} \left( { \nu_t \frac{\partial \overline{U_{j}}} {\partial x_{i}}} \right) ~; ~ i,j = {1,2,3}.
\end{equation}

The above diffusion terms (\ref{eq:1.7}) can be further simplified as,  $ \nu_{eff}  ~=~ \nu ~ + ~ \nu_t$, where, $\nu_t$ is a spatially varying function and $\nu_{eff}$ is the effective viscosity. Diffusivity (also given by $\nu_t$) not being a constant is an attribute of non-Fickian diffsuion equation. This is an important remark, as it bridges the two literature namely, anomalous diffusion using non-Fickian diffusion equation and turbulence modeling using eddy viscosity, where both have non-constant diffusivity. 

The primary goal is to deduce $\nu_t$, the only unknown in the equation to solve the closure problem. It first led to class of linear eddy-viscosity models following a linear stress-strain relationship. Naturally, turbulence is a non-linear and non-local process, thus, it short-coming were soon uncovered when applied to complicated setting. To address the non-linearity in turbulence, thereby mitigating the limitations of linear eddy-viscosity, a non-linear eddy viscosity hypothesis was proposed. This follows a non-linear stress-strain relationship (refer appendix \ref{sec:turb} for a brief review).  

All the eddy-viscosity models use local operators. Towards the use of non-local operators in turbulence modeling; the work of Richardson \cite{richardson1926atmospheric} inspired three directions which today are the corner stone for non-local modeling of Turbulence. 

\begin{itemize}
    \item The first direction was laid by Chen \cite{chen2006speculative} who was the first to propose the use of fractional derivative of a constant order with connections to Richardson's work \cite{richardson1926atmospheric}. Chen's work was furthered in \cite{song2018universal} with a proposition of a variable-order fractional closure model in non-divergence form. Its limitation \cite{song2018universal} was mitigated by formulating a novel model in divergence form \cite{mehta2019discovering}, where it was showed that no-coefficient was required when non-dimensionalised in wall units.

    \item The second direction, was an early attempt by Kraichnan \cite{kraichnan1987eddy}, however it did not use Fractional derivatives, but non-local operators (a further generalisation of fractional calculus, where it admits an arbitrary kernel). It was furthered by Hamba in \cite{hamba1995analysis} \cite{hamba2004nonlocal} \cite{hamba2005nonlocal} \cite{hamba2006mechanism} \cite{hamba2017history} following Kraichnan \cite{kraichnan1987eddy}, which details numerical results. The key inference here is that there is no unique non-local kernel even for turbulent flows, but rather depended on the spatial location. This indicates the presence of multiple-scales and proves that the non-locality is not a constant in turbulent flows as initially proposed by Chen \cite{chen2006speculative} where he proposed the use of fractional derivative of a fixed order. This piece of work indeed shows the same fact, where the use of a variable-order fractional closure model implies that the non-locality is spatially dependent. It is to be noted that, there is a limitation to Kraichnan's and Hamba's work, where the non-local operator constructed cannot approximate the local or laminar regions, thus this model may not be suitable for modeling wall-bounded flows, where the near wall regions are dominated by the viscous actions. However, as we show our model is valid for all regimes of the flow.

    \item The third interesting direction is within the framework of kinetic theory laid in \cite{epps2018turbulence} \cite{hayot1996non}, and furthered in \cite{samiee2020fractional} for Large Eddy Simulations (LES). 
\end{itemize}

In this work, we formulate fractional stress-strain relationship for closure problem using variable-order Caputo fractional derivative. We first elaborate over our previous work of one-sided model developed in \cite{mehta2019discovering} for couette flow, by applying to channel and pipe flow. Mitigating the limitations of one-sided model, we formulated a two-sided model in section \ref{sec:frans} and its extension to tempered fractional calculus in section \ref{sec:temp}. %This paper is structured as follows:

The result presented in this paper are available as arXiv:2105.03646 \cite{mehta2021fractional} since 2021 for one-sided and two-sided model. While our results of tempered fractional model were presented in ICTAM 2020+1, held in Italy (virtually), the book of abstract (available online) has our conference paper \cite{mehta_temp} (this conference was postponed by a year due pandemic, but we submitted our results as early as  January 2020).

\section{Preliminary : Fractional Calculus} \label{sec:FC}

In this section, we define the fractional derivatives and their properties

\subsection{Gr\"{u}nwald-Letnikov Definition}

Inorder to derive the Gr\"{u}nwald-Letnikov fractional derivative \cite{grunwald1867uber} (see also \cite{podlubny1999}), consider,

\begin{equation*}
        f'(x) = {d f \over d x} = \lim_{h \rightarrow 0} {f(x) - f(x-h) \over h}
\end{equation*}

\begin{equation*}
        f''(x) = {d^{2} f \over d x^{2}} = \lim_{h \rightarrow 0} {f(x) - 2f(x-h)  + f(x-2h) \over h^{2}}
\end{equation*}

\begin{equation*}
        f'''(x) = {d^{3} f \over d x^{3}} = \lim_{h \rightarrow 0} {f(x) - 3 f(x-h)  + 3 f(x-2h) - f(x-3h) \over h^{3}}
\end{equation*}

therefore, for a $p^{th}$-order differentiation, where $p \in \mathbb{N}$ (and $f \in C^{p}[a, b] $), we have,

\begin{equation} \label{eq:2.1}
        f^{p}(x) = {d^{p} f \over d x^{p}} = \lim_{h \rightarrow 0} {1 \over h^{p}} \sum_{r=0}^{p} (-1)^{r} \binom{p}{r} f (x - r h)
\end{equation}

where, 
\begin{equation*}
 \binom{p}{r} = {p (p -1) (p-2) (p-3) \dots (p-r+1)  \over r !} 
\end{equation*}

Formula (\ref{eq:2.1}) derived for $p \in \mathbb{N}$, can be equivalently re-written as (\ref{eq:2.2}), by constructing a uniform grid $x \in [a, b]$, with $N = (x-a)/h$. Since the terms after $\binom{p}{p}$, vanishes \cite{podlubny1999}. Although, formula (\ref{eq:2.2}) is derived for $p \in \mathbb{N}$, but it is valid for any non-integer order ($p > 0$). Thus, the Gr\"{u}nwald-Letnikov fractional derivative is defined as (\ref{eq:2.2}) for $p \in \mathbb{R}_+$, 

\begin{equation} \label{eq:2.2}
        f^{p}(x) = {d^{p} f \over d x^{p}} = \lim_{h \rightarrow 0} {1 \over h^{p}} \sum_{r=0}^{N} (-1)^{r} \binom{p}{r} f (x - r h)
\end{equation}

However, for non-integer order the terms after  $\binom{p}{p}$ does not vanish. Thus Diethelm in \cite{diethelmbook} imposed an additional condition on the function ($f$) as (\ref{eq:2.3}), its equivalency is argued in theorem 2.25 of \cite{diethelmbook}.

\begin{equation}  \label{eq:2.3}
    f(x)=\begin{cases}
    f(x), & \text{if $x \in [a, b]$}.\\
    0, & \text{otherwise}.
  \end{cases}
\end{equation}

For negative order of ($p$), consider (\ref{eq:2.4}), 

\begin{equation} \label{eq:2.4}
        f^{-p}(x) = {d^{-p} f \over d x^{-p}} = \lim_{h \rightarrow 0} {1 \over h^{-p}} \sum_{r=0}^{N} (-1)^{r} \binom{-p}{r} f (x - r h)
\end{equation}

where, 
\begin{equation*}
 \binom{-p}{r} = {-p (-p -1) (-p-2) (-p-3) \dots (-p-r+1)  \over r !} = (-1)^r {\left[\begin{matrix} ~p \\ ~r \end{matrix}\right]}
\end{equation*}

The introduced notation ${\left[\begin{matrix} ~p \\ ~r \end{matrix}\right]}$, implies (following \cite{podlubny1999}),

\begin{equation*}
 {\left[\begin{matrix} ~p \\ ~r \end{matrix}\right]} = {p (p +1) (p+2) (p+3) \dots (p+r-1)  \over r !} 
\end{equation*}

Thus, the Gr\"{u}nwald-Letnikov fractional integral is defined as (\ref{eq:2.5}), 

\begin{equation} \label{eq:2.5}
        f^{-p}(x) = {d^{-p} f \over d x^{-p}} = \lim_{h \rightarrow 0} {h^{p}} \sum_{r=0}^{N} {\left[\begin{matrix} ~p \\ ~r \end{matrix}\right]} f (x - r h)
\end{equation}

\begin{remark}
Formula (\ref{eq:2.2}) and (\ref{eq:2.5}) is the unification of fractional derivative and integration, respectively, for arbitrary order ($p \in \mathbb{R}$)
\end{remark}

%%%%%%%%%%%%%%%%%%%%%%%%%%%%%%
\subsection{Riemann-Liouville Definition}

In the section, we give the definition of the Riemann-Liouville operators. Recall the Cauchy's formula for repeated integration  (\ref{eq:2.6}) for $p \in  \mathbb{N}$, 

\begin{equation}  \label{eq:2.6}
   \underbrace{ \int_{a}^{x} \int_{a}^{x_p} \int_{a}^{x_{p-1}} \dots \int_{a}^{x_{2}}}_{\text{p-integrals}} f(x_1) d x_1 d x_2 \dots d x_p = {1 \over (p-1)!}  \int_{a}^{x} {(x - \tau)}^{p-1} f (\tau) d \tau
\end{equation}

Here, $p \in  \mathbb{N}$, to generalise this formula (\ref{eq:2.6}) introduce $\Gamma(p) = (p-1)!$, where $\Gamma(.)$ is the Euler gamma function, thus the Riemann-Liouville fractional integral is defined as (\ref{eq:2.7}) for ($p \in \mathbb{R}_+$)

\begin{equation}  \label{eq:2.7}
   {}_a I^p_x f(x)  ~:=~ {1 \over \Gamma(p)}   \int_{a}^{x} {(x - \tau)}^{p-1} f (\tau) d \tau
\end{equation}

Note the notation introduced to denote Riemann-Liouville fractional integral ($ {}_a I^p_x$), which implies $p-th$ integration performed, with the lower limit as $a$ and upper limit as $x$. Theorem 2.1 of \cite{diethelmbook} gives the existence proof. Further, we state the theorem 2.2 of \cite{diethelmbook}, the semi-group property of Riemann-Liouville fractional integrals as (\ref{eq:2.8}) for $p, q \in \mathbb{R}_+$, 

\begin{equation}  \label{eq:2.8}
   {}_a I^p_x ~ {}_a I^q_x  ~ f(x) ~=~ {}_a I^{p+q}_x ~ f(x)
\end{equation}

Subsequently, the Riemann-Liouville fractional derivative is defined as (\ref{eq:2.9}) for $p \in \mathbb{R}_+$ and $k \in \mathbb{N}$,

\begin{equation}  \label{eq:2.9}
{}_a^{RL} D^p_x f(x)  ~:=~ {1 \over \Gamma(p-k)} {d^k \over d x^k}
\int_{a}^{x} {(x - \tau)}^{k-p-1} f (\tau) d \tau ~~, ~ k-1 \leq p < k 
\end{equation}

here, ($d^k / dx^k$) is the classical integer-order derivative.  The Riemann-Liouville fractional derivative in terms of the Riemann-Liouville fractional integral is denoted as (\ref{eq:2.10}), 

\begin{equation}  \label{eq:2.10}
   {}_a^{RL} D^p_x f(x) ~=~ {d^k \over d x^k} ~ {}_a I^{k-p}_x f(x) ~~, ~ k-1 \leq p < k 
\end{equation}

Lemma 2.12 of \cite{diethelmbook} gives the existence proof for Riemann-Liouville fractional derivative.  Indeed for $p \in \mathbb{N}$ the Riemann-Liouville fractional derivative is the classical integer-order derivative. Further, we state the theorem 2.13 of \cite{diethelmbook}, the semi-group property of Riemann-Liouville fractional derivative as (\ref{eq:2.11}) for $p, q \in \mathbb{R}_+$, 

\begin{equation}  \label{eq:2.11}
   {}_a^{RL} D^p_x ~ {}_a^{RL} D^q_x  ~f(x) ~=~ {}_a^{RL} D^{p+q}_x ~ f(x)
\end{equation}

We further state the theorem 2.14 of \cite{diethelmbook} as (\ref{eq:2.12}), 

\begin{equation}  \label{eq:2.12}
   {}_a^{RL} D^p_x ~ {}_a I^p_x  ~f(x) ~=~  f(x)
\end{equation}

\begin{remark}
     Upon performing the limit in (\ref{eq:2.2}), it can be shown that the Gr\"{u}nwald-Letnikov fractional derivative coincides with Riemann-Liouville fractional derivative. Thus the two definitions are equivalent to each other (refer section 2.3.7 of \cite{podlubny1999} and section 3.3 of \cite{oldham1974fractional})
\end{remark}

%%%%%%%%%%%%%%%%%%%%%%%%%%%%%%%%%%%%%%%%%%%%%%%%%%
\subsection{Caputo Definition}

Although, the Riemann-Liouville fractional derivative is mathematically well established, there are two key problems for application to physical systems (refer section 2.4 of \cite{podlubny1999}).

\begin{itemize}
    \item For a constant function ($f$), the ${}_a^{RL} D^p_x  ~f \neq 0$, in general. It is zero only for a special as $a \rightarrow -\infty$. This is not very convenient for application to physical systems. 
    
    \item The initial conditions are specified as $ \lim_{t \rightarrow a} ~{}_a^{RL} D^p_t ~f(t) = g$. Such a initial condition does not make physical sense, since we measure the function value explicitly as $ f(t) |_{t=0} = g$ (refer section 2.4 of \cite{podlubny1999} its problem with initial condition is further elaborated by taking the Laplace transform)
    
\end{itemize}

As a remedy, Caputo defined a fractional derivative \cite{Caputo1967} as (\ref{eq:2.13}) for $p \in \mathbb{R}_+$ and $k \in \mathbb{N}$, 

\begin{equation}  \label{eq:2.13}
{}_a^{C} D^p_x f(x)  ~:=~ {1 \over \Gamma(p-k)} 
\int_{a}^{x} {(x - \tau)}^{k-p-1}  {d^k \over d x^k} f (\tau) d \tau ~~, ~ k-1 < p \leq  k 
\end{equation}

Indeed, the Caputo fractional derivative of a constant is zero and the initial conditions are defined in a classical way. Theorem 3.1 of \cite{diethelmbook}, expresses the Caputo fractional derivative in terms of Riemann-Liouville fractional derivative by introduction of a Taylor polynomial. We further state theorem 3.7 of \cite{diethelmbook} as (\ref{eq:2.14}) ,

\begin{equation}  \label{eq:2.14}
   {}_a^{C} D^p_x ~ {}_a I^p_x  ~f(x) ~=~  f(x)
\end{equation}

\begin{remark}   
For homogeneous conditions the Caputo fractional derivative equivalent to the Riemann-Liouville fractional derivative. This can be shown by taking repeated integration by parts.
\end{remark}

The semi-group property of Caputo fractional derivative is given as (\ref{eq:2.14_1}) for $p, q > 0$, where $p, q$ can be integers too (refer lemma 3.13 of \cite{diethelmbook}).

\begin{equation}  \label{eq:2.14_1}
   {}_a^{C} D^p_x ~~ {}_a^{C} D^q_x  ~f(x) ~=~  {}_a^{C} D^{p+q}_x ~f(x)
\end{equation}

%%%%%%%%%%%%%%%%%%%%%%%%%%%%%%%%%%%%%%%%%%%%5
\subsection{Left and right sided fractional derivative}

Upon close observation to the definitions introduced in the preceding section, it is evident that the fractional derivative only takes into account the non-local interactions from the left, hence there are termed as the left fractional derivative. A right fractional derivative can be defined too, which takes account of the non-local interactions from the right, which is a also a well defined object. Thus the mathematical theory of fractional calculus is complete. In this paper, we employ both the left and right definitions as physical process are amalgamation of interactions with both left and right boundary.

Again, we state the left Riemann-Liouville fractional derivative as (\ref{eq:2.15}) for a function defined over an interval $[a,b]$, $p \in \mathbb{R}_+$ and $k \in \mathbb{N}$, 

\begin{equation}  \label{eq:2.15}
   {}_a^{RL} D^p_x f(x)  ~:=~ {1 \over \Gamma(p-k)} {d^k \over d x^k}
\int_{a}^{x} {(x - \tau)}^{k-p-1} f (\tau) d \tau ~~, ~ k-1 \leq p < k 
\end{equation}

while, the right Riemann-Liouville fractional derivative is defined as (\ref{eq:2.16}) for $p \in \mathbb{R}_+$ and $k \in \mathbb{N}$, notice the domain of integration. 

\begin{equation}  \label{eq:2.16}
   {}_x^{RL} D^p_b f(x)  ~:=~  { (-1)^k \over \Gamma(p-k)} {d^k \over d x^k}
\int_{x}^{b} {(\tau - x)}^{k-p-1} f (\tau) d \tau ~~, ~ k-1 \leq p < k 
\end{equation}

Similarly, we state the left Caputo fractional derivative as (\ref{eq:2.17}) for a function defined over an interval $[a,b]$, $p \in \mathbb{R}_+$ and $k \in \mathbb{N}$,

\begin{equation}  \label{eq:2.17}
{}_a^{C} D^p_x f(x)  ~:=~ {1 \over \Gamma(p-k)} 
\int_{a}^{x} {(x - \tau)}^{k-p-1}  {d^k \over d x^k} f (\tau) d \tau ~~, ~ k-1 < p \leq  k 
\end{equation}

while, the right Caputo fractional derivative is defined as (\ref{eq:2.18}) for $p \in \mathbb{R}_+$ and $k \in \mathbb{N}$, again notice the domain of integration. 

\begin{equation}  \label{eq:2.18}
{}_x^{C} D^p_b f(x)  ~:=~  { (-1)^k \over \Gamma(p-k)} 
\int_{x}^{b} {(\tau - x)}^{k-p-1}  {d^k \over d x^k} f (\tau) d \tau ~~, ~ k-1 < p \leq k 
\end{equation}

The left and right definitions allow us to explicitly define the Riesz fractional derivative (\ref{eq:2.19}) for a function defined over an interval $[a,b]$ and $c_p = - 0.5 cos (p \pi / 2)$, 

\begin{equation}  \label{eq:2.19}
     {}_a^{RZ} D^p_b f(x) ~:=~  c_{p} \left ( {}_a^{RL} D^p_x f(x) ~+~ {}_x^{RL} D^p_b f(x) \right )
\end{equation}

%%%%%%%%%%%%%%% Section 3 %%%%%%%%%%%%%%%%%%%%%%%
 \section{Fractional Reynolds-averaged Navier-Stokes equations (f-RANS)} \label{sec:frans}

%\setcounter{section}{2} \setcounter{equation}{0} %% to have proper 2-digits numbers of eqs
%% Note that this style produces 1-digit numbering of definitons, statements, exmaples, etc.

In this section, we introduce the fractional closure model for fluid flows. The governing equations are the Reynolds-averaged Navier-Stokes (RANS) equations  (\ref{eq:rans}) \cite{reynolds1895iv} in wall units (refer appendix \ref{sec:rans}), where $\overline{(.)}$ implies that quantities are temporal-averaged. The instantaneous velocity ($U_i$) and pressure ($P$) are given by $U_i = \overline{U_i} + u_i$ and $P = \overline{P} + p$, where $u_i$ and $p$ are the fluctuating components of velocity and pressure about the mean. The time-period, $T$ for the ensemble is large enough to satisfy, $ {1 \over  T} \int_0^T u_i = 0$ and  $ {1 \over  T} \int_0^T p = 0$.

\begin{equation}\label{eq:rans} 
{\partial \overline{U^+_i} \over \partial t^+} + \overline{U^{+}_{j}} \frac{\partial \overline{U^{+}_{i}}} {\partial x^{+}_{j}} = -\frac{\partial \overline{P^{+}}} {\partial x^{+}_{i}} + \frac{\partial} {\partial x^{+}_{j}} \left( \frac{\partial \overline{U^{+}_{i}}} {\partial x^{+}_{j}} - \overline{u_{i}u_{j}}^{+} \right); ~ i, j = {1,2,3}
\end{equation}

The above RANS equations (\ref{eq:rans}) is written in non-dimensional form in wall units (derived in appendix \ref{sec:rans}) using the friction velocity ($U^2_\tau = \tau_w / \rho$) and kinematic viscosity ($\nu$), where $\tau_w$ is the wall shear stress. The notation $(.)^+$  indicates the same and as given in (\ref{eq:non-dim-new-model}). 

\begin{equation}\label{eq:non-dim-new-model}
\overline{{U^{+}_{i}}} = \frac{ \overline{U_{i}}} {U_{\tau}},~  {x^{+}_{i}} = \frac{x_{i} U_{\tau}}{\nu}, ~ t^+ = {t U^2_\tau \over \nu} , ~ \overline{u_{i}u_{j}}^+ = \frac{\overline{u_{i}u_{j}}}{U^{2}_{\tau}},~ \overline{{P^{+}}} = \frac{\overline{P}}{\rho U^{2}_{\tau}},  \mbox{ and } U_{\tau} = \sqrt{ \frac{\tau_{w}} {\rho}} .
\end{equation}

%\begin{equation}\label{eq:rey} 
%    \overline{u_{i}u_{j}}^{+} = \overline{U^+_i U^+_j} - \overline{U^+_i} ~\overline{U^+_j}
%\end{equation}

\begin{hypothesis}
The hypothesis for closure modeling of RANS, "\textit{total shear stress is given by fractional strain rate}" defined as (\ref{eq:frac}).
\end{hypothesis}

\begin{equation}\label{eq:frac} 
     \frac{\partial \overline{U^{+}_{i}}} {\partial x^{+}_{j}} - \overline{u_{i}u_{j}}^{+} ~=~ ^M D_{x^+_j}^{\alpha(x^+_j)} (\overline{U^+_i}) ~;~ i,j = 1,2,3 ~;~ \alpha(x^+_j) \in (0, 1]
\end{equation}

The fractional order ($\alpha$) is spatially dependent function of $x^+_j$, as turbulence exhibits multiple scales, where $\alpha$ is co-related to turbulence length-scale. The notation $^M D_{x^+_j}^{\alpha(x^+_j)} (\overline{U^+_i})$ introduced here is to jointly refer our two model construction, namely one-sided (\ref{eq:one}) and two-sided model (\ref{eq:two}) using Caputo fractional derivatives, which will be subsequently introduced shortly. 

Using the semi-group property for Caputo derivatives (refer section \ref{sec:FC}), the diffusion term is manipulated as (\ref{eq:frac_semi}), 

\begin{equation} \label{eq:frac_semi}
    \frac{\partial} {\partial x^{+}_{j}} \left(  ^M D_{x^+_j}^{\alpha(x^+_j)} (\overline{U^+_i}) \right) ~=~ ^M D_{x^+_j}^\bold{{\alpha(x^+_j) + 1}} (\overline{U^+_i}) ~;~ i,j = 1,2,3 ~;~ \alpha(x^+_j) \in (0, 1]
\end{equation}

Using (\ref{eq:frac_semi}) in (\ref{eq:rans}), we get, the fractional Reynolds-averaged Navier-Stokes equations (f-RANS) as (\ref{eq:frans}), 

\begin{equation}\label{eq:frans} 
{\partial \overline{U^+_i} \over \partial t^+} + \overline{U^{+}_{j}} \frac{\partial \overline{U^{+}_{i}}} {\partial x^{+}_{j}} = -\frac{\partial \overline{P^{+}}} {\partial x^{+}_{i}} ~+~   ^M D_{x^+_j}^{\alpha(x^+_j)+1} (\overline{U^+_i})  ~; i,j = 1,2,3 ;~ \alpha(x^+_j) \in (0, 1]
\end{equation}

The f-RANS equation constructed has similarity with the non-Fickian equation in \cite{richardson1926atmospheric}. For our case, the diffusion term has a fractional derivative, while in case of \cite{richardson1926atmospheric}, the diffusion term has non-constant diffusivity, a non-Fickian diffusion equation, whilst the rest terms remain "as is". Recall, both fractional diffusion equation and non-Fickian diffusion equation are in-principle on-par to each other addressing anomalous diffusion. Further, we note the eddy-viscosity hypothesis for RANS closure has a non-constant diffusivity ($\nu_t$), an attribute of non-Fickain diffusion equation (discussed in section \ref{sec:intro} and briefly summarised in appendix \ref{sec:turb}).

In the forthcoming section, we will investigate the sanity of our hypothesis (\ref{eq:frac}) by studying the behavior of fractional order (numerical computed) applied to channel, couette and pipe flow.

For the model to be physically meaningful, it should address two extreme cases, the first, the absence of turbulence and flow entirely dominated by viscosus action. The second when the flow is highly turbulent (and what happens at $Re_\tau \rightarrow \infty$ ?). If the model leads to physically meaningfully solutions in these two extreme cases, then it will be valid for every other case, as they are all some combination of viscous and turbulent effects.

\begin{itemize}
    \item To demonstrate the first case. Consider the fractional order as unity. In this scenario, the fractional diffusion term in f-RANS (\ref{eq:frans}) reduces to a integer-order Laplacian operator, physically this implies the flow is viscous-dominated (the absence of the Reynolds stresses ($\overline{u_{i}u_{j}}$)). 

    \item The second case of flow being completely dominated by turbulence ($\overline{u_{i}u_{j}}$) and at the limit, $Re_\tau \rightarrow \infty$. This cannot be shown trivially, thus, inorder to investigate this property, we have devised computational experiments. The results of this investigation (amongst others) will be presented in forthcoming section. 
\end{itemize}

Indeed, both of these are an important consistency argument for any construction. Such a consistent model, then will be valid for wall-bounded flow and transitional regimes of the flow, as there are significant regions where the flow is dominated by the viscous actions and turbulence. Indeed, the f-RANS model addresses the amalgamation of both viscous and turbulence effects.

Furthermore, any additional term in (\ref{eq:rans}) to account for any body forces would remain "as is" in (\ref{eq:frans}) and our proposition in  (\ref{eq:frac}) would remain valid as long as (\ref{eq:rans}) remains valid, although we do not prove it here and leave it as future work. However, we justify it by recognising the fact that the non-locality resides in the Reynolds stresses ($\overline{u_{i}u_{j}}$), which may not be unique but (\ref{eq:rans}) remains valid, regardless. Thus, it is expected that the fractional order ($\alpha$) is not unique for different flows, however, (\ref{eq:frac}) remains valid.

\subsection{One- and two-sided f-RANS models}

The hypothesis and consistency arguments presented in previous section will be investigated for two model constructions, namely, one-sided model (\ref{eq:one}) defined over the domain $[-\infty, x^+_j]$ and two-sided model (\ref{eq:two}) defined over the domain $[-\infty, \infty]$. Note, the notation ${_{[-\infty, \infty]}} ^T D_{x^+_j}^{\alpha(x^+_j)} ( \overline{U^+_i})$ introduced for two-sided model in (\ref{eq:two}).

\begin{equation}\label{eq:one} 
    {_{-\infty}} ^C D_{x^+_j}^{\alpha(x^+_j)} ( \overline{U^+_i}) = \frac{\partial \overline{U^{+}_{i}}} {\partial x^{+}_{j}} - \overline{u_{i}u_{j}}^{+}~;~ i,j = 1,2,3 ~;~ \alpha(x^+_j) \in (0, 1]
\end{equation}

\begin{align} \label{eq:two}
\begin{split}
    {_{[-\infty, \infty]}} ^T D_{x^+_j}^{\alpha(x^+_j)} ( \overline{U^+_i}) ~~&:=  {1 \over 2 }~ ({_{-\infty}} ^C D_{x^+_j}^{\alpha(x^+_j)} ~-~~  {_{x^+_j}} ^C D_{\infty}^{\alpha(x^+_j)} ) ~\overline{U^+_i}  \\ ~~&= \frac{\partial \overline{U^{+}_{i}}} {\partial x^{+}_{j}} - \overline{u_{i}u_{j}}^{+} ~;~ i,j = 1,2,3 ~;~ \alpha(x^+_j) \in (0, 1]
\end{split}
\end{align}

\noindent
where, ${_{-\infty}} ^C D_{x^+_j}^{\alpha(x^+_j)} \overline{U^+_i}$ and ${_{x^+_j}} ^C D_{\infty}^{\alpha(x^+_j)} \overline{U^+_i}$ are the left-sided (\ref{eq:left}) and right-sided (\ref{eq:right}) Caputo fractional derivative respectively for $\alpha(x^+_j) \in (0, 1] $.

\begin{equation}\label{eq:left} 
   {_{-\infty}} ^C D_{x^+_j}^{\alpha(x^+_j)}  \overline{U^+_i} = \frac{1}{\Gamma(1-\alpha(x^+_j))}\int_{-\infty}^{x^+_j} (x^+_j-\xi)^{-\alpha(x^+_j)}\frac{ d \overline{U^+_i}(\xi)}{dx_j^+}d\xi
\end{equation}

\begin{equation}\label{eq:right} 
   {_{x^+_j}} ^C D_{\infty}^{\alpha(x^+_j)} \overline{U^+_i} = - \frac{1}{\Gamma(1-\alpha(x^+_j))}\int_{x^+_j}^{\infty} (\xi - x^+_j)^{-\alpha(x^+_j)}\frac{d\overline{U^+_i}(\xi)}{dx_j^+}d\xi
\end{equation}

\begin{remark}
    
 It is to be noted that we model the total shear stress as opposed to the common practice of only modeling the Reynolds stresses. 
\end{remark}

 \begin{remark}
 It is evident whilst modeling in wall units, it is free from all assumptions, thus this model is generlisable to any flow. Since, there appears no coefficient to consider, this model can be readily applied to any complex three-dimensional flows. 

\end{remark}

%%%%%%%%%%%%%%%%%%%%%%%%%%%%%%%%%%%%%%%%%%%%%%%%%%%%%%%%%%%%%%%%%%%%%%%%%%%%%%%%%%%%%%%%%%%%%%%%
\section{Application to couette, channel and pipe flows} \label{sec:application}

%\setcounter{section}{3} \setcounter{equation}{0} %% to have proper 2-digits numbers of eqs
%% Note that this style produces 1-digit numbering of definitons, statements, exmaples, etc.

In this section, we shall outline the application of one- and two- sided f-RANS model (as constructed in section \ref{sec:frans}) to couette, channel and pipe flow.

\subsection{Turbulent couette flow}

Couette flow is characterised by flow between two infinitely long parallel plates \cite{avsarkisov2014turbulent}, where one of them is moving with a constant velocity in the streamwise direction ($x^+$), while the other is at rest with zero pressure gradient. This implies that there is only a finite stream-wise component of velocity ($\overline{U^+}$) and wall-normal ($y^+$) component of velocity ($\overline{V^+}$) is infinitesimally small. The spanwise direction ($z^+$) is treated as homogeneous with little or no-variation in the flow quantities, thus mean spanwise velocity component ($\overline{W^+}$) and all gradients in spanwise direction vanish. After simplifying the RANS equations (\ref{eq:rans}), the governing equation for the turbulent couette flow is given as follows (\ref{eq2.3}).

\begin{align}\label{eq2.3} 
\frac{d} {d y^{+}} \left( \frac{d \overline{U^{+}}} {d y^{+}} - (\overline{uv})^{+} \right) = 0.
\end{align}

\noindent
Here, $\overline{U^+}$ is a function of $y^+$ due to $\frac{\partial \overline{U^+}}{\partial x^+}=0$. Upon integrating (\ref{eq2.3}) with respect to $y^+$, produces (\ref{eq2.4}).

\begin{equation}\label{eq2.4}
  \frac{d \overline{U^{+}}} {d y^{+}} - (\overline{uv})^{+} = C.
\end{equation}

The constant, $C$ is evaluated at the wall (within the viscous sub-layer), where the Reynolds stresses are negligible, namely, $(\overline{uv})^+|_{y^+=0} \approx 0$, and $\overline{U^+} \approx y^+$. Thus, we have $C=(d \overline{U^{+}}/{d y^+})|_{y^+=0} = \tau_w^+=1$. Finally, the simplified RANS equation for the turbulent couette flow is written as (\ref{eq2.5}).

\begin{align} \label{eq2.5}
   \frac{d \overline{U^{+}}} {d y^{+}} - (\overline{uv})^{+} = 1.
\end{align}

The f-RANS one-sided model is defined over the domain $[0, y^+]$, where $ y^+ \in (0 , Re_{\tau}]$ is given as (\ref{eq2.6}), while the two-sided model (\ref{eq2.7}) is defined over the domain $[0, 2Re_\tau]$, where $ y^+ \in (0 ,2Re_{\tau})$ for turbulent couette flow. For numerical computation of the fractional order ($\alpha(y^+)$) we employ the DNS database generated in \cite{avsarkisov2014turbulent} for velocity ($\overline{U^+}$) for each friction Reynolds number.

\begin{equation}  \label{eq2.6}
 ^C_{0}D_{y^+}^{\alpha(y^+)} \overline{U^+} ~=~ \frac{d \overline{U^{+}}} {d y^{+}} - (\overline{uv})^{+} ~=~  1 ~;~ \alpha(y^+) \in (0, 1]
\end{equation}

\begin{equation}  \label{eq2.7}
 {_{[0, ~2Re_\tau]}} ^T D_{y^+}^{\alpha(y^+)} \overline{U^+} ~=~ \frac{d \overline{U^{+}}} {d y^{+}} - (\overline{uv})^{+} ~=~   1 ~;~ \alpha(y^+) \in (0, 1]
\end{equation}

\subsection{Turbulent channel flow}

Channel flow is also flow between two infinity long parallel plates, unlike the couette flow, both the plates remain at rest and flow is driven by streamwise pressure gradient \cite{lee_moser_2015}. All the other remarks remain similar as the couette for simplifying the RANS equation (\ref{eq:rans}), thus the governing equation for the turbulent channel flow is given by (\ref{eq2.3:chal}),

\begin{align}\label{eq2.3:chal} 
\frac{d} {d y^{+}} \left( \frac{d \overline{U^{+}}} {d y^{+}} - (\overline{uv})^{+} \right) = -\frac{\partial \overline{P^{+}}} {\partial x^{+}}.
\end{align}
Here, $\overline{U^+}$ is a function of $y^+$ due to $\frac{\partial \overline{U^+}}{\partial x^+}=0$. Further integrating (\ref{eq2.3:chal}) with respect to $y^+$, produces (\ref{eq2.4:chal})

\begin{equation}\label{eq2.4:chal} 
  \frac{d \overline{U^{+}}} {d y^{+}} - (\overline{uv})^{+} = -\frac{\partial \overline{P^{+}}} {\partial x^{+}} y^+ + C
\end{equation}

The constant $C$ is evaluated at the wall (within the viscous sub-layer), where the Reynolds stresses are negligible, namely, $(\overline{uv})^+|_{y^+=0} \approx 0$, and $\overline{U^+} \approx y^+$. Thus, we have $C=(d \overline{U^{+}}/{d y^+})|_{y^+=0} = \tau_w^+=1$.
Furthermore, at the center line ($y^+ = Re_\tau$), as the flow is turbulent $(d \overline{U^{+}}/{d y^+})|_{y^+=Re_\tau} = 0$. As a consequence of symmetry, we have, $\overline{uv^+} = 0$ ,  this produces $\frac{\partial \overline{P^{+}}} {\partial x^{+}} = {1 / Re_\tau} $, thus the RANS equation for turbulent Channel flow is given by (\ref{eq2.5:chal})
\begin{align} \label{eq2.5:chal}
   \frac{d \overline{U^{+}}} {d y^{+}} - (\overline{uv})^{+} = -{y^+ \over Re_\tau} + 1
\end{align}

The f-RANS one-sided model is defined over the domain $[0, y^+]$, where $ y^+ \in (0 ,Re_{\tau}]$ is given as (\ref{eq2.6:chal}), while the two-sided model (\ref{eq2.7:chal}) is defined over the domain $[0, 2Re_\tau]$, where $ y^+ \in (0 ,2Re_{\tau})$ for turbulent channel flow. For numerical computation of the fractional order ($\alpha(y^+)$) we employ the DNS database generated in \cite{lee_moser_2015} for velocity ($\overline{U^+}$) for each friction Reynolds number.

\begin{equation}  \label{eq2.6:chal}
  ^C_{0}D_{y^+}^{\alpha(y^+)} \overline{U^+} ~=~ \frac{d \overline{U^{+}}} {d y^{+}} - (\overline{uv})^{+} ~=~   -{y^+ \over Re_\tau} + 1 ~;~ \alpha(y^+) \in (0, 1]
\end{equation}

\begin{equation}  \label{eq2.7:chal}
  {_{[0, ~2Re_\tau]}} ^T D_{y^+}^{\alpha(y^+)} \overline{U^+} ~=~ \frac{d \overline{U^{+}}} {d y^{+}} - (\overline{uv})^{+} =   -{y^+ \over Re_\tau} + 1 ~;~ \alpha(y^+) \in (0, 1]
\end{equation}

\subsection{Turbulent pipe flow}

The arguments for the pipe flow \cite{wu2008direct} follows that of the channel, however we use the transformed equations in cylindrical co-ordinates \cite{batchelor_2000}. Since, the azimuthal direction for this case is regarded as the homogeneous direction besides the stream-wise direction. Thus, all dependence for the pipe follows a radial direction. The stream-wise pressure gradient is also non-zero in this case. Thus the simplified RANS equation for turbulent pipe flow is given in (\ref{eq2.3:pipe}),

\begin{align}\label{eq2.3:pipe} 
\frac{d} {d r^{+}} \left( \frac{d \overline{U_z^{+}}} {d r^{+}} - (\overline{u_r u_z})^{+} \right) = -\frac{d \overline{P^{+}}} {d z^{+}}.
\end{align}

\noindent
Here, $\overline{U_z^{+}}$ is a function of the radial direction, $r^+$. Upon Integrating (\ref{eq2.3:pipe})  with respect to $r^+$ produces (\ref{eq2.4:pipe}),

\begin{align}\label{eq2.4:pipe} 
  \frac{d \overline{U_z^{+}}} {d r^{+}} - (\overline{u_r u_z})^{+}  = -\frac{d \overline{P^{+}}} {d z^{+}} r^+ + C.
\end{align}

At the centre of the pipe ($r^+ = 0$), we have $ d\overline{U_z^{+}}/ {d r^{+}} \approx 0 $ as the flow is dominated by turbulence. As a result of symmetry we have, $\overline{u_r u_z}^{+} = 0$, this results in $C = 0$. Furthermore, near the wall, we have $ d\overline{U_z^{+}}/ {d r^{+}} = \tau_w  =1$, since the flow is dominated by the viscous interactions, we have $\overline{u_r u_z}^{+} \approx 0$, thus (\ref{eq2.5:pipe}), where $R^+$ is the Karman number, which plays the similar role as the friction Reynolds number ($Re_\tau $)

\begin{align}\label{eq2.5:pipe} 
  \frac{d \overline{P^{+}}} {d z^{+}} = {1 \over R^+}
\end{align}

Furthermore, we express the RANS equation for turbulent pipe flow from the wall to the center of the pipe by $ r^+ = ((1- r)/R)^+$, where $r \in [0, R]$, with the radius of the pipe, $R = 1$, thus we have (\ref{eq2.6:pipe}). Upon examining, the right hand-side the similarity with Channel flow is obvious.

%Upon examining, the right hand-side the similarity with Channel in the preceding section is obvious where the equivalence can be seen explicitly by $ r^+ = R^+ - r^+$.

\begin{align} %\label{eq2.6:pipe} 
  \frac{d \overline{U_z^{+}}} {d ( 1- r)^+} - \overline{u^+_{(1-r)^+}  u^+_z}  = -{(1 - r)^+ \over R^+} 
\end{align}

The f-RANS one-sided model is defined over the domain $[0, (1-r)^+]$, where $ (1-r)^+ \in (0 ,R^+]$ is given as (\ref{eq2.6:pipe}), while the two-sided model (\ref{eq2.7:pipe}) is defined over the domain $[0, 2R^+]$, where $ (1-r)^+ \in (0 ,2R^+)$ for turbulent pipe flow. For numerical computation of the fractional order ($\alpha((1-r)^+)$) we employ the DNS database generated in \cite{el2013direct} for velocity ($\overline{U_z^+}$) for each Karman number.

\begin{align}  \label{eq2.6:pipe}
\begin{split}
^C_{0}D_{(1-r)^+}^{\alpha((1-r)^+)} \overline{U_z^+} ~&=~ \frac{d \overline{U_z^{+}}} {d ( 1- r)^+} - \overline{u^+_{(1-r)^+}  u^+_z}  \\ &=~ -{(1 - r)^+ \over R^+}  ~~;~ \alpha((1-r)^+) \in (0, 1]
 \end{split}
\end{align}

\begin{align}  \label{eq2.7:pipe}
\begin{split}
 {_{[0, ~2R^+]}} ^T D_{(1-r)^+}^{\alpha((1-r)^+)} \overline{U_z^+} ~&=~ \frac{d \overline{U_z^{+}}} {d ( 1- r)^+} - \overline{u^+_{(1-r)^+}  u^+_z}  \\ &=~  -{(1 - r)^+ \over R^+}  ~~;~ \alpha((1-r)^+) \in (0, 1]
 \end{split}
\end{align}

%%%%%%%%%%%%%%%%%%%%%%%%%%%%%%%%%%%%%%%%%%%%%%%%%%%%%%%%%%%%%%%%%%%%%%%%%%%%%%%%%%%%%%%%%%%%%%%%%%%%%%%%%%%%%%%5

\section{Numerical Scheme: Physics-informed Neural Network} \label{sec:fpinns}

In this section we outline the numerical implementation using Physics-informed Neural Networks (PINNs). PINNs are deep neural network where the loss comprises of two parts, namely, data and equations along with weights, $\lambda$ (hyperparameter) to balance the two components of the loss \cite{raissi2019physics} \cite{fpinns}, to accelerate the convergence, as shown in the expression below (\ref{eq:loss}). 
\begin{equation} \label{eq:loss}
    L = L_d + \lambda L_e
\end{equation}
The term $L_d$ is the data part, where we supply the boundary and initial conditions, while the term $L_e$ is for the governing equations \cite{raissi2019physics}, where in this case we supply fractional models. The input of the neural network are spatio-temporal points. Furthermore, we employ adaptive activation functions to further accelerate convergence substantially and avoid bad minima~\cite{Ameya_AAF}.

\subsection{Inverse problem: computing fractional order (pointwise)}

For the inverse problem the goal is to find the fractional order ($\alpha(y^+)$) given the velocity profile and total shear stress from DNS database or otherwise as outlined in the preceding section for each case. Here we developed a pointwise strategy, this implies that we train the neural network for a single training point at a time. Naturally, the neural network is much smaller consisting of at most two to four layers and two to three neurons per hidden layer. 

\begin{figure}
\centering
\includegraphics[width=0.9\textwidth]{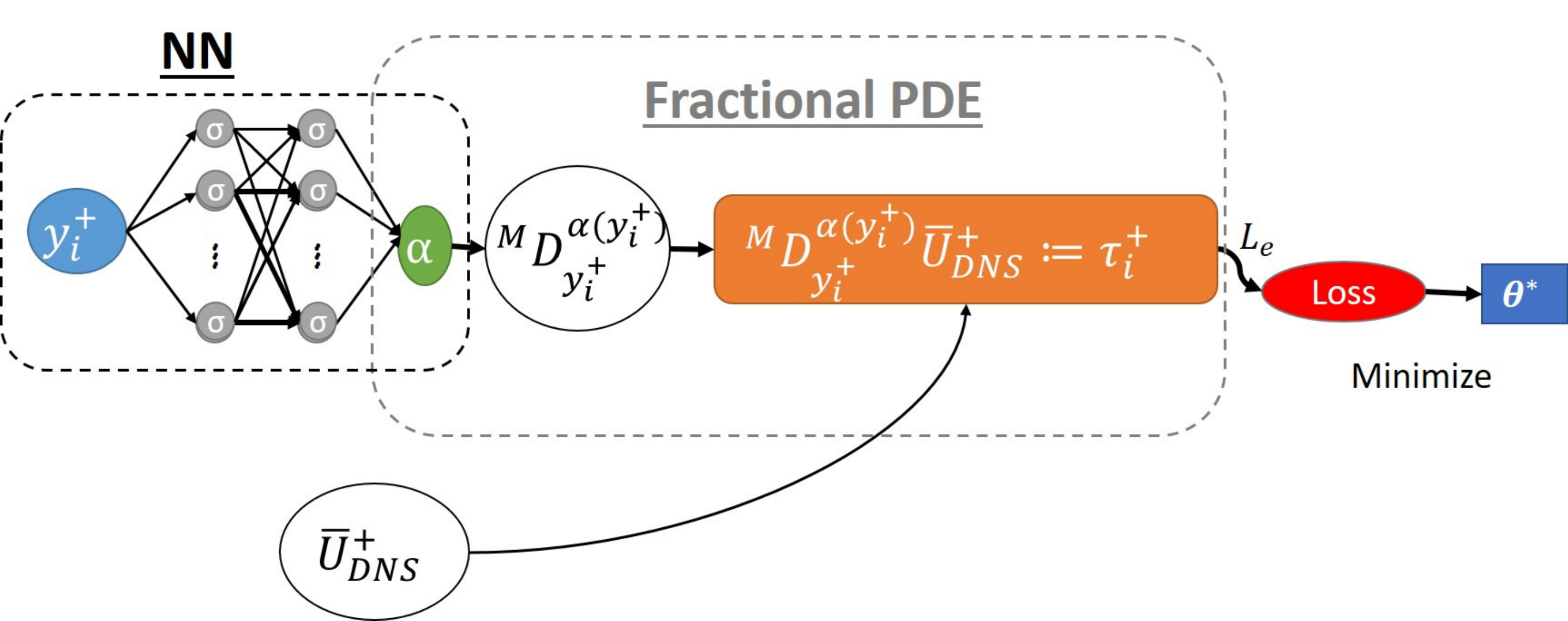}%
\caption{A schematic representation for the inverse modeling, here the spatial location ($y^+$) is the input of the feed forward neural network, while the fractional order ($\alpha$) is the output, which is used to computed the loss function which comprises of f-RANS model using velocity from DNS databases}
\label{fig:fpinns}
\end{figure}

The input of feed forward neural network is single location $y^+_i$ (only one training point), where the output is the fractional order at that location ($y^+_i$). The loss, $L$ (\ref{eq:loss}) only comprises of the equation term ($L_e$) given in (\ref{eq:loss_e_al}) written for the $i^{th}$ training point. Recall that $^M D_{y^+_i}^{\alpha_{NN}(y^+_i)} (\overline{U^+_{DNS}})$ implies either a one-sided or a two-sided f-RANS model, depending upon the case. A schematic diagram is shown in fig.~\ref{fig:fpinns}. A hyper-parameter ($\lambda$) may be added to normalize too high or low numbers, however, it does not affect the training process, since it depends on the gradient of the loss rather than its absolute value.  

\begin{equation} \label{eq:loss_e_al}
    Loss = L_e = \left[ ^M D_{y^+_i}^{\alpha_{NN}(y^+_i)} (\overline{U^+_{DNS}}) - \tau^+(y^+_i) \right ]^2 
\end{equation}

For numerical discretisation  of the fractional derivative we employ an {\it L1} scheme \cite{yang2010numerical} for the left- (\ref{eq:left_unn}) and right-sided (\ref{eq:right_unn}) derivative. A uniform finite difference grid is generated by dividing the domain in $M$ points such that the step size, $h_l$ was kept constant for all the training points, where, $l = \{1, 2, ..., M-1 \}$. It is to be noted that the domain for one-sided model is $[0, y^+], y^+ \in (0, Re_\tau]$, while the domain for the two-sided model is $[0, 2Re_\tau], y^+ \in (0, 2Re_\tau)$. Furthermore, we exploit the fact that the Riemann–Liouville and Caputo fractional derivative varies in one term for the fractional order, $\alpha \in (0,1]$.

\begin{align}  \label{eq:left_unn}
\begin{split}
    &^C_0 D_{y^+}^{\alpha_{NN}(y^+_i)} (\overline{U^+)} \approx \\ & -{h^{-\alpha_{NN}(y^+_i)} \over \Gamma(2 - {\alpha_{NN}(y^+_i)})}  \sum_{j=0}^{l-1} (\overline{U^+}_{{l-j}} -  \overline{U^+}_{l-j-1}) [ (j+1)^{1-{\alpha_{NN}(y^+_i)}} -  j^{1-{\alpha_{NN}(y^+_i)}} ]
\end{split}
\end{align}

\begin{align} \label{eq:right_unn}
\begin{split}
    &^C_{y^+} D_{2Re_\tau}^{{\alpha_{NN}(y^+_i)}} (\overline{U^+)} \approx \\& -{h^{{-\alpha_{NN}(y^+_i)}} \over \Gamma(2 - {\alpha_{NN}(y^+_i)})} \sum_{j=0}^{M-l-1} (\overline{U^+}_{{l+j}} -  \overline{U^+}_{{l+j+1}}) [ (j+1)^{1-{\alpha_{NN}(y^+_i)}} -  j^{1-{\alpha_{NN}(y^+_i)}} ]
\end{split}
\end{align}

\noindent
The merit of pointwise strategy are summarised as follows:

\begin{itemize}
    \item Given that the fractional order exists, it does not require boundary conditions, which earlier required to impose unity at the wall \cite{mehta2019discovering}. Thus for more complex flows where the near wall region is not viscous dominated, we can find a physically consistent fractional order.
    \item There are cases, where the fractional order is a discontinuous function or the neural network collapses during training process; a pointwise algorithm mitigates all the limitations. However, for solving for the mean velocity, we rarely encounter any problems during its training, since, the mean velocity is smooth and continuous function as demonstrated in \cite{mehta2019discovering} for one-sided f-RANS model.
\end{itemize}

%%%%%%%%%%%% Section 5 %%%%%%%%%%%%%%%%%%%%%%%%%%%%%%%%%%%%%%%%%%%%%%%

\section{Results : fractional order of one-sided f-RANS model applied to turbulent couette, channel and pipe flow} \label{sec:one-sided}

In this section we describe our findings for one-sided model applied to turbulent couette, channel and pipe flow. 

%For all three flows, in the viscous sub-layer (near wall region) the mean velocity follows the scaling, $\overline{U^+} \approx y^+$. While away from the wall the mean velocity follows a logarithmic law, where, $\overline{U^+} \approx \kappa ~log (y^+)$, also known as the log-law of the wall. In between, these two regions is the buffer layer. Beyond, the  logarithmic region is a wake region, where mean velocity no longer flows in a logarithmic fashion. Following Coles \cite{coles_1956}, a wake function can be introduced to capture the departure of mean velocity from log-law, it is now known as, law of the wake. All these regions contributes to our fundamental understanding of turbulence, and modeler's often exploit these scaling features to build a model. 

\begin{figure} [!htb]
\centering
    \subfloat[Couette]{{\includegraphics[trim=0.3cm 0.3cm 0.3cm 0.3cm,clip,    width=0.48\textwidth]{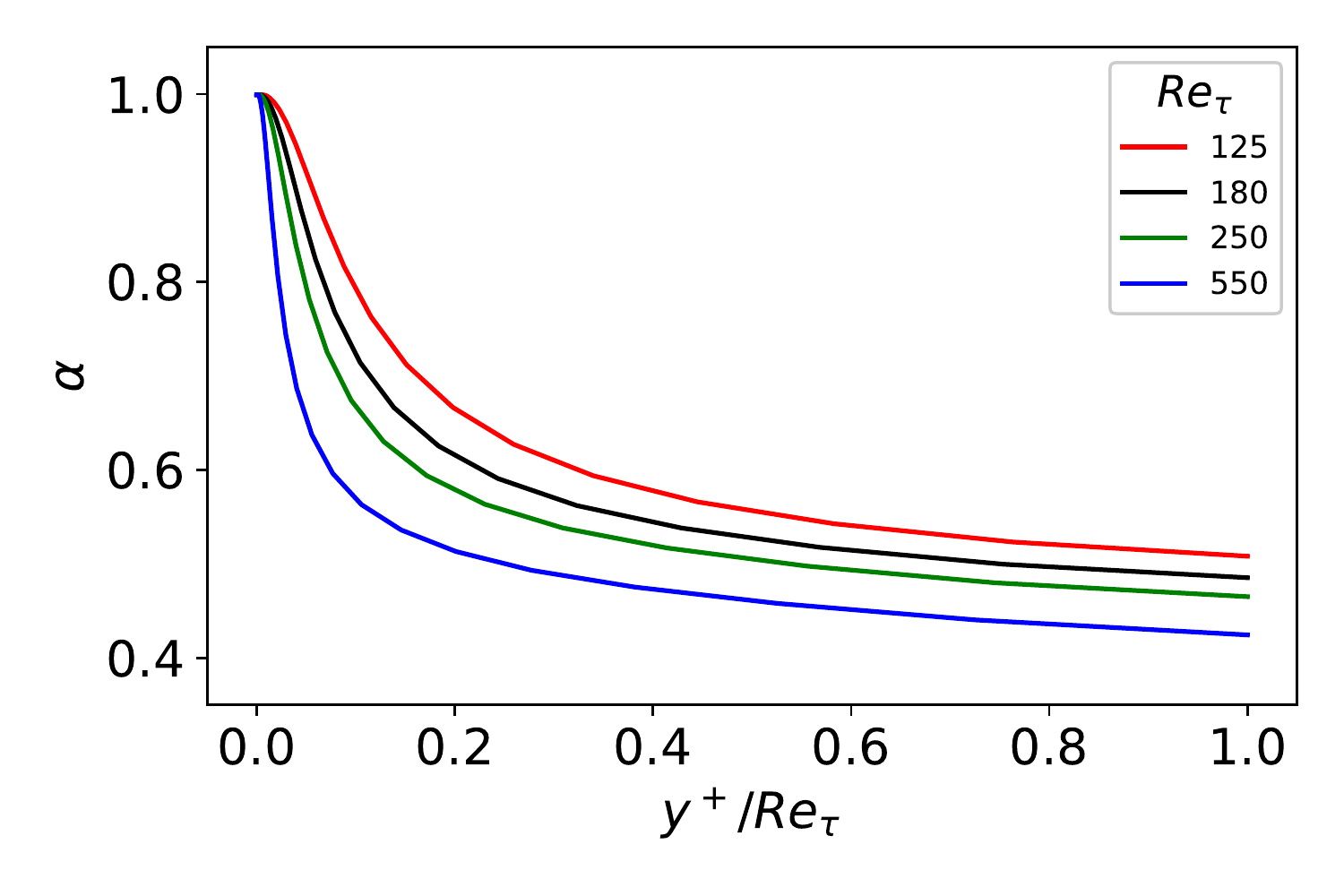}}}
   \subfloat[Channel]{{\includegraphics[trim=0.3cm 0.3cm 0.3cm 0.3cm,clip,    width=0.48\textwidth]{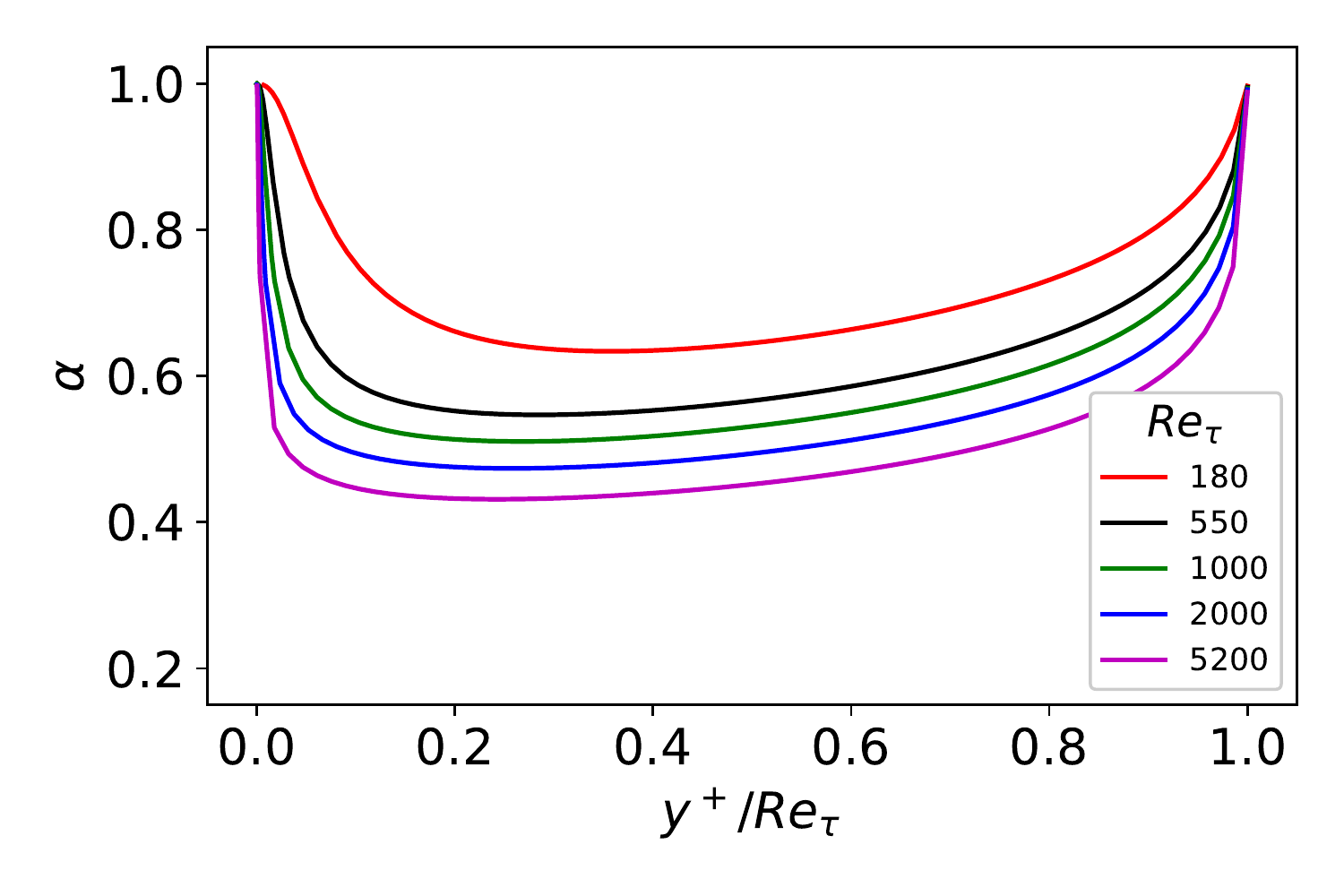}}} \\
   \subfloat[Pipe]{{\includegraphics[trim=0.3cm 0.3cm 0.3cm 0.3cm,clip,    width=0.48\textwidth]{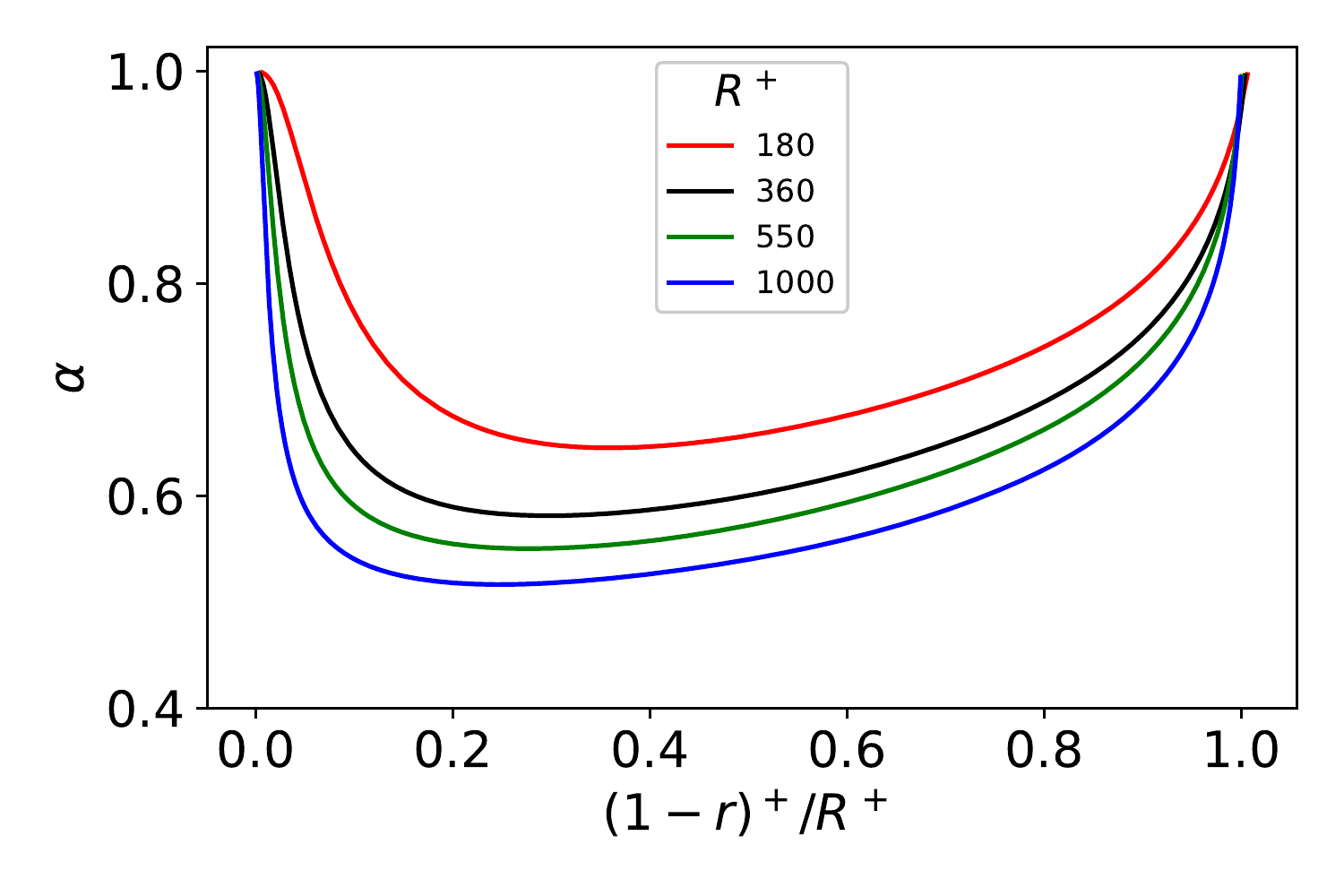}}}%
    \caption{Fractional order of one-sided f-RANS model for channel, couette and pipe flow. Here the x-axis of the plot: channel and couette: $y^+/Re_\tau$, pipe: $(1-r)^+/R^+$. It is observed as the Reynolds / Karman number increases the fractional order lowers corresponding to higher turbulence intensity. The channel and pipe show an artifact due to symmetry, where a fractional order if unity is not a physical solution at the center-line, merely a numerical solution of our constructed model}
    \label{fig:frac_ydel_1sided}
\end{figure}

\begin{figure} [!htb]
\centering
    \subfloat[Couette]{{\includegraphics[trim=0.3cm 0.3cm 0.3cm 0.3cm,clip, width=0.48\textwidth]{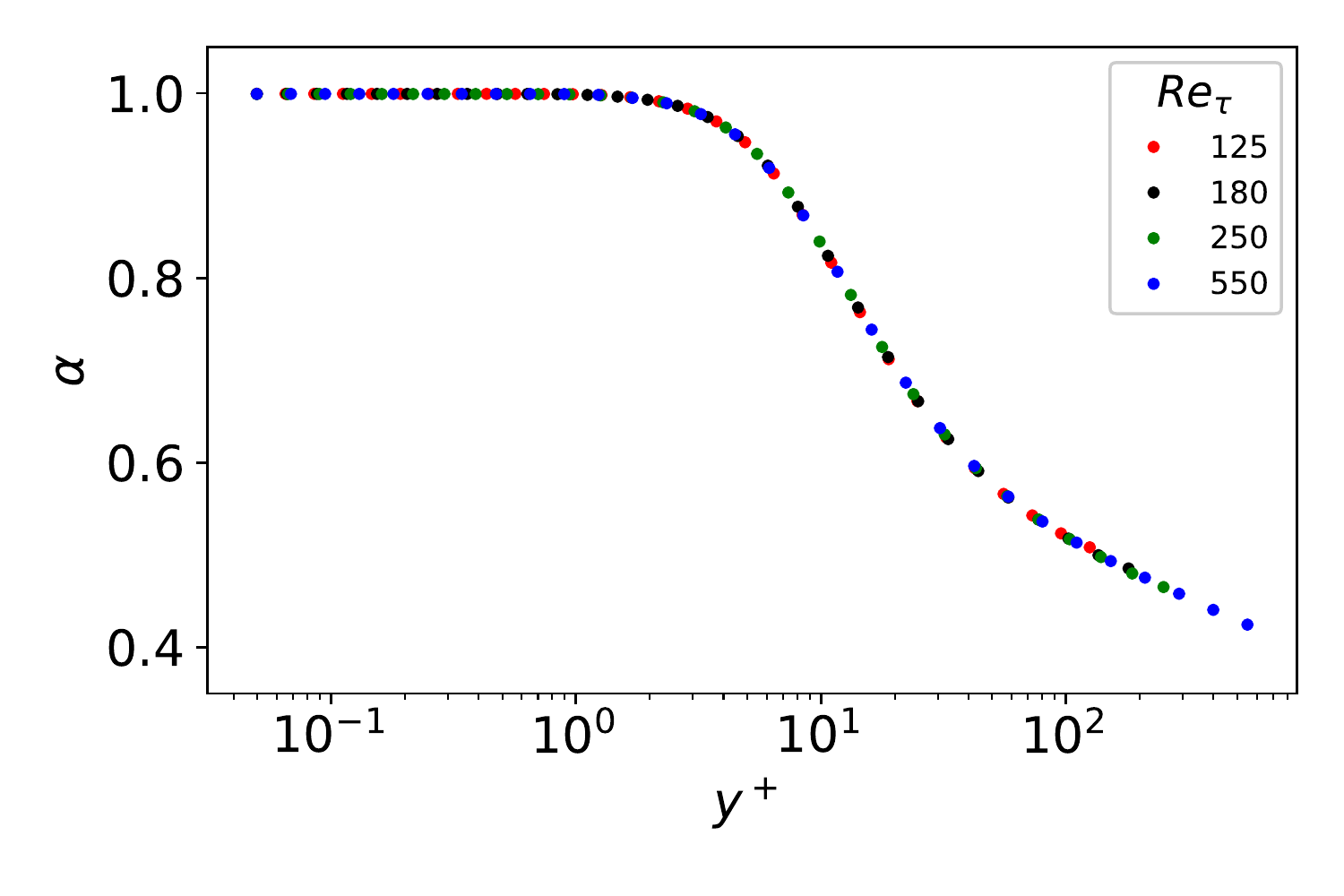}}}%
   \subfloat[Channel]{{\includegraphics[trim=0.3cm 0.3cm 0.3cm 0.3cm,clip,    width=0.48\textwidth]{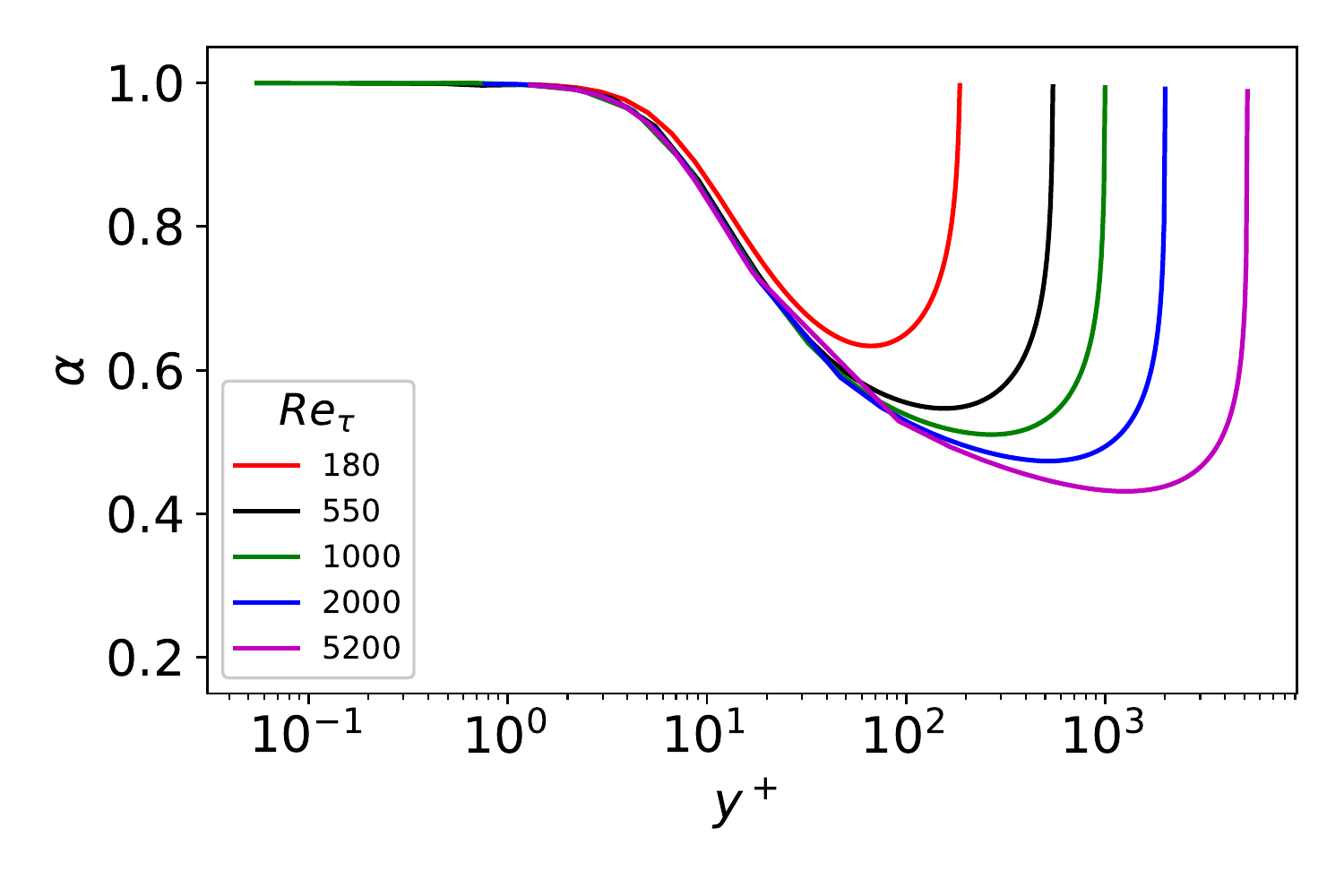}}} \\
   \subfloat[Pipe]{{\includegraphics[trim=0.3cm 0.3cm 0.3cm 0.3cm,clip,    width=0.48\textwidth]{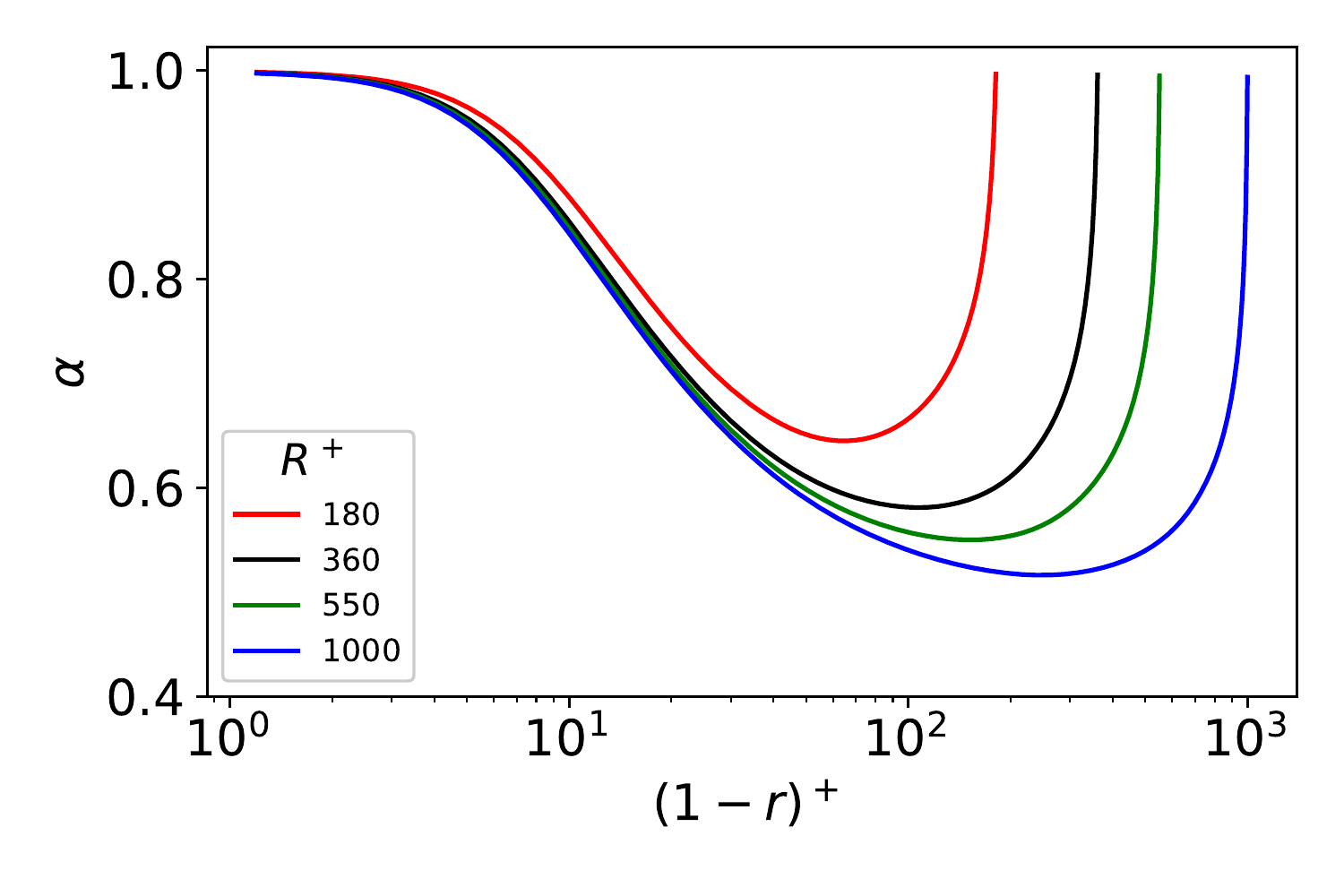}}}%
    \caption{Fractional order of one-sided f-RANS model for channel, couette and pipe flow. Here the x-axis of the plot: channel and couette: $y^+$, pipe: $(1-r)^+$. Remarkably, the fractional order for couette flow shows universality. The channel and pipe shows an artifact due to symmetry, exposing the limitation of the one-sided model. For all three cases, the the fractional order is universal/overlaps in the viscous sub-layer and to some extent the buffer layer.}
    \label{fig:frac_yplus_1sided}
\end{figure}

Turbulence is the cause of non-locality, a super-diffusion process governed by the fractional derivatives, while pure molecular diffusion (absence of turbulence) is a diffusion process which is entirely governed by the Laplacian (integer-order). 

In the viscous sub-layer (near wall region), the flow is dominated by the viscous action and Reynolds stress are negligible. Thus, in this viscous dominated region, the fractional order is unity implying that the flow is entirely governed by local operators, also confirmed by computation of our fractional order for the one-sided model shown in fig.~\ref{fig:frac_yplus_1sided}. 

Indeed, as we move from the viscous sub-layer onto the buffer layer, and all the way to outer region, the non-locality gradually increases (as turbulence intensity increases) indicated by lowering of the fractional order as seen in fig.~\ref{fig:frac_ydel_1sided}(a) and ~\ref{fig:frac_yplus_1sided}(a). 

Furthermore, across the Reynolds number or Karman number the turbulence intensity varies too. For higher Reynolds number or Karman number has higher turbulence intensity. Thus more consideration to non-locality is given by lower fractional order across the Reynolds number or Karman number (infer fig.~\ref{fig:frac_ydel_1sided}).

In fig.~\ref{fig:frac_ydel_1sided} (b, c) and fig.~\ref{fig:frac_yplus_1sided} (b, c) the channel and pipe flow shows an anomaly where at the center-line the fractional order is unity implying a local region. On the contrary, the non-locality is the strongest at the center-line. The reason for this anomaly is due to symmetry, here the Reynolds stress's value is identically equal to zero as it changes sign. Thus the construction of our one-sided model falsely interprets as a viscous dominated region. Indeed the fractional order unity is a numerical solution of our one-sided model at the center-line and produces the correct total shear stress value, but physically is absurd. Motivated by this anomaly we formulated the two-sided model, which will be the subject of discussion of the next section. 

It is worth noting that the one-sided f-RANS model for couette flow shows universality as seen in fig.~\ref{fig:frac_yplus_1sided} (a) (see also \cite{mehta2019discovering}), while the other two cases do not show any universality beyond the buffer layer.

%%%%%%%%%%%%%%%%%%%%%%%%%%%%%%%%%%%%%%%%%%%%%%%%%%%%%%%%%%%%%%%%%%%%%%%%%%%%%%%%%%%%%%%%%%%%%%%%%%%%%%%%%%%%%%%%%%%%%%%%%%%%%%%%%%%%%%%%%%%%%%%%%%%%%%%%%%%%%%%%%%%%%%%%%%%%%%%%%%%%%%%%%%%%%%%%%%%%%%%%%%%%%%%%%%%%%%%%%%%%%%%%%%%%%%%%%%%%%%%%%%%%%%%%%%%%%%%%%%%%%%%%%%%%%%%%%%%%%%%%%%%%%%%%%%%%%%%%%%%%%%%%%%%%%%%%%%%%%%%

\section{Results : fractional order of two-sided f-RANS model applied to turbulent couette, channel and pipe flow} \label{sec:two-sdied}

Non-locality at a given point is an aggregate effect of all directions. By this consideration, we formulated the two-sided model. Indeed, for the case of two-sided model, like the one-sided model, the fractional order is unity within the viscous-dominated regions and lower fractional order corresponding to higher non-locality as a result of higher turbulence intensity.

\begin{figure} [!htb]
\centering
    \subfloat[Couette]{{\includegraphics[trim=0.3cm 0.3cm 0.3cm 0.3cm,clip, width=0.48\textwidth]{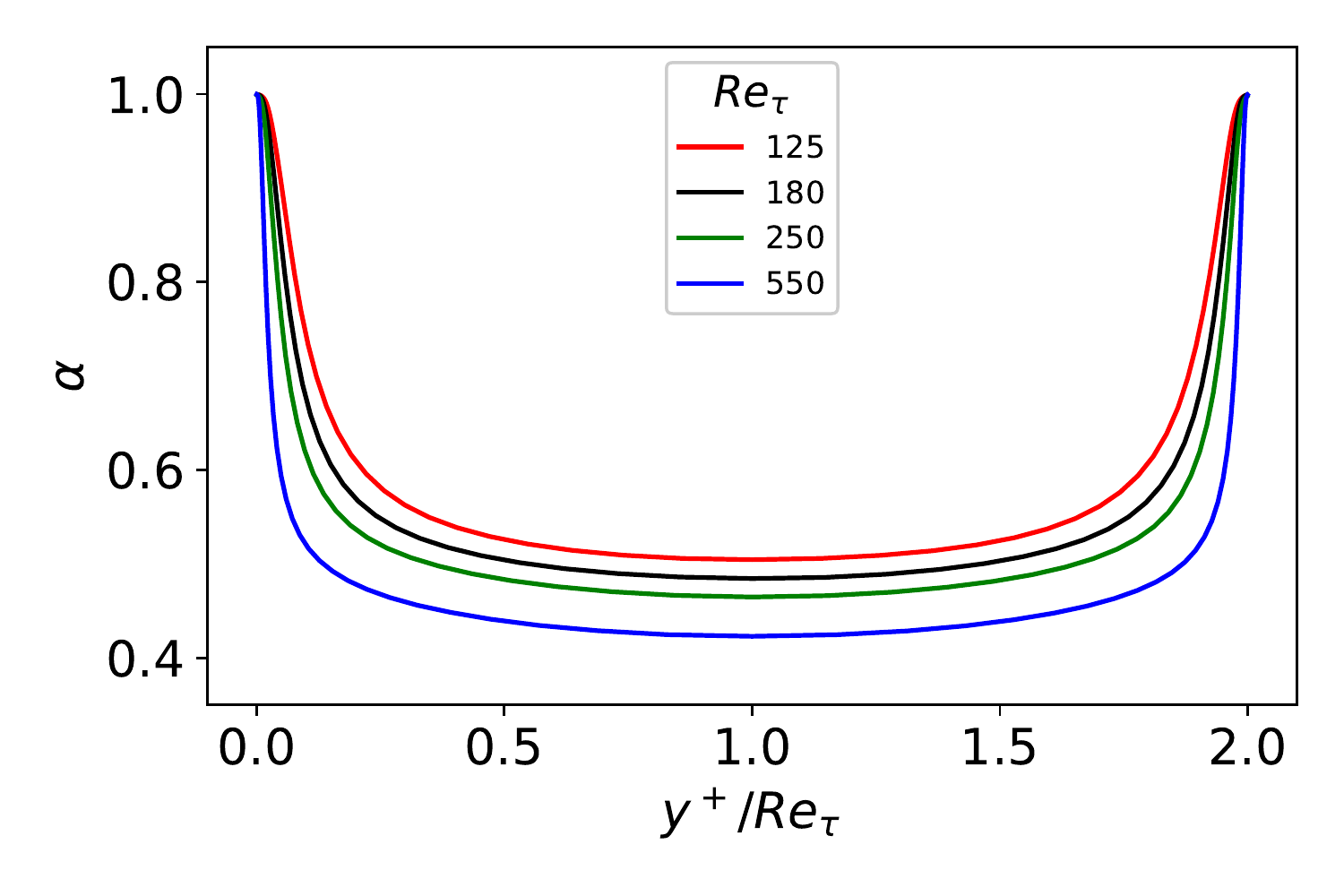}}}%
   \subfloat[Channel]{{\includegraphics[trim=0.3cm 0.3cm 0.3cm 0.3cm,clip,    width=0.48\textwidth]{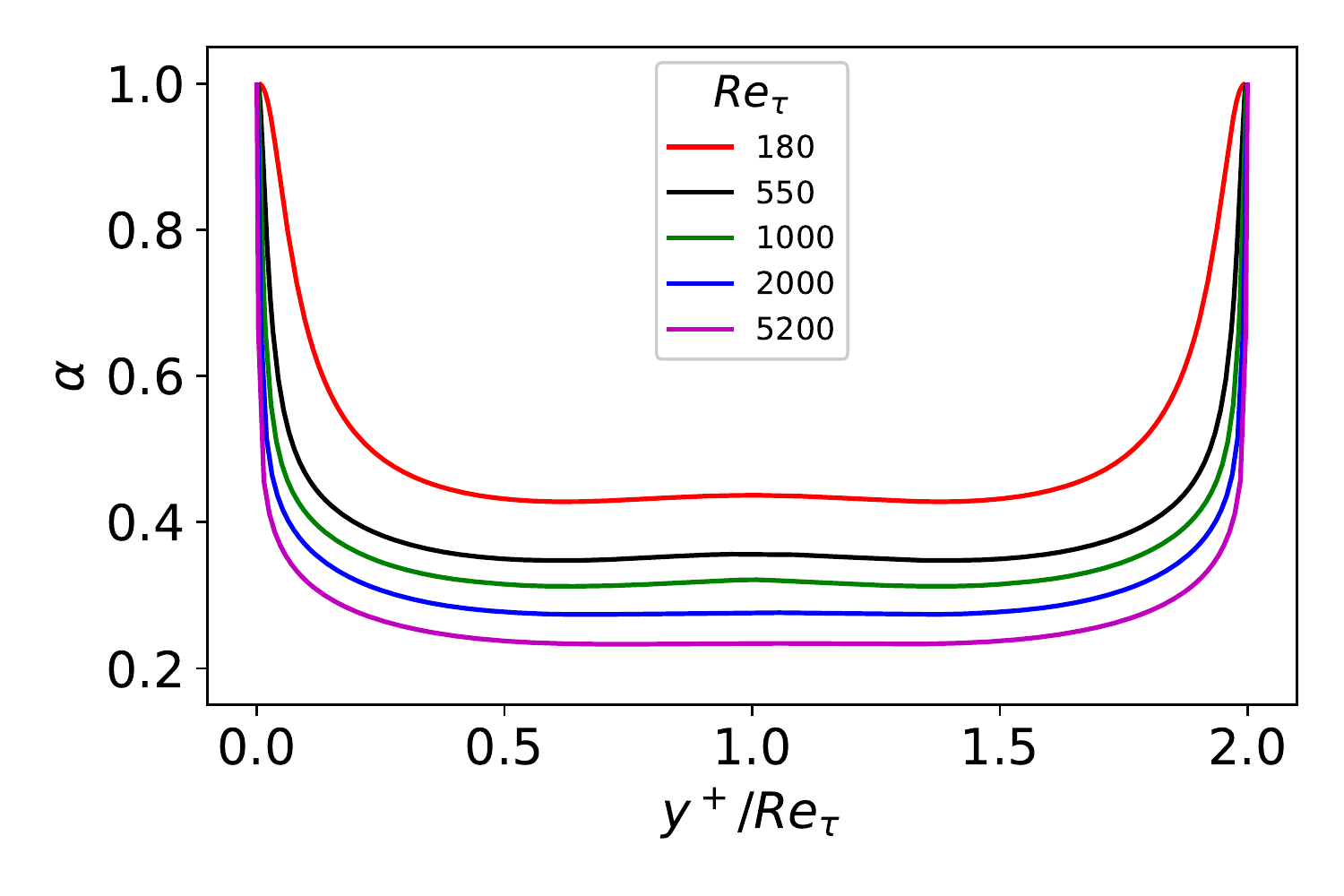}}} \\
   \subfloat[Pipe]{{\includegraphics[trim=0.3cm 0.3cm 0.3cm 0.3cm,clip,    width=0.48\textwidth]{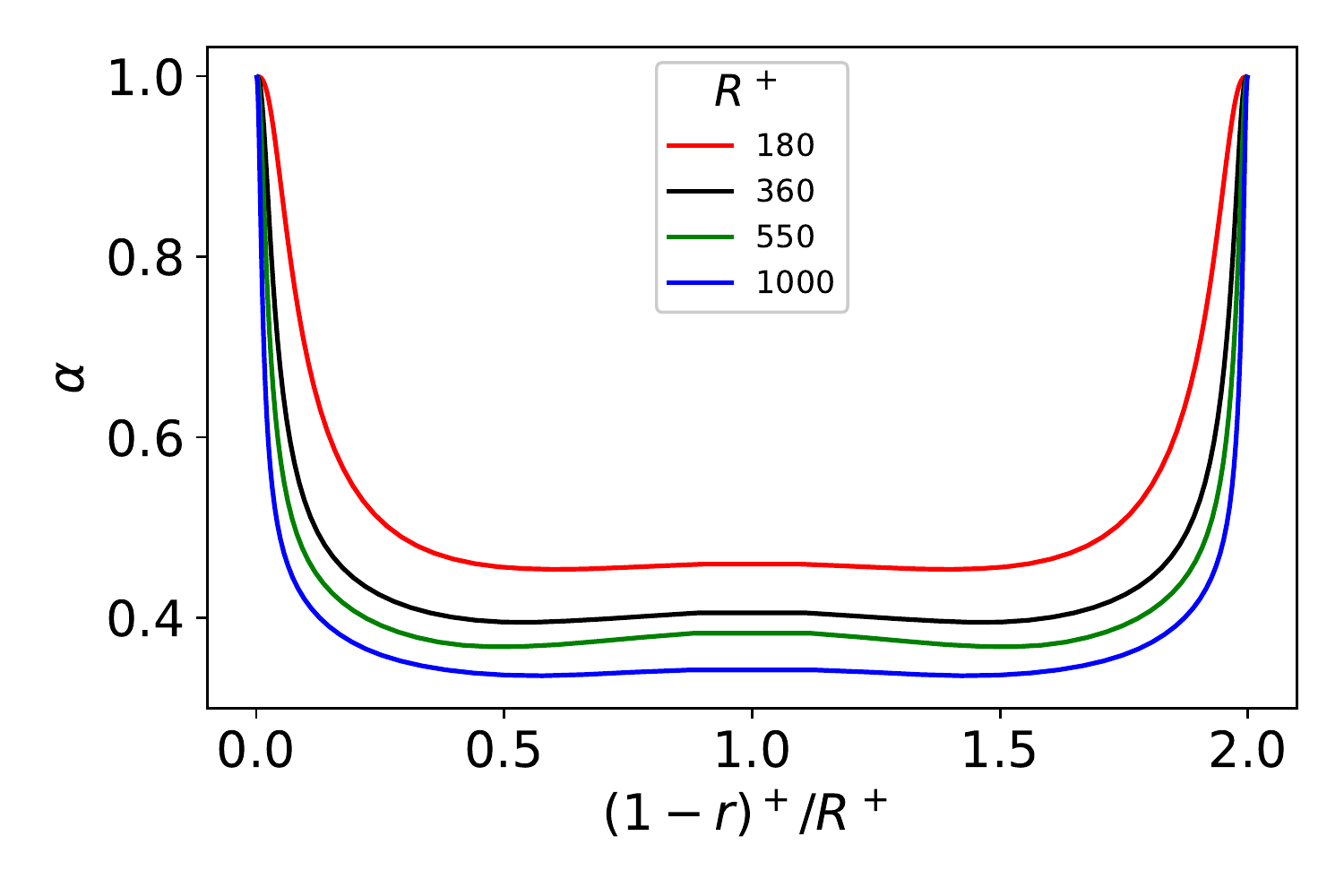}}}%
    \caption{Fractional order of two-sided f-RANS model for channel, couette and pipe flow. Here the x-axis of the plot: channel and couette: $y^+/Re_\tau$, pipe: $(1-r)^+/R^+$. It is observed as the Reynolds / Karman number increases the fractional order lowers corresponding to higher turbulence intensity. The artifact due symmetry of one-sided model is no longer present in this case, as the non-locality is considered in a physical manner.}
    \label{fig:frac_ydel_2sided}
\end{figure}

Although, in fig.~\ref{fig:frac_ydel_2sided} (b, c) the two-sided model seems to have solved the anomaly of one-sided model in fig.~\ref{fig:frac_ydel_1sided} (b, c) for the channel and pipe flow. The symmetry of the channel and pipe flow has led to another anomaly (a minor one), where any value of $\alpha \in (0,1]$ is a numerical solution at the center-line by construction of the two-sided model. This anomaly is only present at the center-line and not observed in immediate neighbourhood, thus we report the mean value computed from its immediate neighbourhood of the center-line, which we argue it is a physical result, than the anomaly observed in the one-sided model (refer fig.~\ref{fig:frac_ydel_1sided} (b, c) and fig.~\ref{fig:frac_yplus_1sided} (b, c)). The case of couette in fig.~\ref{fig:frac_ydel_2sided} (a) supports our argument, where we have physically consistent results. To reiterate, the physically consistency argument, it needs to have the following features, 

\begin{itemize}
    \item In the viscous dominated region (and absence of turbulence), the fractional order has to be unity (governed by local operators).
    
    \item Onset of the buffer layer, the fractional order should gradually decrease corresponding to the increase in turbulence intensity.
    
    \item The effect of turbulence intensity should also be seen across Reynolds or Karman number.
        
\end{itemize}

All these features can be inferred in both fig.~\ref{fig:frac_ydel_2sided} and fig.~\ref{fig:frac_yplus_2sided}, thus the two-sided model is physically consistent.  

\begin{figure} [!htb]
\centering
    \subfloat[Couette]{{\includegraphics[trim=0.3cm 0.3cm 0.3cm 0.3cm,clip, width=0.48\textwidth]{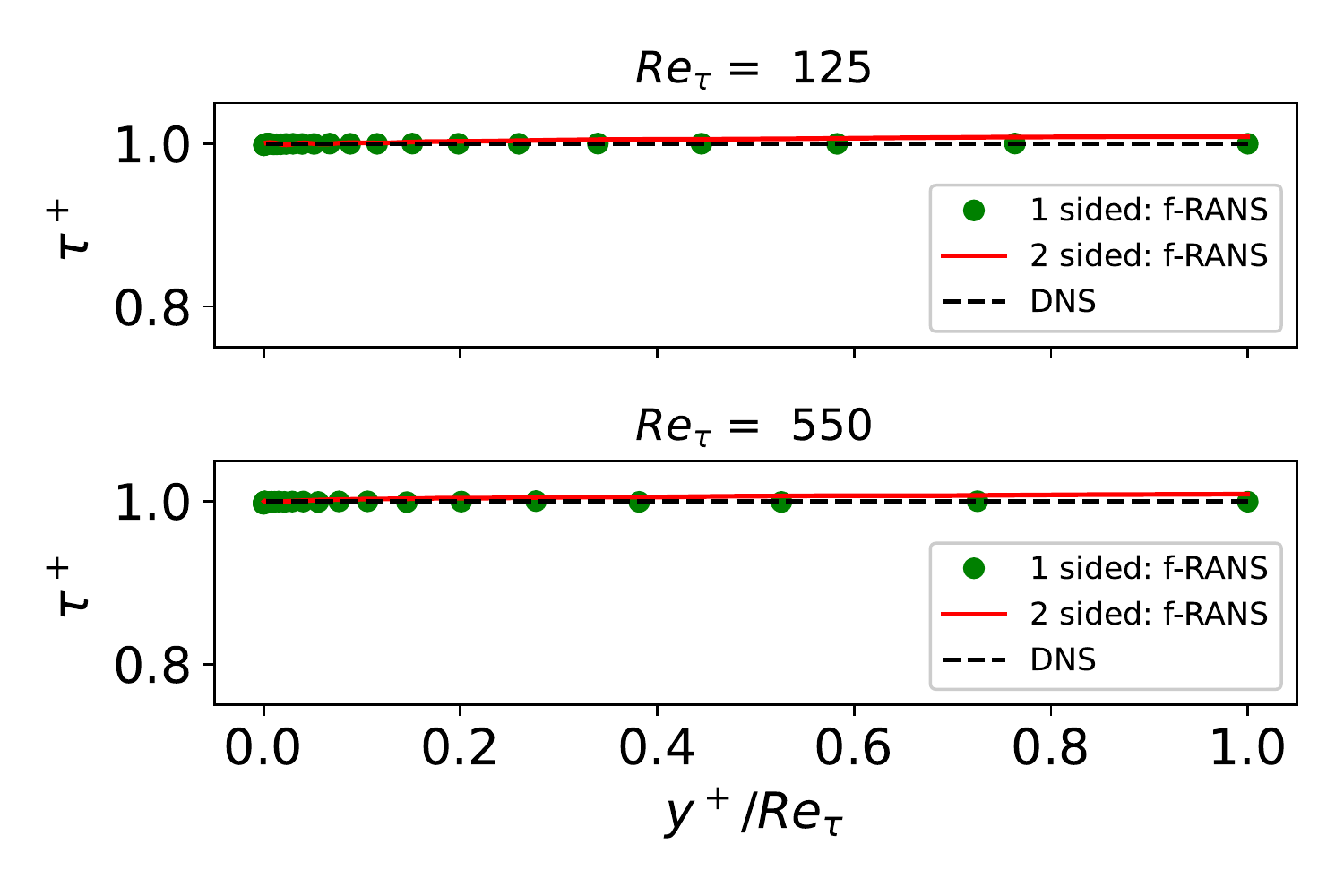}}}%
   \subfloat[Channel]{{\includegraphics[trim=0.3cm 0.3cm 0.3cm 0.3cm,clip,    width=0.48\textwidth]{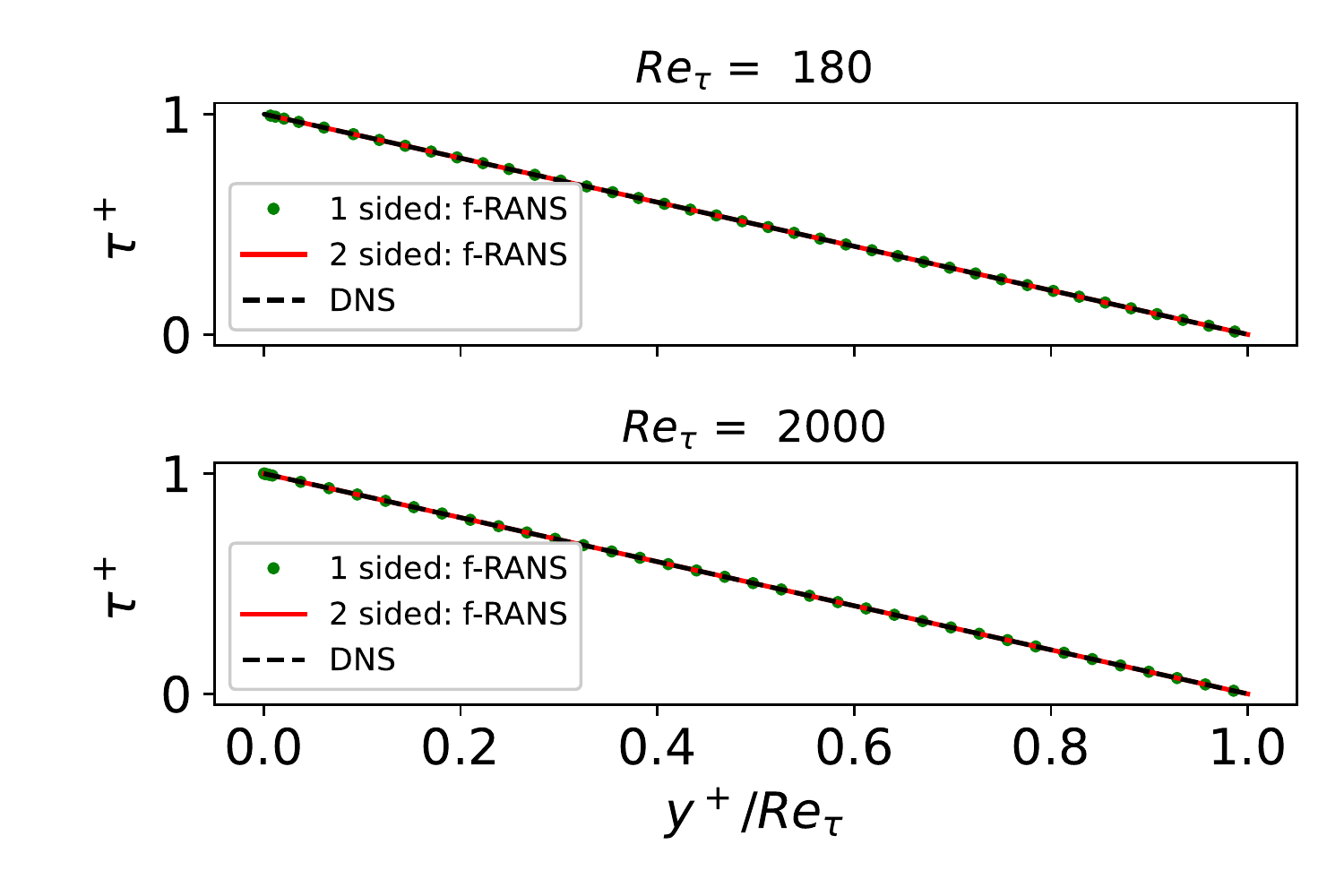}}} \\
    \subfloat[Pipe]{{\includegraphics[trim=0.3cm 0.3cm 0.3cm 0.3cm,clip,    width=0.48\textwidth]{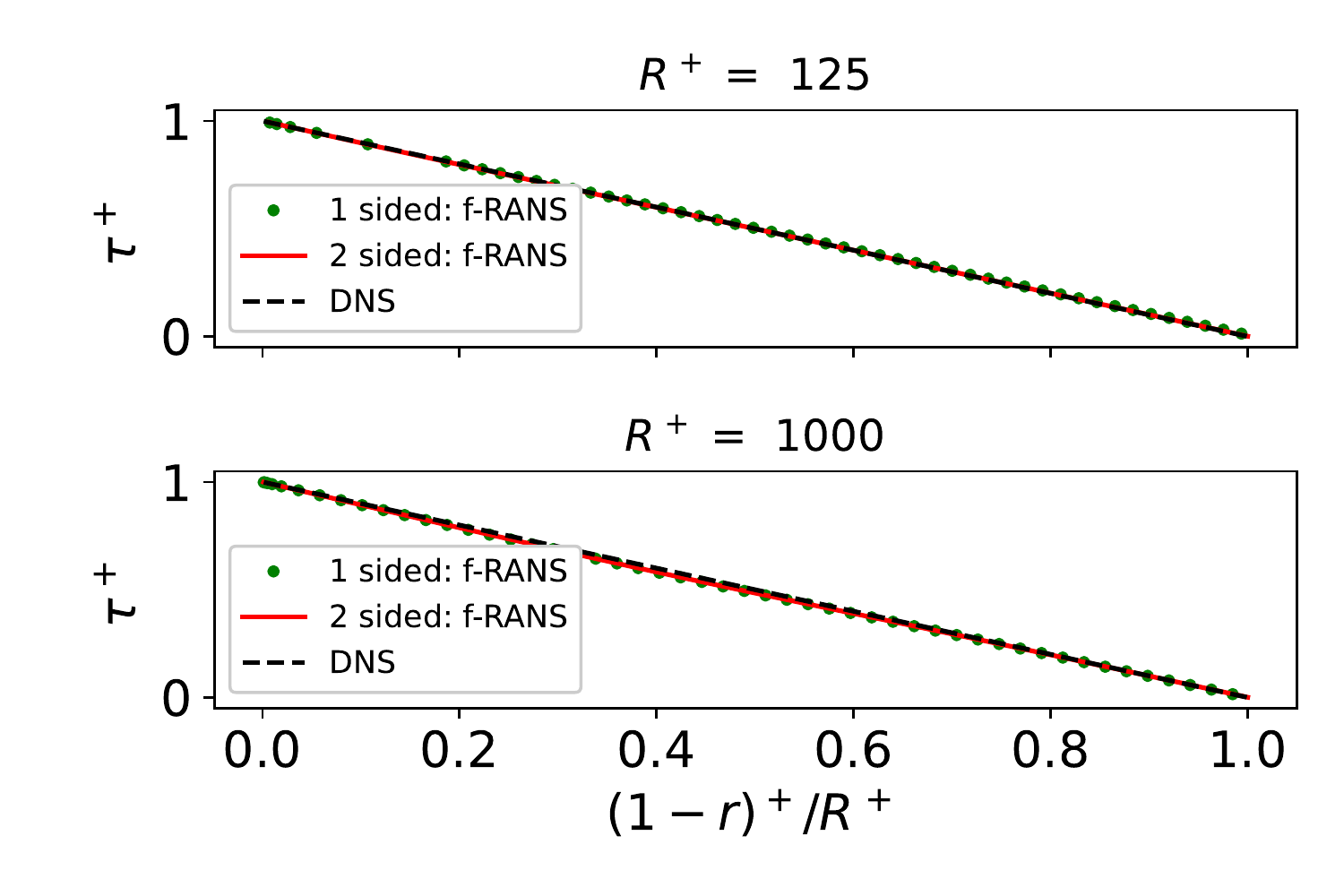}}}%
    \caption{Comparison of total shear stress obtained using one- and two-sided f-RANS model with DNS databases for couette, channel and pipe flow.  Here the x-axis of the plot: channel and couette: $y^+/Re_\tau$, pipe: $(1-r)^+/R^+$. The error in either model is less than 1\%}
    \label{fig:tau_dns}
\end{figure}

A small "bump" is seen at center-line of the channel and pipe as result of symmetry at lower Reynolds number, which flattens for higher Reynolds number or Karman number. Further, we show in fig.~\ref{fig:tau_dns} that the fractional order computed for either one- or two- sided model gives an error less than $1 \%$ in total shear stress.

We further fit the fractional order of the two-sided model. Thus, the analytical expression for couette, channel and pipe flow are given in (\ref{eq:anal_couette}), (\ref{eq:anal_channel}) and (\ref{eq:anal_pipe}) respectively, which are written for one-half of the domain, and the fractional order being symmetric about the center-line ($y^+/Re_\tau = 1$ or $(1-r)^+ / R^+ = 1$) can be mirrored to the other half (refer fig.~\ref{fig:frac_ydel_2sided}). Clearly from (\ref{eq:anal_couette}), (\ref{eq:anal_channel}) and (\ref{eq:anal_pipe}), there are terms which depend only on $y^+$ or $(1-r)^+$, the inner-units. While the last term (contributing to wake) is depended on $y^+ / Re_\tau$ or $(1-r)^+ / R^+ $, the outer units. 

It is to be noted that, (\ref{eq:anal_couette}), (\ref{eq:anal_channel}) and (\ref{eq:anal_pipe}) solves the closure problem in (\ref{eq:frans}) for couette, channel and pipe flow, respectively.
  
\begin{align} \label{eq:anal_couette}
    \begin{split}
         ^{couette} &\alpha ~=~ tanh[(6.9/y^+)^{1.116}] ~+~ 0.644~(1- tanh[(6.9/y^+)^{1.116})]~(y^+)^{-0.0805} \\&+ 0.1646~\exp[-(y^+/Re_\tau)^{-0.6694}]~(y^+)^{-0.0805} ~~~; \forall ~ y^+/Re_\tau \in [0,1]
    \end{split}
\end{align}

\begin{align} \label{eq:anal_channel}
    \begin{split}
         ^{channel} &\alpha ~=~ tanh[(6.907/y^+)^{1.5}] ~+~ 0.908~(1- tanh[(6.907/y^+)^{1.5})]~(y^+)^{-0.175} \\&+ 0.418~\exp[-(y^+/Re_\tau)^{-1.634}]~(y^+)^{-0.175} ~~~; \forall ~ y^+/Re_\tau \in [0,1]
    \end{split}
\end{align}

\begin{align} \label{eq:anal_pipe}
    \begin{split}
         ^{pipe} &\alpha ~=~ tanh[(7.988/(1-r)^+)^{1.07}] \\&+~ 0.645~(1- tanh[(7.988/(1-r)^+)^{1.07})]~((1-r)^+)^{-0.12} \\&+ 0.409~\exp[-((1-r)^+/R^+)^{-1}]~((1-r)^+)^{-0.12} ~~; \forall ~ (1-r)^+/R^+ \in [0,1]
    \end{split}
\end{align}

\subsection{Law of the wake of fractional order (of two-sided f-RANS model)}

\begin{figure} [!htb]
\centering
    \subfloat[Couette]{{\includegraphics[trim=0.3cm 0.3cm 0.3cm 0.3cm,clip, width=0.48\textwidth]{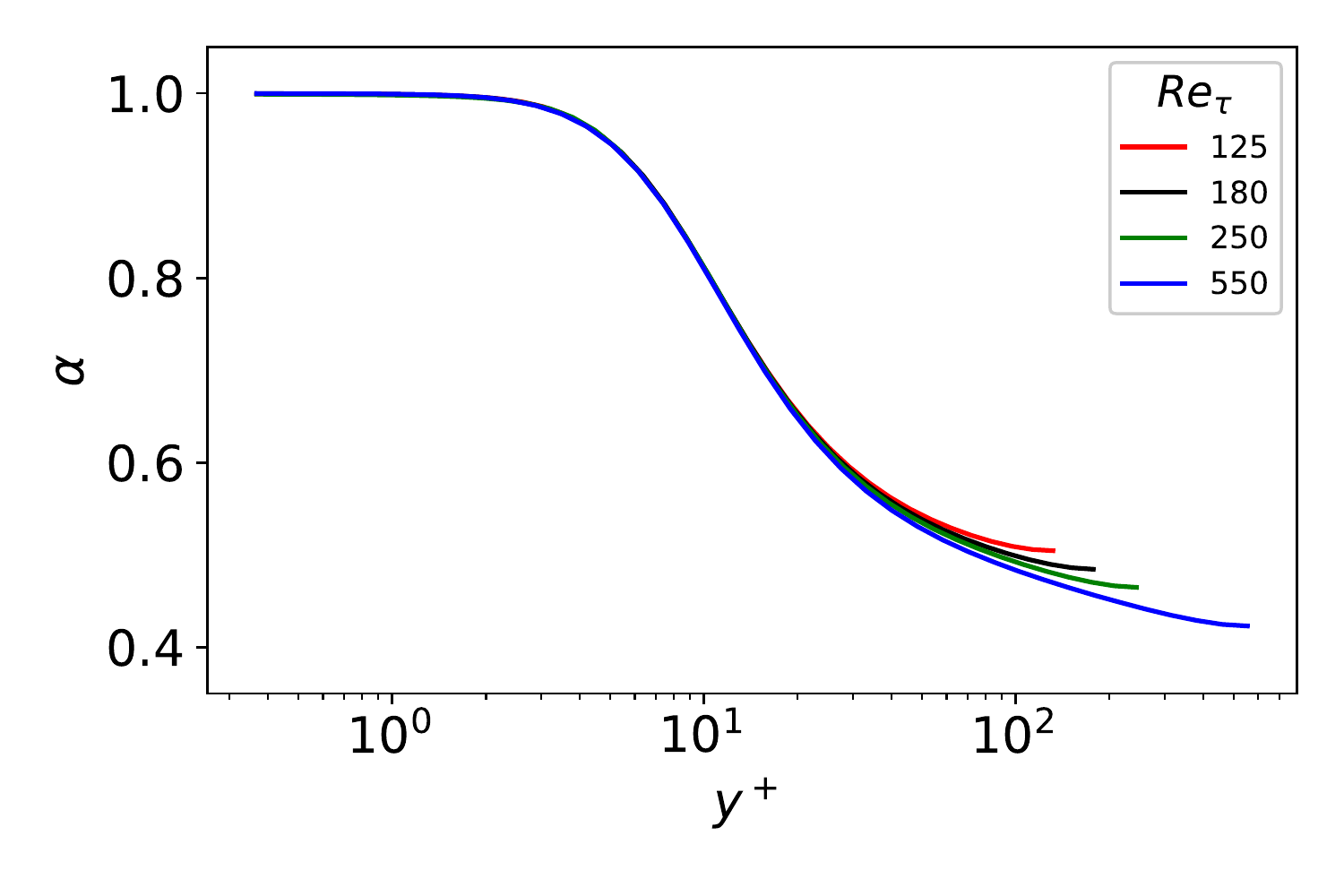}}}%
   \subfloat[Channel]{{\includegraphics[trim=0.3cm 0.3cm 0.3cm 0.3cm,clip,    width=0.48\textwidth]{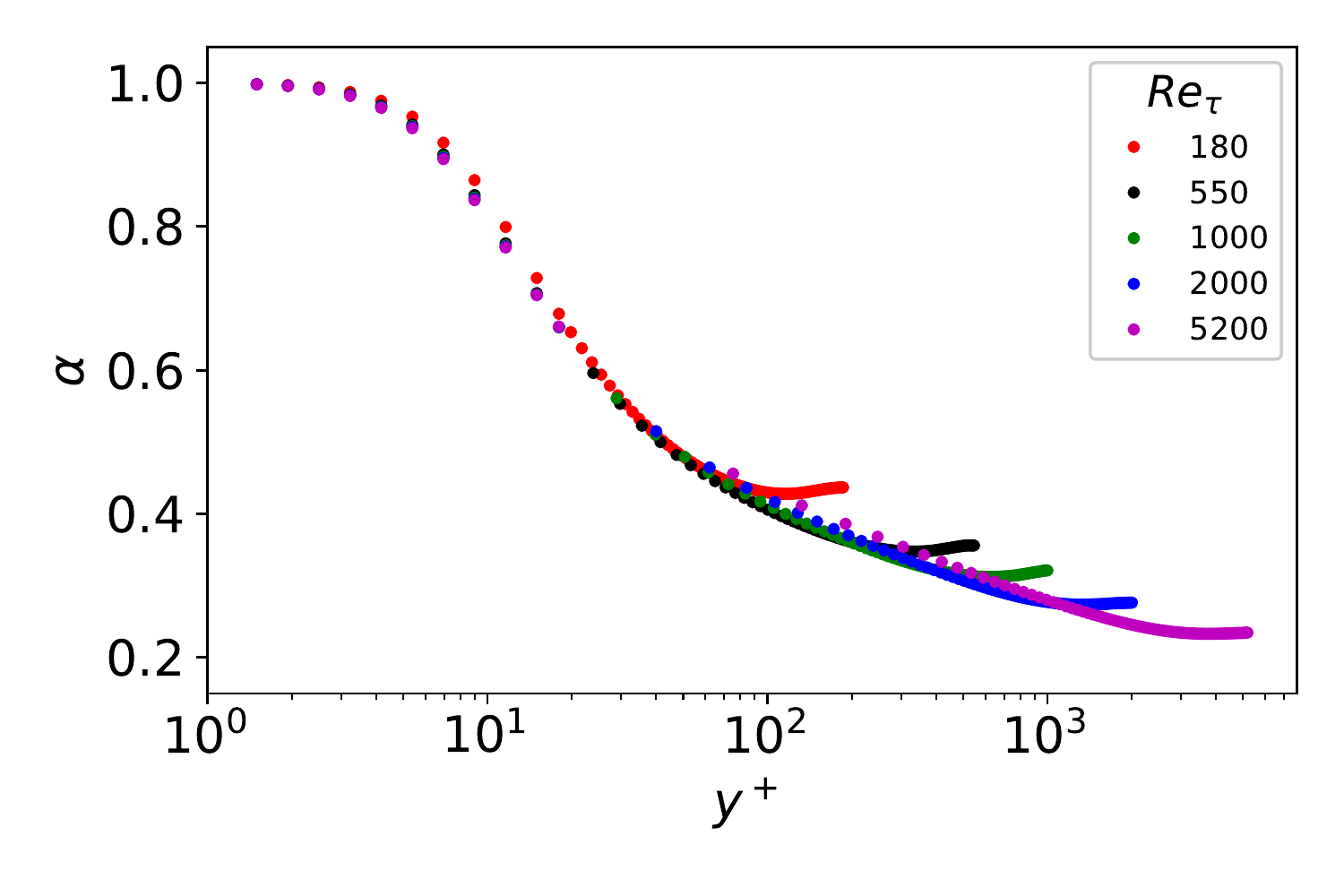}}} \\
   \subfloat[Pipe]{{\includegraphics[trim=0.3cm 0.3cm 0.3cm 0.3cm,clip,    width=0.48\textwidth]{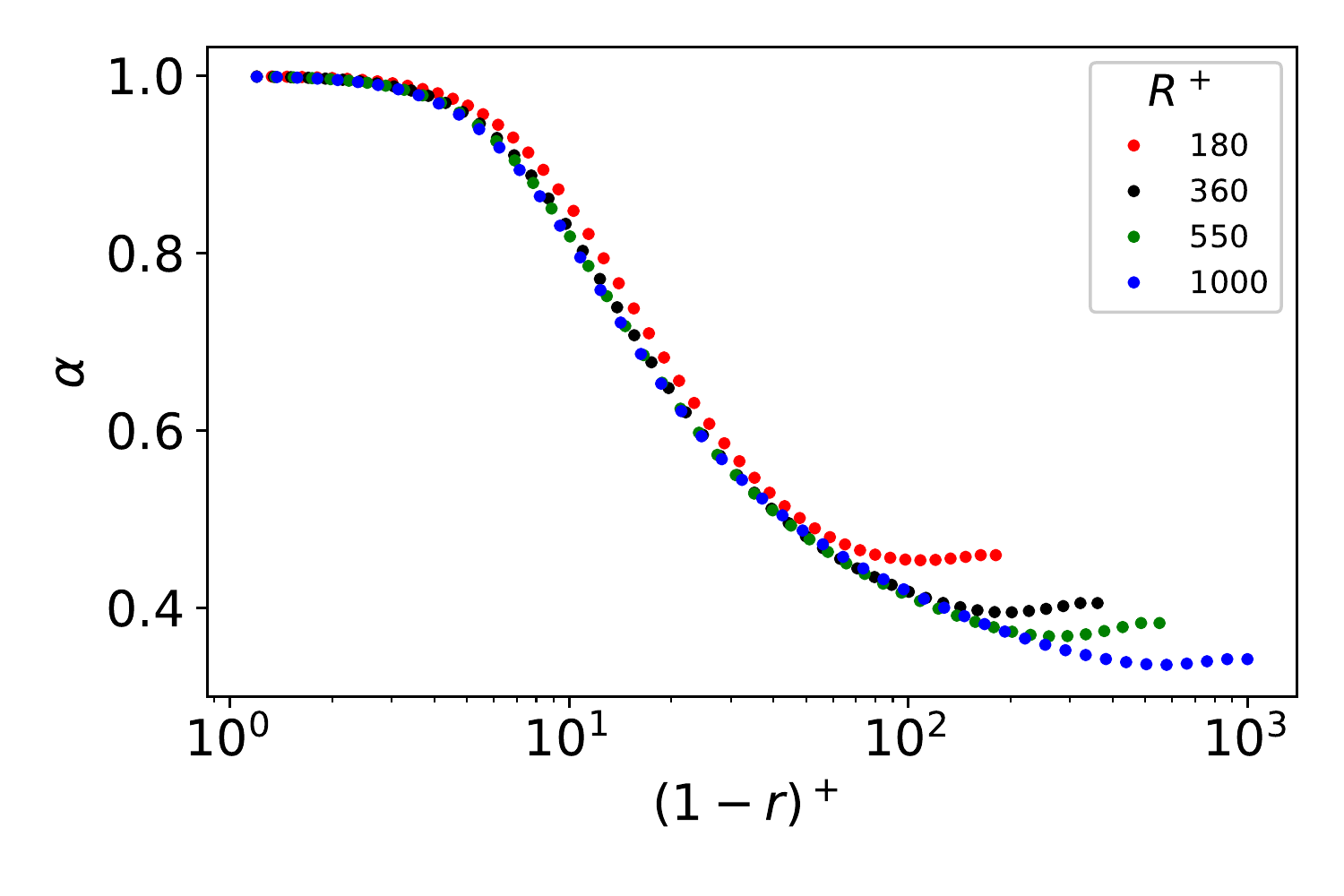}}}%
    \caption{Fractional order of two-sided f-RANS model for channel, couette and pipe flow, plotted for one-half of the domain ($y^+/Re_\tau \le 1$ or $(1-r)^+ / R^+ \le 1$). Here the x-axis of the plot: channel and couette: $y^+$, pipe: $(1-r)^+$. A universal region and wake region is observed in all three cases. The universal region spans across the viscous sub-layer, buffer and logarithmic region of the mean velocity.}
    \label{fig:frac_yplus_2sided}
\end{figure}

\begin{figure} [!htb]
\centering
    \subfloat[Couette]{{\includegraphics[trim=0.3cm 0.3cm 0.3cm 0.3cm,clip, width=0.48\textwidth]{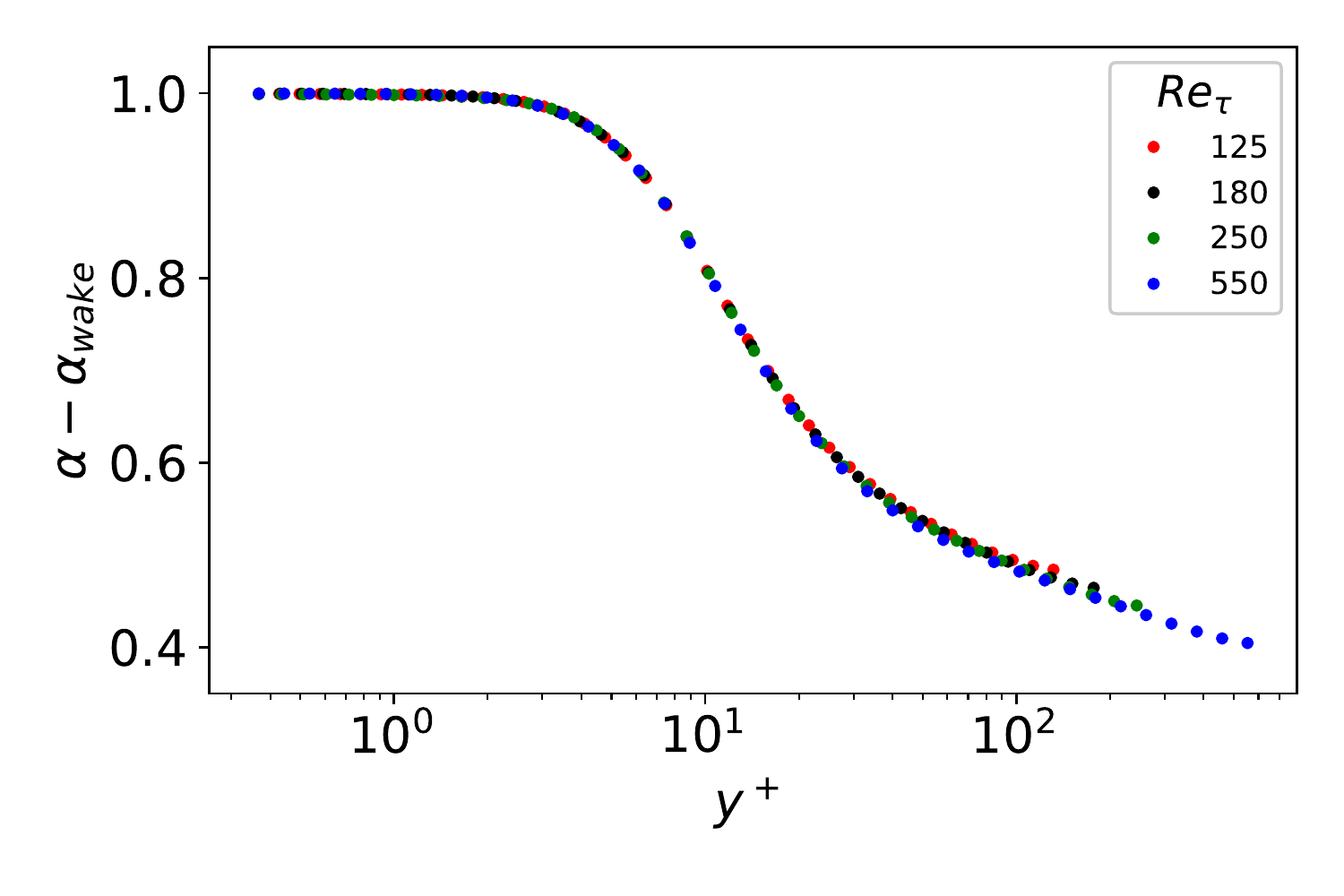}}}%
   \subfloat[Channel]{{\includegraphics[trim=0.3cm 0.3cm 0.3cm 0.3cm,clip,    width=0.48\textwidth]{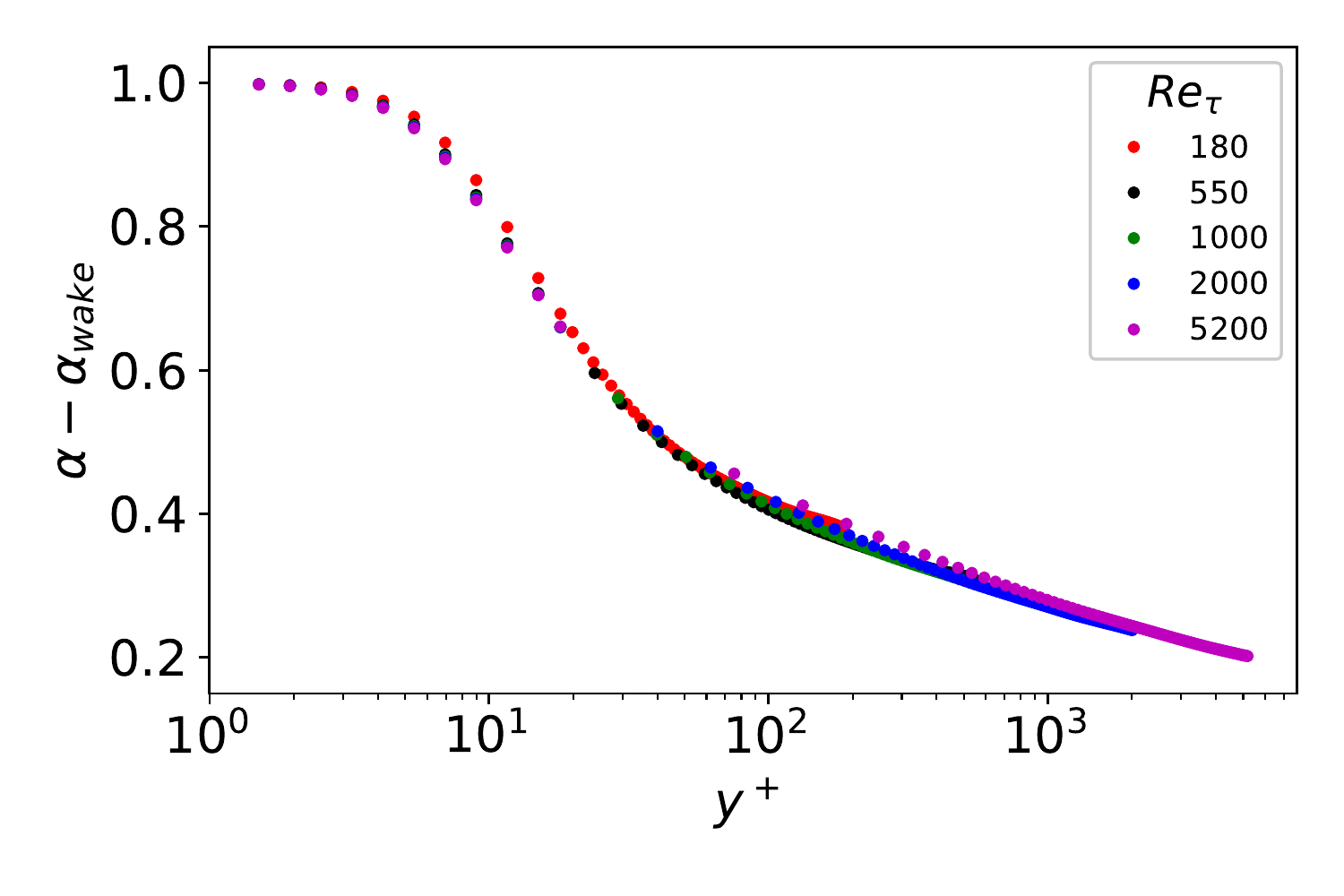}}} \\
    \subfloat[Pipe]{{\includegraphics[trim=0.3cm 0.3cm 0.3cm 0.3cm,clip,    width=0.48\textwidth]{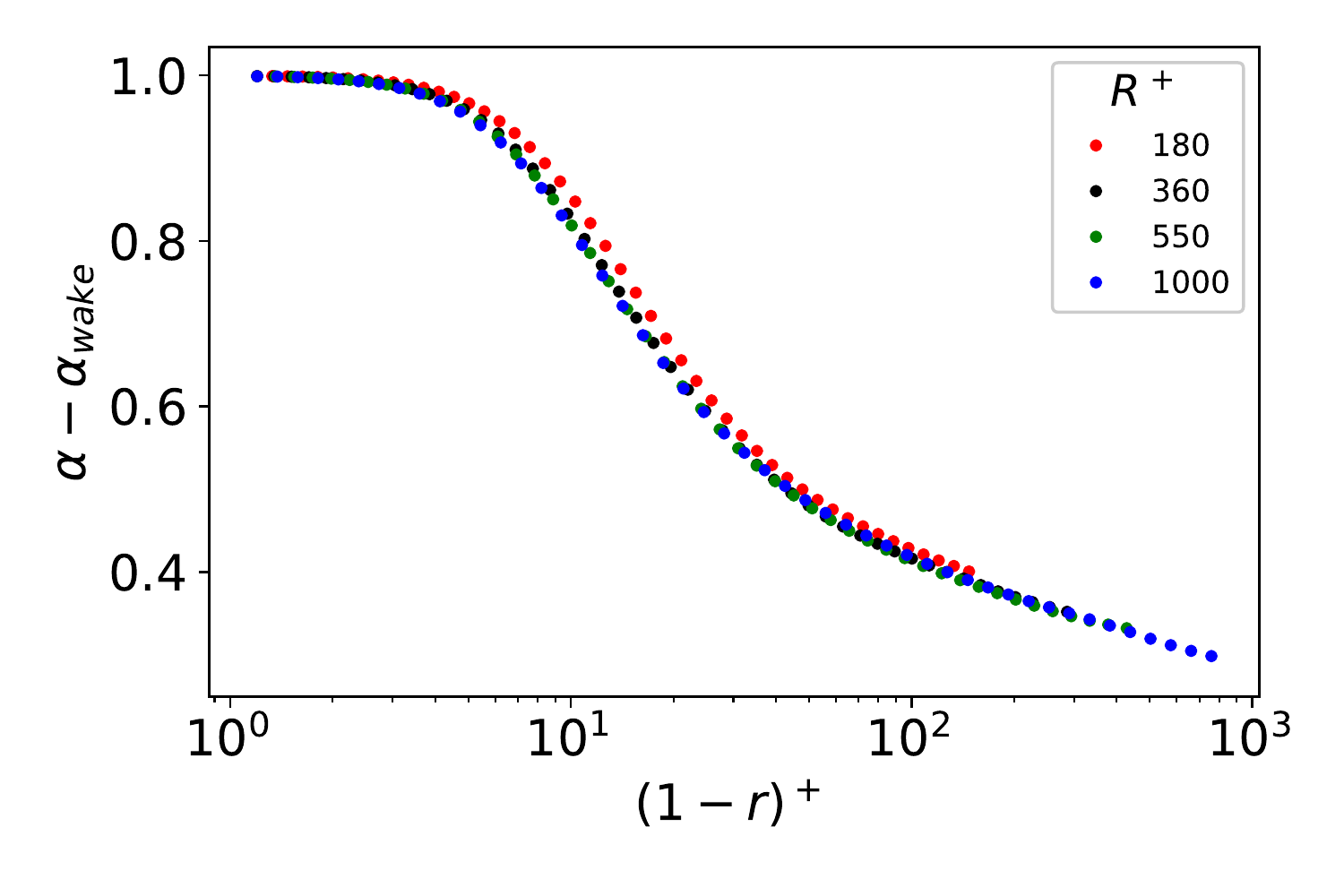}}}%
    \caption{Fractional order for two-sided model shown for the universal part of the curve obtained by subtracting the wake contribution. Fractional order plotted for one-half of the domain ($y^+/Re_\tau \le 1$ or $(1-r)^+ / R^+ \le 1$) Here the x-axis of the plot: channel and couette: $y^+$, pipe: $(1-r)^+$.}
    \label{fig:alp_wake}
\end{figure}

Upon close observation in the log-scale in fig.~\ref{fig:frac_yplus_2sided}, a notion of universality can be seen in the fractional order of two-sided model, where in the wake region, a non-universal scaling is present. We decompose the universal and the wake parts of the curve as (\ref{eq:wake1}). The universal component is a function purely of $y^+$ or $(1-r)^+$, however, the wake region is a function of $y^+ / Re_\tau$ or $(1-r)^+/R^+$. It is to be noted that this wake behavior is analogous to law of the wake for mean velocity \cite{coles_1956}. An analytical expression is deduced by fitting the fractional order for the wake part given in (\ref{eq:wake_couette}), (\ref{eq:wake_channel}) and (\ref{eq:wake_pipe}) for couette, channel and pipe flow respectively. It immediately follows, under the limit $Re_\tau \rightarrow \infty$ or $R^+ \rightarrow \infty$, the wake component vanishes in (\ref{eq:wake1}) and the fractional order is truly universal for each case as shown in fig.~\ref{fig:alp_wake}.

\begin{equation} \label{eq:wake1}
    \alpha ~=~ \alpha_{universal} ~+~ \alpha_{wake}
\end{equation} 

\begin{equation} \label{eq:wake_couette}
    ^{couette} \alpha_{wake} = 0.08052~(y^+)^{-0.075}~\exp[-(y^+/Re_\tau)^{-1}]
\end{equation} 

\begin{equation} \label{eq:wake_channel}
    ^{channel} \alpha_{wake} = 0.36461~(y^+)^{-0.165}~\exp[-(y^+/Re_\tau)^{-1.5}]
\end{equation} \

\begin{equation} \label{eq:wake_pipe}
    ^{pipe} \alpha_{wake} = 0.3838~(y^+)^{-0.125}~\exp[-(y^+/Re_\tau)^{-1.25}]
\end{equation}

\subsection{Wall behavior of fractional order (fractional power-law)} \label{sec:asym}

In fig.~\ref{fig:alp_wake} the universal part of the fractional order is obtained by subtracting the wake contribution. The universal curve spans over viscous sub-layer, buffer layer and logarithmic region of the mean velocity. A part of the universal curve is analogous to log law of the wall for velocity, where in the absence of the wake contribution, it would extend to center-line or infinity. Further, we show a power-law behaviour. This power-law behaviour of fractional order is analogous to the log-law for mean velocity.

In order to show the power-law and asymptotic behaviour of fractional order we take two approaches, computational and asymptotic analyses of (\ref{eq:anal_couette}), (\ref{eq:anal_channel}) and (\ref{eq:anal_pipe}). 

\subsubsection{Computational Approach} \label{sec:comp_power_law}

Here, the computation of $\alpha_{min} :=  \{\alpha : \min \alpha , Re_\tau = C \}$ is sufficient, where $C$ is a constant. Note that, $\alpha_{min}$ occurs at the centre-line for each Reynolds number, since channel and pipe has an anomaly at the centre-line, these values where computed at $y^+/Re_\tau \approx 0.75$ and $ (1-r)^+/R^+ \approx 0.75 $, respectively but this should not be a problem, since all curves are flat at high Reynolds numbers. We employ the Spalding's curve \cite{Spalding} (with different constants) to generate velocity profiles for the case of couette and channel at higher Reynolds number, while we use the Superpipe data \cite{zagarola1998mean} along with Spalding's curve for near wall region for the case of pipe.

\begin{figure}[!htb]
\centering
    \subfloat[Couette]{{\includegraphics[trim=0.3cm 0.3cm 0.3cm 0.3cm,clip, width=0.48\textwidth]{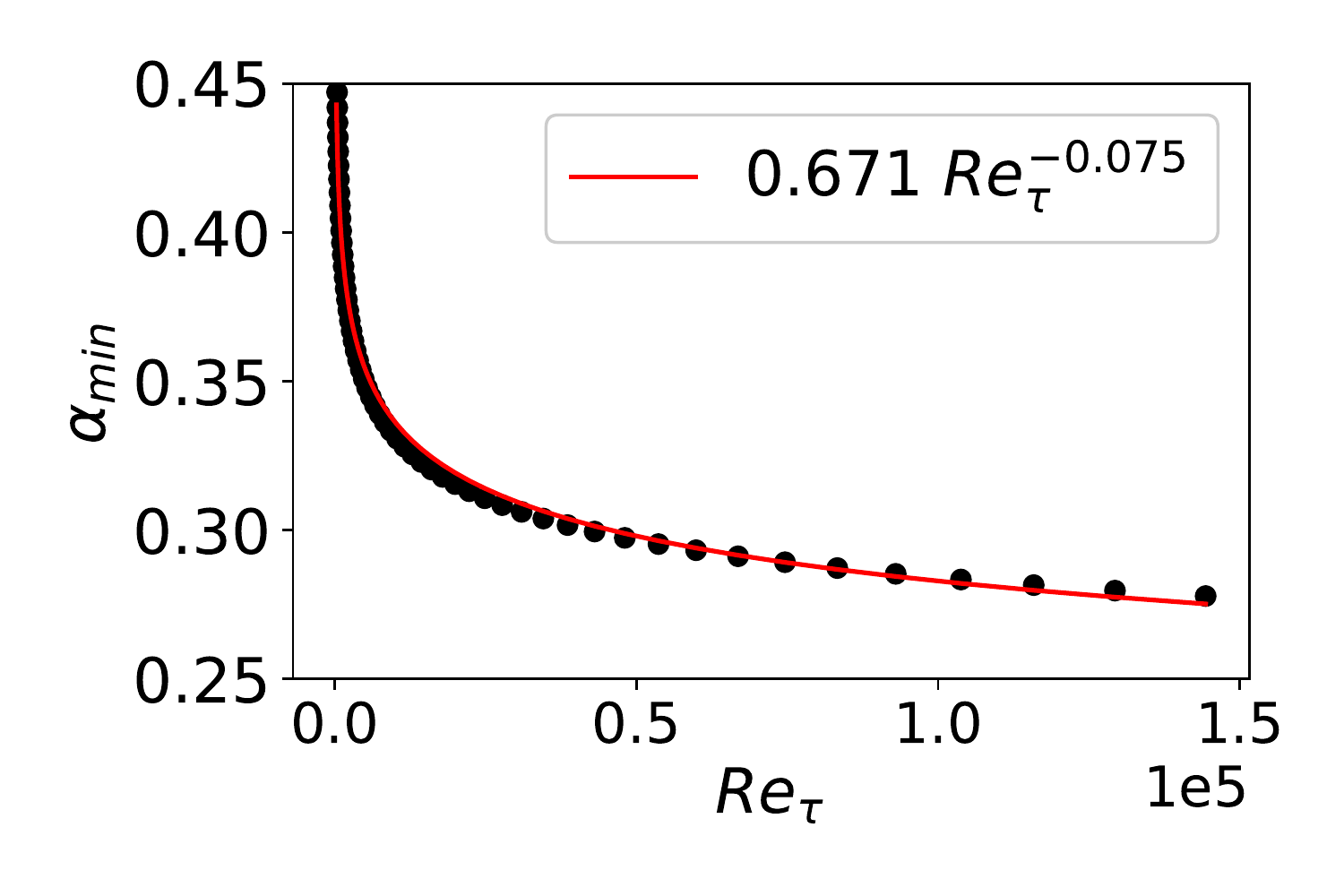}}}%
   \subfloat[Channel]{{\includegraphics[trim=0.3cm 0.3cm 0.3cm 0.3cm,clip,    width=0.48\textwidth]{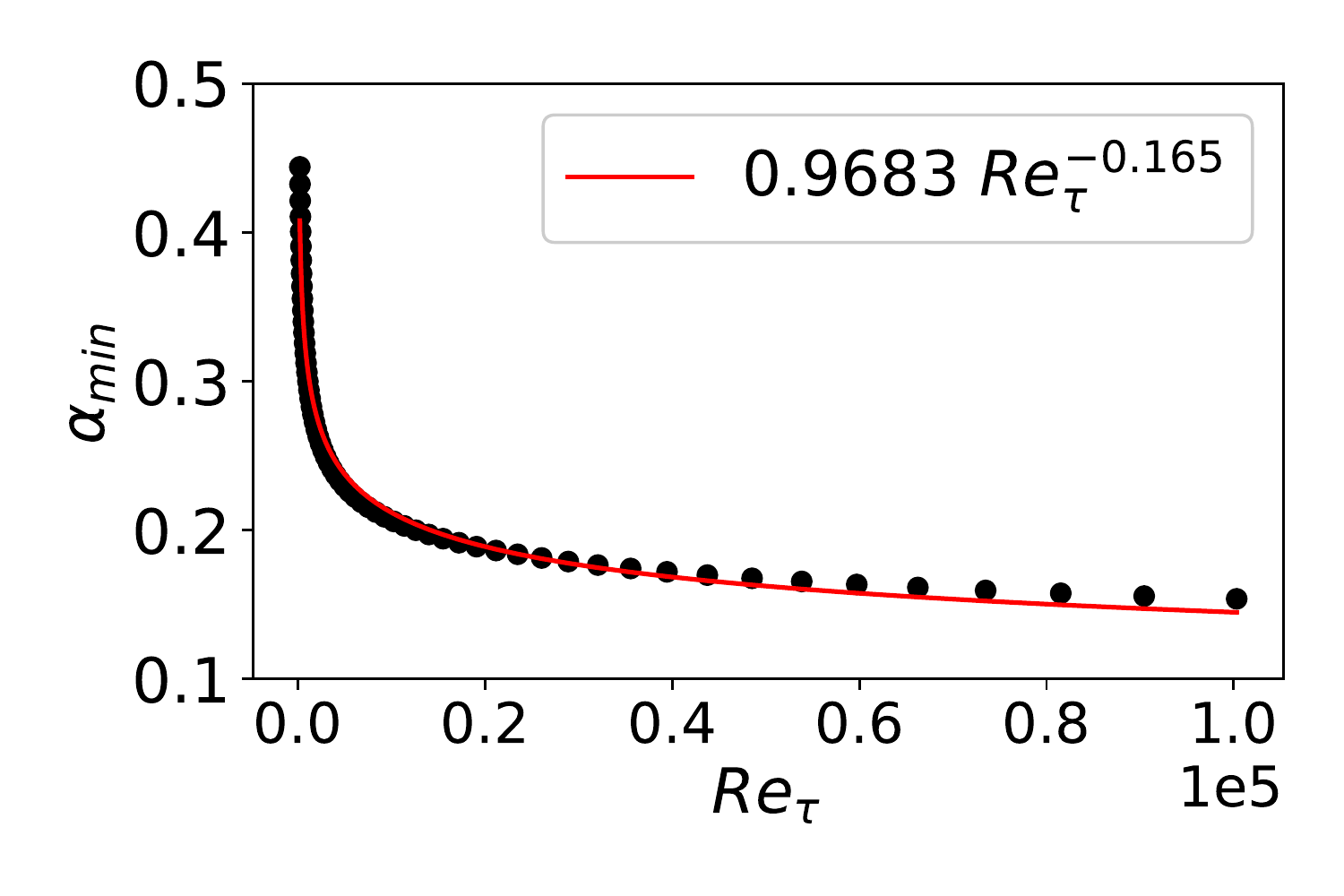}}} \\
    \subfloat[Pipe]{{\includegraphics[trim=0.3cm 0.3cm 0.3cm 0.3cm,clip,    width=0.48\textwidth]{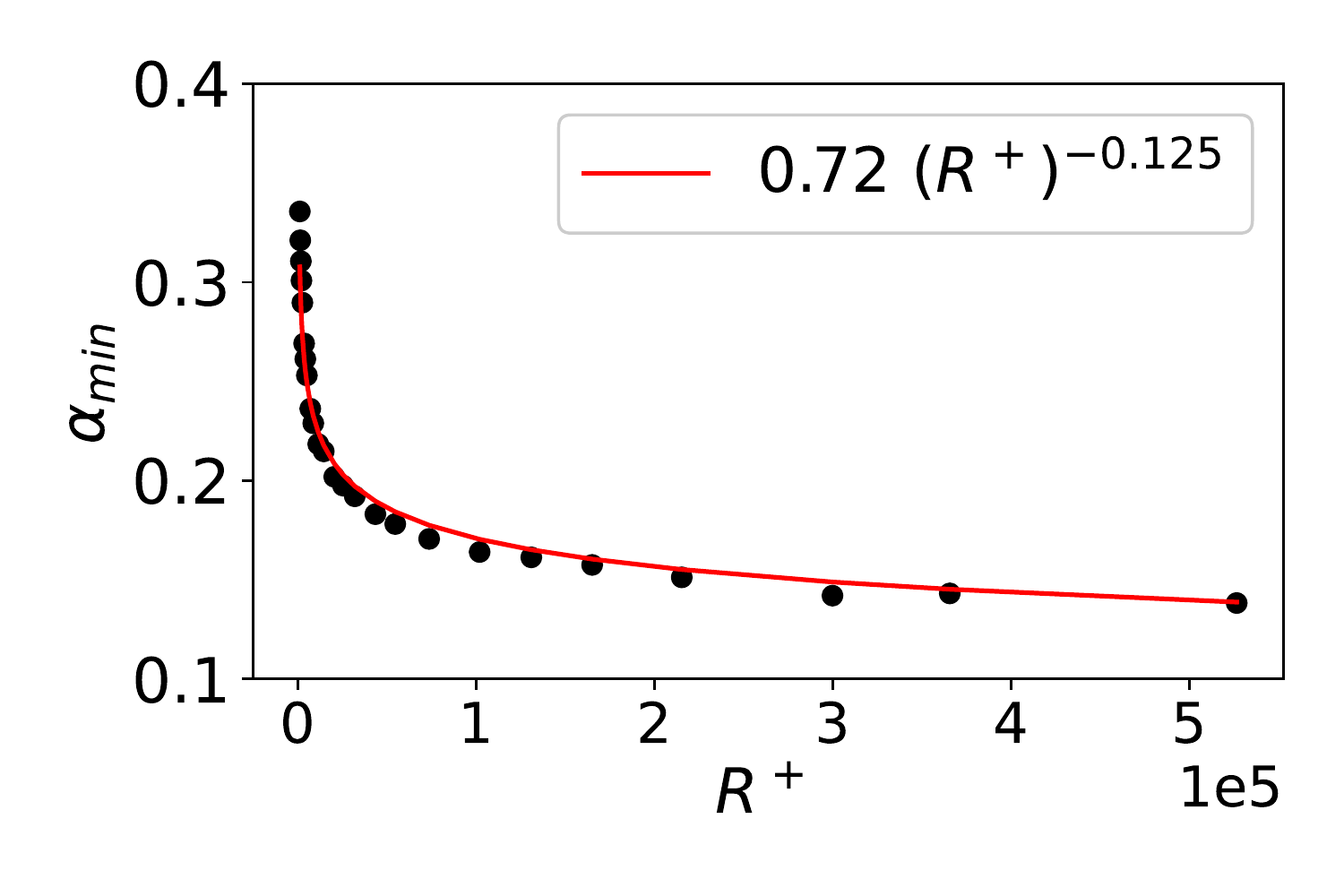}}}%
    \caption{Power law fitting for channel, couette and pipe shown in red-dashed line for minimum fractional order shown as black dots for two-sided model. }
    \label{fig:alp_asym}
\end{figure}

Fig.~(\ref{fig:alp_asym}) shows a power-law relationship, which asymtopdes. We report the $\alpha_{min}$ from our computations, which are, couette: $Re_\tau \approx 144338, \alpha_{min} \approx 0.28 $, channel: $Re_\tau \approx 100319, \alpha_{min} \approx 0.15 $ and pipe: $R^+ \approx 526360, \alpha_{min} \approx 0.13$.

This exercise could be repeated at any location (at a constant value of $y^+ / Re_\tau$ or $ (1-r)^+ / R^+$) and it would lead to asymptotic (power-law) curve with different constants. This can also be deduced from  (\ref{eq:anal_channel}), (\ref{eq:anal_couette}) and (\ref{eq:anal_pipe}), by fixing  $y^+ / Re_\tau$ or $ (1-r)^+ / R^+$ and letting $y^+ \rightarrow \infty$ or $(1-r)^+ \rightarrow \infty$ gives a finite value within our considered range of $\alpha \in (0,1]$ as we show in the next section.

\subsubsection{Asymptotic Analysis}

We perform the asymptotic analysis of (\ref{eq:anal_channel}), (\ref{eq:anal_couette}) and (\ref{eq:anal_pipe}), where we suppress, $(.)^+$ to avoid any confusion with the signs in powers.

\begin{corollary} \label{col:1}
    For $y \rightarrow 0$, within the viscous sub-layer, the fractional order asymptotes to 1.

Consider (\ref{eq:anal_couette}) written for couette flow. We suppress, $(.)^+$ to avoid any confusion with the signs in power. We consider each term individually.  The first term in (\ref{eq:anal_couette}) tends to 1. 

For second term, we notice as $y \rightarrow 0$ implies $y^{-0.0805} \rightarrow \infty$, however, its rate is slower than $[1 - tanh ((6.9/y)^{1.116})] \rightarrow 0$. Thus, as $y \rightarrow 0$ implies $ \left [ 1 - tanh ((6.9/y)^{1.116}) \right ] y^{-0.0805} \rightarrow 0$. 

Similarly, the third term can be treated. Again, exponential decays faster than $y^{-0.0805} \rightarrow \infty$, thus, $ \exp[-(y/Re_\tau)^{-0.6694}]~(y)^{-0.0805} \rightarrow 0$ as $y \rightarrow 0$. Thus, we conclude, $\alpha \rightarrow 1$ as $y \rightarrow 0$.

Further, we want to explicitly compute $y$ for which  $\left [ 1 - tanh ((6.9/y)^{1.116}) \right ] \approx 0$. Consider, $tanh [(6.9/y_{c_1})^{1.116}] = 1 $, or, $(6.9/y_{c_1})^{1.116} \approx 2.5$. This gives us $y_{c_1} \approx 3$. For most cases $y \approx 3$ is the viscous sub-layer, and almost on the onset of the buffer layer. Therefore, $ \forall y \le y_{c_1}$, the first term in (\ref{eq:anal_couette}), $tanh [(6.9/y_{c_1})^{1.116}]$ is the leading order term, as the other terms vanishes. Therefore, $ \forall y \le y_{c_1}$, the fractional order is given by, 

\begin{equation}
    \alpha \approx tanh [(6.9/y)^{1.116}] \approx 1 ~~; ~ \forall y \le 3.
\end{equation}

Since (\ref{eq:anal_channel}) and (\ref{eq:anal_pipe}) for channel and pipe, respectively, has a similar structure as (\ref{eq:anal_couette}), repeating our analysis would lead to the same conclusion. 

\end{corollary} 

\begin{remark}
    The result of corollary \ref{col:1} conclusively shows that in the viscous dominated near-wall region (in the absence of turbulence), the fractional order is unity, implying a local operator. 
\end{remark}

\begin{corollary} \label{col:2}
    The fractional order uniformly asymptotes ($\forall ~ y/Re_\tau \in [0,2]$) as $Re_\tau \rightarrow \infty$. 

Again, consider, (\ref{eq:anal_couette})  written for couette flow. We suppress, $(.)^+$ to avoid any confusion with the signs in power. Before, we consider the infinite limit, we explicitly compute $y$ for which  $tanh [(6.9/y)^{1.116}] \rightarrow 0$. Consider, $tanh [(6.9/y_{c_2})^{1.116}] \approx 0 $, or, $(6.9/y_{c_2})^{1.116} \approx 0$. To a sufficient numerical accuracy, $y_{c_2} \approx 100$ is acceptable (within 5\% accuracy). Thus,  $ \forall y > y_{c_2} \approx 100$ can be regarded a sufficiently large $y$ giving term $tanh [(6.9/y)^{1.116}] \approx 0$. 

Now consider, $y$ to be sufficiently large, which implies $tanh [(6.9/y)^{1.116}] \approx 0$ in (\ref{eq:anal_couette}), we have, 

\begin{equation}
    \alpha \approx 0.664 ~ y^{-0.0805} + 0.1646 ~ \exp[-(y/Re_\tau)^{-0.6694}]~(y)^{-0.0805}
\end{equation}

Let, $\gamma = y/Re_\tau \in (0, 1]$, and $\theta = 0.1646 ~ \exp[-(\gamma)^{-0.6694}]$, therefore, we have, 

\begin{equation} 
    \alpha \approx [0.664 + \theta]~(y)^{-0.0805}
\end{equation}

$\theta$ is a number. For each value of $\gamma \in (0, 1]$, $\theta$ is a constant, thus, $c =  0.664 + \theta$ is a constant to each value of $\gamma \in (0, 1]$. 

\begin{equation} \label{eq:power_law}
    \alpha \approx c~(y)^{-0.0805}
\end{equation}

Let $y \rightarrow \infty$ keeping $\gamma$ constant, implies $Re_\tau \rightarrow \infty$, thus for each value of $\gamma \in (0, 1]$ the fractional order ($\alpha$) asymptotes. Supplemented with result from corollary \ref{col:1} of $y \rightarrow 0$ implies $\gamma \rightarrow 0$.  It can be conclusively, said the fractional order asymptotes uniformly $\forall \gamma \in [0,1]$. Note, that for $\forall \gamma \in [1,2]$, the fractional order is symmetric. Thus, the fractional order asymptotes uniformly over the whole domain ($\forall \gamma \in [0,2]$).

Since (\ref{eq:anal_channel}) and (\ref{eq:anal_pipe}) for channel and pipe, respectively, has a similar structure as (\ref{eq:anal_couette}), repeating our analysis would lead to same conclusion. 

\end{corollary}

\begin{remark}
    The remark of corollary \ref{col:2} of the fractional order curves uniformly asymptotes as $Re_\tau \rightarrow \infty$, implies, for any fixed value of $y^+ / Re_\tau \in [0,2]$, the fractional order asymptotes. Thus, the fractional order asymptotes both as a function of $y^+ / Re_\tau$ and $Re_\tau$ as $Re_\tau \rightarrow \infty$
\end{remark}

\begin{remark}
The result in (\ref{eq:power_law}) of corollary \ref{col:2}  is for sufficiently large $y^+$, it is clear, analogous to logarithmic layer for the mean velocity, is a power-law in the fractional order. Such an analysis can be repeated for channel and pipe flow, and similar conclusions will drawn. Table \ref{tab:summary} summaries  the results obtained in this section for two-sided model. 
\end{remark}

\begin{table}[h!]
\centering
\begin{tabular}{ p{0.3\textwidth} p{0.3\textwidth} p{0.3\textwidth} }  
 \hline
 Flow region & Mean velocity scaling & Fractional order scaling \\
 \hline
 Viscous sub-layer & $\overline{U}^+ \approx y^+$ & universal ($\alpha \approx 1$) \\
 Buffer Layer  & -- & universal \\sufficient
 Outer region  & log-law  ($\overline{U}^+ \approx \kappa log (y^+)$)   & power law  ($ \alpha \approx c ~ ({y^+}) ^p  $  )\\
 Wake & law of wake \cite{coles_1956} & fractional law of wake  \\ \hline
\end{tabular} 
\caption{Analogy between velocity and fractional order (of two-sided model) for different regions of the flow for all three cases. Here $c$, $p$ and $\kappa$ are arbitrary constants depending upon the individual case.} \label{tab:summary}
\end{table}

\begin{remark}
A special case of (\ref{eq:power_law}) of corollary \ref{col:2} is, $\gamma = 1$, this gives $c = 0.72455$. Since $\gamma =1$, we have, $y = Re_\tau$, and for $Re_\tau = 144338$, the fractional order $ \alpha \approx 0.27844$, which is very close to the value obtained via our computational approach for couette flow (refer section \ref{sec:comp_power_law}). Similarly, we could repeat the analysis for channel and pipe flow. Table \ref{tab:com_aysmp} summaries it for all three flows. 
\end{remark}

\begin{table}[h!]
\centering
\begin{tabular}{ p{0.3\textwidth} p{0.3\textwidth} p{0.3\textwidth} }  
 \hline
 Case  & Computational Approach & Asymptotic analysis \\
 \hline
 Couette ($Re_\tau \approx 144338$) & 0.28 & 0.27844 \\
 Channel ($Re_\tau \approx 100319$) & 0.15 & 0.14151 \\
 Pipe ($R^+ \approx 526360$) & 0.13 & 0.1637 \\ \hline
\end{tabular} 
\caption{Comparing the fractional order obtained at center-line using the computational experiments with asymptotic analysis for all three flows.} \label{tab:com_aysmp}
\end{table}

%%%%%%%%%%%%%%%%%%%%%%%%%%%%%%%%%%%%%%%%%%%%%%%%%%%%%%%%%%%%%%%%%%%%%%%%%%%%%%%%%%%%%%%%%%%%%%%%%%%%%%%%%
\section{Extension to tempered fractional derivatives} \label{sec:temp}

For a fractional derivative defined over an unbounded domain, the second moment is infinite. This implies that the particle can have infinite jump lengths, which is not physical. Thus, in-order for the stochastic process to have finite moments, tempering L\'evy flight was introduced in \cite{mantegna1994stochastic} and its characteristic function in \cite{koponen1995analytic}. A complete theory for tempered $\alpha$-stable L\'evy process is established in \cite{rosinski2007tempering}. Its connection to tempered fractional diffusion equations was shown in \cite{del_castillo}, and a tempered fractional calculus was formally introduced in \cite{mark_tfc}.

For a  tempered $\alpha$-stable L\'evy process, theorem 3.1 of \cite{rosinski2007tempering} has two result, namely, the short and long time behavior. For a short time behavior, the stochastic process  remains heavy-tailed. While for long time, tempered $\alpha$-stable L\'evy process converges (in distribution) to a normal distribution (or Brownian motion). These two results seem contradictory. How do we physically interpret it? Consider a scenario, where an aircraft passing by, produces trails as result of its wake. Immediately measured, the disturbances can be still be observed, since the fluid continues to retain the energy. However, for a long enough time, the air settles down (as a result of energy dissipation, which is slow process) and eventually no memory of the aircraft is retained (assume no other source), indeed the fluid is now performing Brownian motion.

\subsection{Tempered fractional models for RANS}

In this section, we introduce the tempered fractional closure model as a generalisation of fractional closure model introduced in section \ref{sec:frans} for Reynolds-averaged Navier-Stokes equations (RANS). Recall that, a tempered fractional derivative only allows finite jump length. Further, a tempered fractional derivative definition allows cases defined over infinite or semi-infinite domains computationally feasible. For example, a boundary layer over a flat plate is unbounded, as it is open to atmosphere. Although mathematically the notion of infinity is well defined, but computationally impossible. Thus one often truncates or tempers the domain with suitable boundary conditions which mimics the infinite extent of atmosphere.

We shall introduce two tempering parameters, exponential term and Heaviside function, and will refer it as tempered f-RANS and truncated f-RANS, respectively. We shall make the following main investigations for case of channel and couette flow, 

\begin{itemize}
\item Influence over fractional order by the introduced tempering parameters. Computational experiments are designed to investigate if the fractional order continues to be physically meaningful.

\item  We study the equivalence between the truncated and tempered fractional models. The need for such a study is motivated by computational expense. The domain of integration for tempered fractional is the whole (or original) domain, while for a truncated definition, its domain is a subset of the original domain. Integrating over a smaller domain reduces the computational cost. Further, to implement domain decomposition techniques over heterogeneous computing systems, the communication overhead will be smaller too. Thus, the truncated fractional models have a computational advantage over its counterpart tempered fractional model. Within the literature of fractional calculus, stochastic process and probability theory, the tempered definition is more rigorously studied, by way of equivalence the results can be translated to truncated definitions.
\end{itemize}

For the tempered fractional derivative, we introduce the exponential term, with $\lambda$ as the tempering parameter in left-sided Caputo fractional derivative as (\ref{eq:left
-temp}).

\begin{equation} \label{eq:left
-temp}
    {_{-\infty}} ^C D_{x_j^+} ^{(\alpha,  {\lambda})} ~~ \overline{U^+} = \frac{1}{\Gamma(1-\alpha(x_j^+))}\int_{-\infty}^{x_j^+} (x_j^+-\zeta)^{-\alpha(x_j^+)}   { e^{-\lambda  \frac{ |x_j^+ - \zeta|} { Re_\tau} }} \frac{ d \overline{U^+}}{dx_j^+}   d\zeta
\end{equation} 

Similarly for the right-sided tempered definition as (\ref{eq:right_temp}), 

\begin{equation} \label{eq:right_temp}
    {_{x_j^+}} ^C D_{\infty} ^{(\alpha,  {\lambda})} ~~ \overline{U^+} = \frac{-1}{\Gamma(1-\alpha(x_j^+))}\int_{x_j^+}^{\infty} (\zeta - x_j^+)^{-\alpha(x_j^+)}   { e^{-\lambda  \frac{ |\zeta - x_j^+|} { Re_\tau} }} \frac{ d \overline{U^+}}{dx_j^+}   d\zeta
\end{equation}

Thus the two-sided tempered fractional model for RANS is given by (\ref{eq:two-temp}) and shall be refereed as tempered f-RANS model.

\begin{align} \label{eq:two-temp}
\begin{split}
{_{[-\infty, ~ \infty]}} ^T D_{x_j^+}^{\alpha(x_j^+),   {\lambda})} \overline{U^+} ~&:=~
    {1 \over 2} \left ( _{-\infty} ^C D_{x_j^+} ^{(\alpha(x_j^+),  {\lambda})}~ \overline{U^+} ~ - ~  _{x_j^+} ^C D_{\infty} ^{(\alpha(x_j^+),  {\lambda})} ~\overline{U^+} \right ) ~ \\ &= ~\frac{\partial \overline{U^{+}_{i}}} {\partial x^{+}_{j}} - \overline{u_{i}u_{j}}^{+} ~;~ i,j = 1,2,3 ~;~ \alpha(x^+_j) \in (0, 1]
\end{split}
\end{align}

For the truncated fractional derivative, we introduce the Heaviside function ($H[.]$), with $\delta^+ \in (0, +\infty]$ as the truncating parameter in left-sided Caputo fractional derivative as (\ref{eq:left-trun}),

\begin{align} \label{eq:left-trun}
\begin{split}
        {_{-\infty}} ^C D_{x_j^+} ^{(\alpha(x_j^+),  {\delta^+})} ~~ \overline{U^+} &= \frac{1}{\Gamma(1-\alpha(x_j^+))}\int_{-\infty}^{x_j^+} (x_j^+-\zeta)^{-\alpha(x_j^+)}   { H [ x_j^+ - \delta^+  ]} \frac{ d \overline{U^+}}{d x_j^+}   d\zeta \\
        & = \frac{1}{\Gamma(1-\alpha(x_j^+))}\int_{x_j^+ - \delta^+}^{x_j^+} (x_j^+-\zeta)^{-\alpha(x_j^+)}   \frac{ d \overline{U^+}}{d x_j^+}   d\zeta \\
        & =~~  {_{x_j^+ -\delta^+}} ^C D_{x_j^+} ^{\alpha(x_j^+)} ~~ \overline{U^+}
\end{split}
\end{align} 

thus, the domain of integration is no longer $[-\infty, x^+_j]$ but is now truncated (with a sharp cutoff) to  $[x_j^+ - \delta^+, x_j^+]$, where $\delta^+ \in (0, +\infty]$. Note that, for $\delta^+ = \infty$ is the classical left-sided Caputo fractional derivative.  

Similarly, the right-sided truncated definition as (\ref{eq:right-trun}), 

\begin{align} \label{eq:right-trun}
\begin{split}
        {_{x_j^+}} ^C D_{\infty} ^{(\alpha(x_j^+),  {\delta^+})} ~~ \overline{U^+} &= \frac{-1}{\Gamma(1-\alpha(x_j^+))}\int_{x_j^+}^{\infty} (\zeta - x_j^+)^{-\alpha(x_j^+)}   { H [ - x_j^+ + \delta^+ ]} \frac{ d \overline{U^+}}{d x_j^+}   d\zeta \\
        & = \frac{-1}{\Gamma(1-\alpha(x_j^+))}\int_{x_j^+}^{x_j^+ + \delta^+} (x_j^+-\zeta)^{-\alpha(x_j^+)}   \frac{ d \overline{U^+}}{d x_j^+}   d\zeta \\
         & =~~  {_{x_j^+}} ^C D_{x_j^+ + \delta^+} ^{\alpha(x_j^+)} ~~ \overline{U^+}
\end{split}
\end{align}

thus, the domain of integration is no longer $[x_j^+, \infty]$ but is now truncated (with a sharp cutoff) to  $[x_j^+, x_j^+ + \delta^+]$, where $\delta^+ \in (0, +\infty]$. Note that, for $\delta^+ = \infty$ is the classical right-sided Caputo fractional derivative.   

Thus the two-sided truncated fractional model for RANS is given by (\ref{eq:two-trunc}) and shall be referred as truncated f-RANS model.

\begin{align}  \label{eq:two-trunc}
\begin{split}
{_{[-\infty, ~ \infty]}} ^T D_{y^+}^{(\alpha((x_j^+),   {\delta^+})} \overline{U^+} ~&:=~
    {1 \over 2} \left ( _{-\infty} ^C D_{x_j^+} ^{(\alpha(x_j^+),  {\delta^+})}~ \overline{U^+} ~ - ~  _{x_j^+} ^C D_{\infty} ^{(\alpha(x_j^+),  {\delta^+})} ~\overline{U^+} \right ) ~ \\ &= ~\frac{\partial \overline{U^{+}_{i}}} {\partial x^{+}_{j}} - \overline{u_{i}u_{j}}^{+} ~;~ i,j = 1,2,3 ~;~ \alpha(x^+_j) \in (0, 1]
\end{split}
\end{align}

\subsection{Application to Channel flow}

The two-sided model (\ref{temp:chal}) is defined over the domain $[0, 2Re_\tau]$, where $ y^+ \in (0 ,2Re_{\tau})$ for turbulent channel flow. For numerical computation of the fractional order ($\alpha(y^+)$) we employ the DNS database generated in \cite{lee_moser_2015} for velocity ($\overline{U^+}$) for each friction Reynolds number and the $\tau^+$ as derived in section \ref{sec:application}.

\begin{equation}  \label{temp:chal}
 {_{[0, ~2Re_\tau]}} ^T D_{y^+}^{(\alpha(y^+),   {\xi})} ~ \overline{U^+} ~=~ \frac{d \overline{U^{+}}} {d y^{+}} - (\overline{uv})^{+}   ~=  -{y^+ \over Re_\tau} + 1 ~;~ \alpha(y^+) \in (0, 1]
\end{equation}

where, $ \xi = \{\lambda, \delta^+\}$ for tempered factional and truncated fractional two-sided model, respectively. In-order to incorporate the presence of the wall, $\delta^+$ is a composite function, to eliminate the possibility of domain of integration going beyond the wall. 

For the left-sided truncated Caputo fractional derivative, we have, 

\begin{equation}    
y^+ - \delta^+ = \begin{cases}
   0 ~,~ \text{if} ~~ y^+ - \delta^+ < 0 \\
    y^+ - \delta^+ ~,~ \text{if}~~ y^+ - \delta^+ \geq 0
\end{cases}
\end{equation}

For the right-sided truncated Caputo fractional derivative, we have, 

\begin{equation}    
y^+ + \delta^+ = \begin{cases}
   y^+ + \delta^+ ~,~ \text{if} ~~ y^+ + \delta^+ < 2Re_\tau \\
   2Re_\tau ~,~ \text{if}~~ y^+ + \delta^+ \geq 2Re_\tau
\end{cases}
\end{equation}

\subsection{Application to Couette flow}

The two-sided model (\ref{temp:cou}) is defined over the domain $[0, 2Re_\tau]$, where $ y^+ \in (0 ,2Re_{\tau})$ for turbulent couette flow. For numerical computation of the fractional order ($\alpha(y^+)$) we employ the DNS database generated in \cite{avsarkisov2014turbulent} for velocity ($\overline{U^+}$) for each friction Reynolds number and the $\tau^+$ as derived in section \ref{sec:application}.

\begin{equation}  \label{temp:cou}
 {_{[0, ~2Re_\tau]}} ^T D_{y^+}^{(\alpha(y^+),   {\xi})} ~ \overline{U^+} ~=~ \frac{d \overline{U^{+}}} {d y^{+}} - (\overline{uv})^{+}   ~=   1 ~;~ \alpha(y^+) \in (0, 1]
\end{equation}

where, $ \xi = \{\lambda, \delta^+\}$ for tempered fractional and truncated fractional two-sided model, respectively. In-order to incorporate the presence of the wall, $\delta^+$ is a composite function, to eliminate the possibility of domain of integration going beyond the wall. 

For the left-sided truncated Caputo fractional derivative, we have,

\begin{equation}    
y^+ - \delta^+ = \begin{cases}
   0 ~,~ \text{if} ~~ y^+ - \delta^+ < 0 \\
    y^+ - \delta^+ ~,~ \text{if}~~ y^+ - \delta^+ \geq 0
\end{cases}
\end{equation}

For the right-sided truncated Caputo fractional derivative, we have, 

\begin{equation}    
y^+ + \delta^+ = \begin{cases}
   y^+ + \delta^+ ~,~ \text{if} ~~ y^+ + \delta^+ < 2Re_\tau \\
   2Re_\tau ~,~ \text{if}~~ y^+ + \delta^+ \geq 2Re_\tau
\end{cases}
\end{equation}

%%%%%%%%%%%%%%%%%%%%%%%%%%%%%%%%%%%%%%%%%%%%%%%%%%%%%%%%%%%%%%%%%%%%%%%%%%%%%%%%%
\section{Algorithms for tempered and truncated fractional Reynolds-averaged Navier-Stokes equations} \label{sec:comp}

In this section, we present two numerical algorithms, the first to compute the fractional order given the tempering parameter $\xi = \{\lambda, \delta^+ \}$ formulated in (\ref{temp:chal}, \ref{temp:cou}) for channel and couette flow. For this we extended the pointwise fractional physics-informed neural network (section \ref{sec:fpinns}). While second investigations aims at finding the equivalence between the two definitions. We note that algorithm \ref{alg:2} can be used to find the horizon of non-local interactions. 

\subsection{Investigating the influence of tempering parameter over fractional order}

In the first investigation, we evaluate the influence of fractional order for tempered or truncated fractional two-sided model, associated with the tempering parameter $\lambda$ and $\delta^+$, respectively. Algorithm \ref{alg:1} is the modified pointwise fractional physics-informed neural network algorithm, where we additionally supply the  tempering parameter $\lambda$ or $\delta^+$. Please note, again for this pointwise algorithm, no boundary conditions are required for solving the inverse problem of determining the fractional order.

Similar to section \ref{sec:fpinns},  we employ an {\it L1} scheme \cite{yang2010numerical} for numerical discretisation of the fractional derivative for the left- (\ref{eq:temp-left_unn}) and right-sided (\ref{eq:temp-right_unn}) derivative for the $i^{th}$ training point. A uniform finite difference grid was generated by dividing the domain in $M$ points such that the step size, $h_l$ was kept constant for all the training points, where, $l = \{1, 2, ..., M-1 \}$. The domain for the two-sided model is $[0, 2Re_\tau], ~ y^+ \in (0, 2Re_\tau)$.

\begin{align}  \label{eq:temp-left_unn}
\begin{split}
    &^C_0 D_{y^+_i}^{(\alpha(y^+_i), \lambda)} (\overline{U^+)} =~ \frac{1}{\Gamma(1-\alpha_i)}\int_{0}^{y^+_i} (y^+_i-\zeta)^{-\alpha_i}  { e^{- \frac{ \lambda} { Re_\tau} (y^+_i - \zeta)}} (\overline{U^+} (\zeta))^{'}   d\zeta \\
    &=~ - \frac{1}{\Gamma(1-\alpha_i)}\int_{0}^{y^+_i} (\zeta)^{-\alpha_i}  { e^{-\frac{ \lambda } { Re_\tau} \zeta }}  (\overline{U^+} (y^+_i - \zeta))^{'}     d\zeta \\
     &\approx~ - \frac{1}{\Gamma(1-\alpha_i)}  \sum_{j=0}^{l-1} { \overline{U^+}(y^+_i - jh_l) -  \overline{U^+}(y^+_i - (j+1) h_l)   \over h_l }   { e^{- \frac{ \lambda} { Re_\tau} (jh_l) }}   \int_{jh_l}^{(j+1)h_l} (\zeta)^{-\alpha_i}    d\zeta \\
      &=~ - \frac{h_l^{-1}}{\Gamma(1-\alpha_i)}  \sum_{j=0}^{l-1} { (\overline{U^+}_{l-j} -  \overline{U^+}_{l-j-1}) }   { e^{- \frac{ \lambda} { Re_\tau} (jh_l) }}  \left [ {\zeta^{1-\alpha_i} \over  1 - \alpha_i} \right ]^{(j+1)h_l}_{jh_l}      \\
      &=~ - \frac{h^{-\alpha_i}}{\Gamma(2-\alpha_i)}  \sum_{j=0}^{l-1} { (\overline{U^+}_{l-j} -  \overline{U^+}_{l-j-1}) }   { e^{- \frac{ \lambda} { Re_\tau} (jh_l) }}  \left [ {(j+1)}^{1-\alpha_i} -  {j}^{1-\alpha_i}  \right ]
\end{split}
\end{align}

\noindent
Similarly, we can discretise for the right-sided  tempered fractional derivative as (\ref{eq:temp-right_unn}),

\begin{align}  \label{eq:temp-right_unn}
\begin{split}
    &^C_{y^+_i} D_{2Re_\tau}^{{(\alpha(y^+_i), \lambda)}} (\overline{U^+)} =~ \frac{-1}{\Gamma(1-\alpha_i)}\int_{y_i^+}^{2 Re_\tau} (\zeta - y^+_i)^{-\alpha_i}  { e^{-  \frac{\lambda} { Re_\tau} (\zeta - y^+_i) }} (\overline{U^+} (\zeta))^{'}   d\zeta \\
       & \approx ~ \frac{- h_l^{-\alpha_i}}{\Gamma(2-\alpha_i)}  \sum_{j=0}^{M-l-1} (\overline{U^+}_{l+j} -  \overline{U^+}_{l+j+1}) { e^{- \frac{ \lambda} { Re_\tau} (jh_l) }} \left [ { (j+1)^{1-\alpha_i}   - (j)^{1- \alpha_i}} \right ]
\end{split}
\end{align}

The loss function for the feed forward neural network uses the discrete form for the left- (\ref{eq:temp-left_unn}) and right-sided (\ref{eq:temp-right_unn}) definition. Similar to section \ref{sec:fpinns}, the loss function  does not have a data part and solely comprises of the equation part, as the fractional order is found in a pointwise sense.  The input of the neural network is the $i-th$ spatial location ($y^+_i$) and the loss function for $i-th$ training point is written as (\ref{eq:loss_temp}). The velocity is supplied from the DNS database \cite{wu2008direct, avsarkisov2014turbulent}. Algorithm \ref{alg:1} gives more details for finding the fractional order, given a tempering parameter $\lambda$.

\begin{equation} \label{eq:loss_temp}
    Loss = L_e = \left [ {1 \over 2} \left ( {}^C_0 D_{y^+_i}^{(\alpha(y^+_i), \lambda)} (\overline{U^+)} ~-~  {}^C_{y^+_i} D_{2Re_\tau}^{{(\alpha(y^+_i), \lambda)}} (\overline{U^+)} \right ) ~ - ~ \tau^+(y^+_i) \right ]^2
\end{equation}

For the truncated definition, we employ the same finite difference scheme, over the same grid for 
truncated left-sided as (\ref{eq:trunc-left_unn}) and  right sided as (\ref{eq:trunc-right_unn}). The only difference the domain, for the fractional derivative the domain was $[0, 2 Re_\tau]$, while for the truncated definition the domain is $[y^+ - \delta^+, y^+ + \delta^+]$, where $\delta^+$ is a composite function (refer section \ref{sec:temp}) and $y^+ \in (0, 2Re_\tau)$. A uniform grid is generated with $N$ points such that the step size, $h_k$ was kept constant and $k = \{1, 2, 3, 4 \dots N-1 \}$. For an efficient implementation, a fraction of the grid from the previous case with step size $h_l$ is used, by choosing $h_k = h_l$, thus $0 < N \leq M$.

\begin{align}  \label{eq:trunc-left_unn}
\begin{split}
    &^C_0 D_{y^+}^{(\alpha(y^+_i), \delta^+)} (\overline{U^+)} \approx \\ & -{h_k^{-\alpha(y^+_i)} \over \Gamma(2 - {\alpha(y^+_i)})}  \sum_{j=0}^{k-1} (\overline{U^+}_{{k-j}} -  \overline{U^+}_{k-j-1}) [ (j+1)^{1-{\alpha(y^+_i)}} -  j^{1-{\alpha(y^+_i)}} ]
\end{split}
\end{align}

\begin{align} \label{eq:trunc-right_unn}
\begin{split}
    &^C_{y^+} D_{2Re_\tau}^{({\alpha(y^+_i)}, \delta^+)} (\overline{U^+)} \approx \\& -{h_k^{{-\alpha_(y^+_i)}} \over \Gamma(2 - {\alpha(y^+_i)})} \sum_{j=0}^{N-k-1} (\overline{U^+}_{{k+j}} -  \overline{U^+}_{{k+j+1}}) [ (j+1)^{1-{\alpha(y^+_i)}} -  j^{1-{\alpha(y^+_i)}} ]
\end{split}
\end{align}

The loss function for the feed forward neural network uses the discrete form for the left- (\ref{eq:trunc-left_unn}) and right-sided (\ref{eq:trunc-right_unn}) definition. Similar to section \ref{sec:fpinns}, the loss function  does not have a data part and solely comprise of the equation part, as the fractional order is found in a pointwise sense.  The input of the neural network is the $i-th$ spatial location ($y^+_i$) and the loss function for $i-th$ training point is written as (\ref{eq:loss_trunc}). The velocity is supplied from the DNS database  \cite{wu2008direct, avsarkisov2014turbulent}. Algorithm \ref{alg:1} gives more details for finding the fractional order, given a truncating parameter $\delta^+$.

\begin{equation} \label{eq:loss_trunc}
    Loss = L_e = \left [ {1 \over 2} \left ( ^C_0 D_{y^+}^{(\alpha(y^+_i), \delta^+)} (\overline{U^+)} ~-~  ^C_{y^+} D_{2Re_\tau}^{({\alpha(y^+_i)}, \delta^+)} (\overline{U^+)} \right ) ~ - ~ \tau^+(y^+_i) \right ]^2
\end{equation}

\begin{algorithm} 
\caption{Pointwise computation of fractional order $\alpha(y^+)$ for tempered or truncated fractional two-sided model}\label{alg:1}
%\DontPrintSemicolon
%\SetAlgoLined
\KwData{Domain, $\Omega = [0, 2Re_\tau]$} 
\KwData{Velocity, $\overline{U}^+(y^+)$} 
\KwData{Fix tempering or truncating parameter, $\xi = \{\lambda, \delta^+\}$} 
\KwResult{$\alpha (y^+)$}
\SetKwFunction{Frac}{compute\_fractional\_order}
\SetKwFunction{Res}{compute\_residual}
\SetKwProg{FnR}{Function}{:}{\KwRet $residual$}
\SetKwProg{FnA}{Function}{:}{\KwRet $\alpha$}
\SetKwProg{Mn}{Main}{:}{End}

\FnR{\Res{$y^+$, $\overline{U}^+(y^+)$, $\xi$}}
    {
        \Comment{Compute left-sided derivative for a fixed $\xi$} 
        $D^\xi_l \gets$ Use (\ref{eq:temp-left_unn}) for tempered and (\ref{eq:trunc-left_unn}) for truncated formulation \;
        \Comment{Compute right-sided derivative for a fixed $\xi$} 
        $D^\xi_r \gets$ Use (\ref{eq:temp-right_unn}) for tempered and (\ref{eq:trunc-right_unn}) for truncated formulation \;
        \Comment{Residual for two-sided model} 
        $residual \gets 0.5 ~ (D^\xi_l - D^\xi_r)$ \;
    }
\FnA{\Frac{$y^+$, $\overline{U}^+(y^+)$, $\xi$}}
    {
        \Comment{Computes $\alpha$ for supplied $y^+$} 
        \Comment{Given: a fixed $\xi$ and $\overline{U}^+(y^+)$ }
        initialize neural network \;
        \While{ $loss > threshold$ }
        {
            \Comment{get residual value of equation}
            $residual \gets$ \Res{{$y^+$, $\overline{U}^+(y^+)$, $\xi$}} \;
            $loss \gets residual$\;
            train neural network\;
        }
    }
%initialization\;
\Mn{}{
    $y\_train \gets$ discrete points in $(\Omega)$\;
    \For { $y^+ \in$ y\_train}
    {
        \Comment{iterates over the list $y\_train$}
        $\alpha (y^+) \gets$ \Frac{$y^+$, $\overline{U}^+(y^+)$, $\xi$}   \;
    }
}
\end{algorithm}

\subsection{Horizon of non-local interactions and finding the equivalence between the truncated and tempered fractional derivative} \label{sec:horizon}

Recognise, the parameter $\delta^+$ associated with the truncated definition is also the horizon of non-local interactions. 

\begin{definition}
The horizon of non-local interactions ( $\delta^+$) is a {\it distance beyond which no long-range interactions occur}. 
\end{definition}

If there exist such a distance $\delta^+$,  such that the fractional derivative defined over a infinite (or semi-infinite) domain is equivalent to its truncated definition with a compact support, then (\ref{eq:nl_left_horzon}) holds. Following our notations, $ {}_{x - \delta^+} {}^C D_{x} ^{\alpha} f(x) ~=~ {}_{-\infty} {}^C D_{x} ^{(\alpha,  {\delta^+})} f(x)$.  

\begin{equation} \label{eq:nl_left_horzon}
    \left |  _{-\infty} {}^C D_{x} ^{\alpha } f (x) ~~ - ~~    {}_{x - \delta^+} {}^C D_{x} ^{\alpha} f(x) \right | < \epsilon
\end{equation}

Note that the kernel is power-law, thus for any arbitrary function $f(x)$, (\ref{eq:nl_left_horzon}) may not be true. However, if the function decays then there exists a $\delta^+$ (horizon of non-local interactions) which can be found  to an arbitrary accuracy ($\epsilon$). Physical process such a turbulence decays after a long-enough distance, indeed the function (such as velocity) decays too.  Similarly, we can extend this argument for the right-sided derivatives.

\begin{remark}
    The horizon of non-local interaction ($\delta^+$) significantly reduces the computations by finding an equivalent smaller domain, without loss of rigour. 
\end{remark}

For the case of tempered and truncated definitions, we can define equivalence between the two definitions such that the fractional order has to be same, then (\ref{eq:eqi_left}) holds. Note that, $\delta^+$ may not represent the horizon of the non-local interactions. Since, we do not allow the function (in this case velocity) to decay naturally (as a result of dissipation).

\begin{equation} \label{eq:eqi_left}
    \left |~  _{0} {}^C D_{y^+} ^{(\alpha,  {\lambda})} f (x) ~~ - ~~    {}_{0} {}^C D_{y^+} ^{(\alpha,  {\delta^+})} f(x) ~ \right | < \epsilon
\end{equation}

Similarly for the right-sided definitions (\ref{eq:eqi_right}),

\begin{equation} \label{eq:eqi_right}
    \left |~  _{y^+} {}^C D_{2 Re_\tau} ^{(\alpha,  {\lambda})} f (x) ~~ - ~~    {}_{y^+} {}^C D_{2 Re_\tau} ^{(\alpha,  {\delta^+})} f(x) ~ \right | < \epsilon
\end{equation}

In order to compute $\delta^+$, we start with the tempered f-RANS, compute the fractional order for a given $\lambda$, then use this fractional order to compute the parameter $\delta^+$ (of truncated f-RANS) by gradually increment $\delta^+$ until (\ref{eq:eqi_left}) or (\ref{eq:eqi_right}) holds. The discrete form is used for this purpose. Note that, for this exercise, we will have left $\delta^+_l$ and right $\delta_r^+$ associated with left- and right-sided truncated definition, respectively. Algorithm \ref{alg:2} is used for finding $\delta^+_l$ and $\delta_r^+$. It is not assumed a-prior that $\delta^+_l$ and $\delta_r^+$ are a constant function, but treated as spatially varying function, as we find the $\delta^+_l$ and $\delta_r^+$ in a pointwise sense. For  $\delta^+$ being constant function or $\delta^+_l = \delta_r^+$ are just a special case of our consideration. 

\begin{algorithm} 
\caption{To find $\delta^+$ of truncated f-RANS by supplying $\alpha$ of tempered f-RANS for a given $\lambda$.}\label{alg:2}
%\DontPrintSemicolon
%\SetAlgoLined
\KwData{Domain, $\Omega = [0, 2Re_\tau]$} 
\KwData{Velocity, $\overline{U}^+(y^+)$} 
\KwData{Get $\alpha(y^+)$ for a given $\lambda$ of tempered f-RANS from Algorithm \ref{alg:1}}
\Comment{$\delta^+_l (y^+)$ and $\delta^+_r (y^+)$ are associated with left- (\ref{eq:trunc-left_unn}) and right-sided (\ref{eq:trunc-right_unn}) truncated fractional derivative, respectively.} 
\KwResult{truncating parameter, $\delta^+_l (y^+)$ and $\delta^+_r (y^+)$}
\SetKwFunction{FracL}{compute\_truncated\_left\_der}
\SetKwFunction{FracR}{compute\_truncated\_right\_der}
\SetKwFunction{FracC}{compute\_temp\_fractional\_der}
\SetKwProg{FnC}{Function}{:}{\KwRet  {$D^\lambda_l$, $D^\lambda_r$}}
\SetKwProg{FnL}{Function}{:}{\KwRet  $D_l^{\delta^+_l}$}
\SetKwProg{FnR}{Function}{:}{\KwRet $D_r^{\delta^+_r}$}
\SetKwProg{Mn}{Main}{:}{End}
\FnC{\FracC{$y^+$, $\overline{U}^+(y^+)$, $\alpha$, $\lambda$}}
    {
        \Comment{Compute left-sided tempered derivative with fixed $\lambda$} 
        $D^\lambda_l \gets$ Use (\ref{eq:temp-left_unn}) for left-sided tempered derivative derivative \;
        \Comment{Compute right-sided tempered derivative with fixed $\lambda$} 
        $D^\lambda_r \gets$ Use (\ref{eq:temp-right_unn}) for right-sided tempered derivative derivative \;
    }
\FnL{\FracL{$y^+$, $\overline{U}^+(y^+)$, $\alpha(y_i^+)$,  $\delta^+_l$} }
    {
        \Comment{Compute left-sided derivative for the supplied $\delta^+_l$} 
        $D_l^{\delta^+_l} \gets$ Use (\ref{eq:trunc-left_unn}) for left-sided truncated formulation \;
    }
\FnR{\FracR{$y^+$, $\overline{U}^+(y^+)$, $\alpha(y_i^+)$,  $\delta^+_r$} }
    {
        \Comment{Compute right-sided derivative for the supplied $\delta^+_r$} 
        $D_r^{\delta^+_r} \gets$ Use (\ref{eq:trunc-right_unn}) for right-sided truncated formulation \;
    }
\Mn{}
{
    set default $threshold$, $\delta^+_l$ and $\delta^+_r$\ \;    
    \For{ $y^+_i \in y\_train = $ discrete-points $(\Omega)$ }
    {
        \Comment{Get the value for left- and right-sided tempered fractional derivative}
        $D^\lambda_l, ~ D^\lambda_r \gets$ \FracC{$y^+$, $\overline{U}^+(y^+)$, $\alpha(y_i^+)$, $\lambda$}\;
        \While{ $error > threshold$}
        {
            \Comment{This while-loop computes the min. $\delta_l^+$}
            \Comment{Get the value for left-sided truncated fractional derivative}
            $D_l^{\delta^+_l} \gets$ \FracL{$y^+$, $\overline{U}^+(y^+)$, $\alpha(y_i^+)$,  $\delta^+_l$} \;
            $error \gets |D^\lambda_l - D_l^{\delta^+_l}|$ or $|D^\lambda_l - D_l^{\delta^+_l}| / |D^\lambda_l |$\;
            increment $\delta^+_l$ \;
        }
        \While{ $error > threshold$}
        {
            \Comment{Similarly for the right definition, compute min. $\delta_r^+$}
            %\Comment{Get the value for right-sided truncated fractional derivative}
            $D_r^{\delta^+_r} \gets$ \FracR{$y^+$, $\overline{U}^+(y^+)$, $\alpha(y_i^+)$, $\delta^+_r$} \;
            $error \gets |D^\lambda_r - D_r^{\delta^+_r}|$ or $|D^\lambda_r - D_r^{\delta^+_r}| / |D^\lambda_r |$\;
            increment $\delta^+_r$ \;
        }
    }
}
\end{algorithm}

%%%%%%%%%%%%%%%%%%%%%%%%%%%%%%%%%%%%%%%%%%%%%%%%%%%%%%%%%%%%%%%%%%%%%%%%%%%%%%%%%%%%%%%%%%%%%%%%%%%%%%%%%%%%%%%%%%%%%%%%%%%%%%%%%%%%%%%%%%%%
\section{Discussion for tempered and truncated f-RANS models} \label{sec:temp_results}

Having introduced tempered and truncated definition in the previous sections, we discuss the result of our computational experiments for channel and couette flow. The first two sub-section discusses about the behavior of fractional order for these definition. While, the third sub-section, discusses about the equivalence between the two.  Note that, the horizon of non-local interaction was defined in section \ref{sec:horizon}.

\subsection{Discussion for tempered f-RANS}

\begin{figure} [hbt!]
\centering
    \subfloat[Couette]{{\includegraphics[ width=0.5\textwidth]{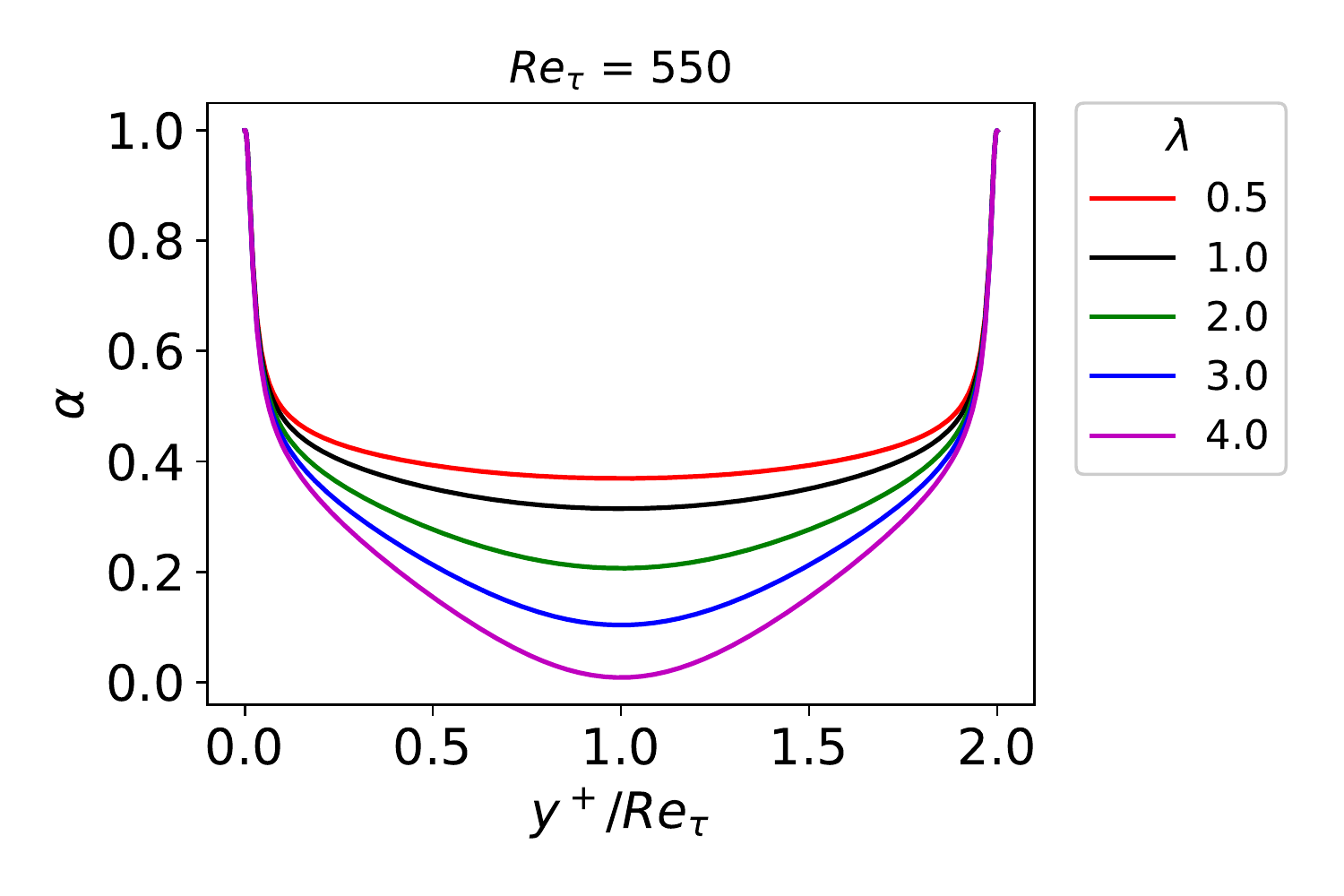}}}%
   \subfloat[Channel]{{\includegraphics[    width=0.5\textwidth]{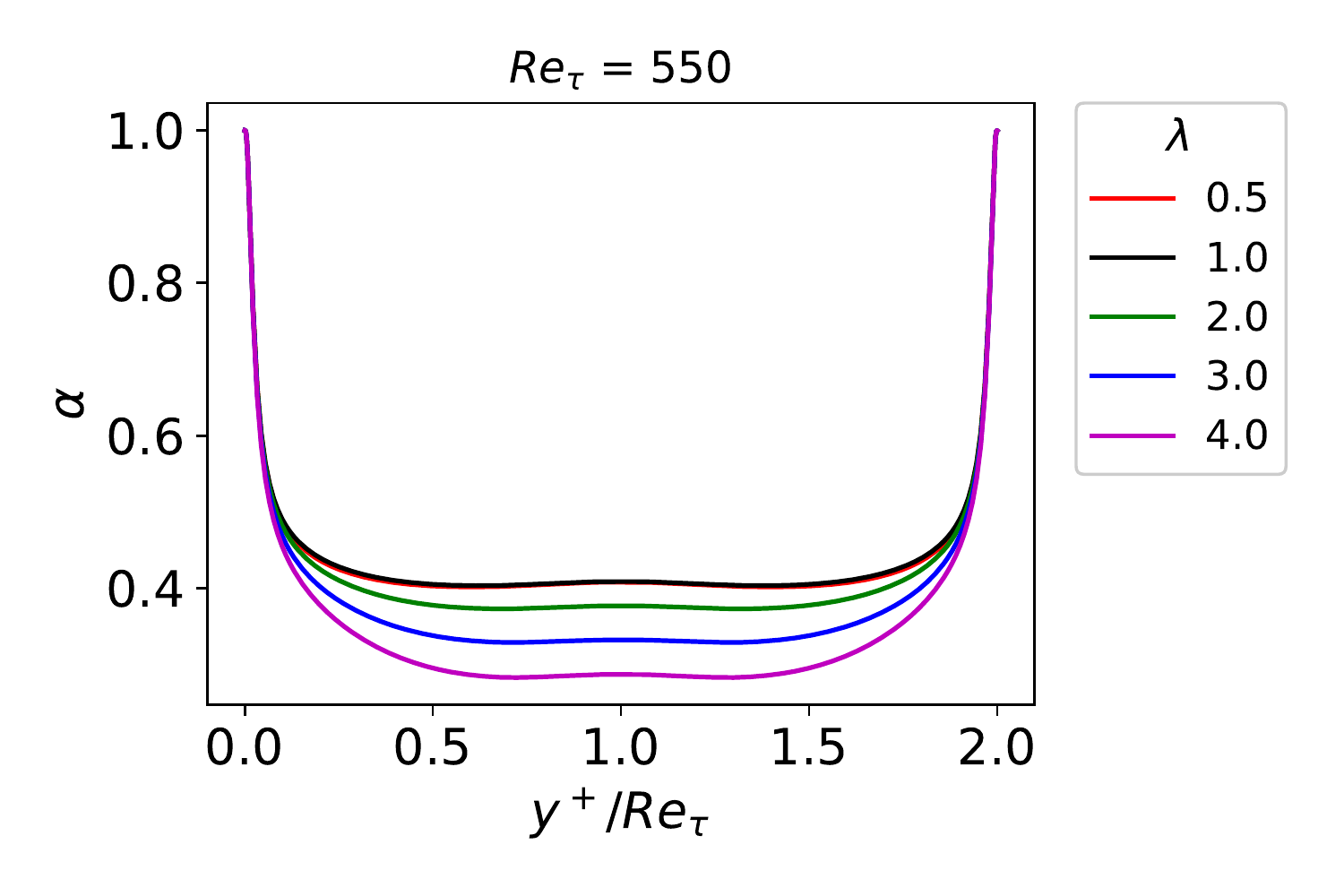}}}%
    \caption{Fractional order of tempered f-RANS model for Channel and Couette for the given tempering parameter $\lambda$.}
    \label{fig:frac_ydel_2sided_temp}
\end{figure}

We first look at the fractional order obtained for the tempered f-RANS model (fig.~\ref{fig:frac_ydel_2sided_temp}). For a fixed tempering parameter $\lambda$,  similar results are obtained as in the case of two-sided f-RANS model as discussed in section , where we can see it's two main attributes in fig.~\ref{fig:frac_ydel_2sided_temp}, 

\begin{itemize}
    \item The fractional order near the wall (viscous sub-layer) is unity. This implies the tempered fractional derivative coincides with the classical integer-order derivative, which is a local operator. This is a physically consistent result as the flow is dominated by the viscous action and with in the absence of turbulence within the viscous-sub layer, the flow is laminar and no non-locality is present.
    \item As we move away from the viscous sub-layer onto the buffer region and outer turbulent region, the fractional order gradually decreases to address the turbulence (indeed, for non-integer values the fractional derivative are essentially non-local). This is a physically consistent result. 
    \item Essentially in the buffer region both local and non-local process exists due to viscous and turbulent actions, respectively. For such regions, the fractional order addresses the amalgamation of both local and non-local process. Indeed, for our discussion here, the tempered f-RANS model address the duality of local and non-local process.  
\end{itemize}

Since the fractional-order of tempered f-RANS (fig.~\ref{fig:frac_ydel_2sided_temp}) has similar characteristics as two-sided f-RANS model, the results obtained for the two-sided model in section \ref{sec:two-sdied}, it will apply for these case too, 

\begin{itemize}
    \item As $Re_\tau \rightarrow \infty$, the fractional order asymptotes. 
    
    \item $\forall y^+ / Re_\tau \in [0,2]$, the fractional order will uniformly asymptotes as $Re_\tau \rightarrow \infty$. In order words, if we fix $ y^+ / Re_\tau \in [0,2]$ and the tempering parameter ($\lambda$), the fractional order will reach an asymptotic limit as $Re_\tau \rightarrow \infty$.
\end{itemize}

Thus the tempered f-RANS model is valid for all Reynolds number. Further, a power law and wake region can be deduced in a similar fashion as in the section \ref{sec:two-sdied}. 

We also note from (fig.~\ref{fig:frac_ydel_2sided_temp}), as we increase the value of tempering parameter $\lambda$, the fractional order is lowered for the same Reynolds number and $y^+/Re_\tau$. 

\begin{itemize}
    \item Of course, this is not an indication of non-locality across $\lambda$, but it shows there are multiple solutions. Thus, the true measure of non-locality is not just the fractional order but also the tempering parameter
    \item If we continue to increase the value of tempering parameter, $\lambda$, after a certain value there are no solution to be found. This indicates, there exist a minimum domain, but no maximum.
\end{itemize}

The above two points would be great interest to future computational studies, where they can balance the coupling of the fractional order and tempering parameter, $\lambda$ for modelling, to mimic the non-locality present in the flow without the loss of rigour.

\subsection{Discussion for truncated f-RANS}

\begin{figure} 
\centering
    \subfloat[]{{\includegraphics[ width=0.5\textwidth]{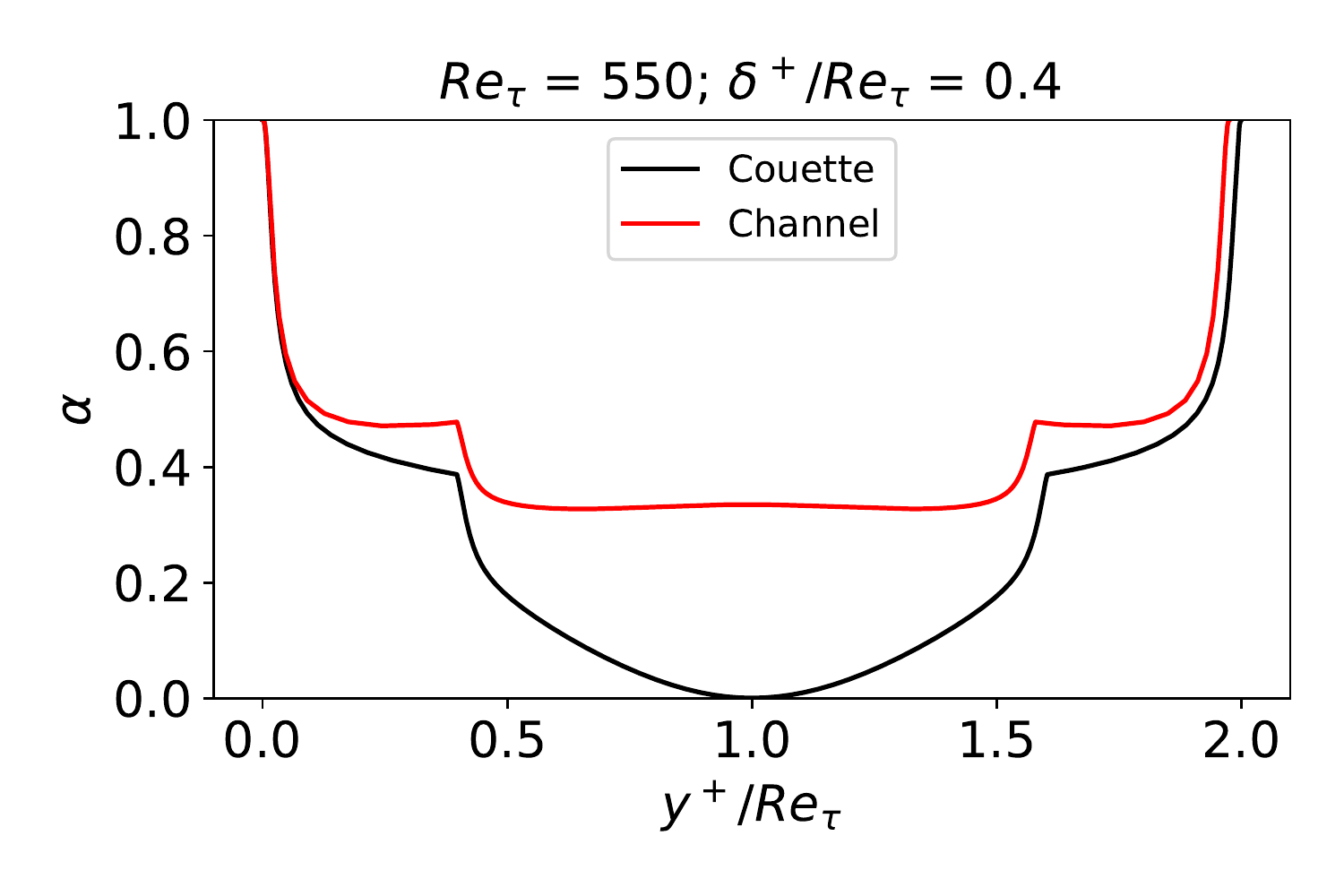}}}%
    \subfloat[]{{\includegraphics[ width=0.5\textwidth]{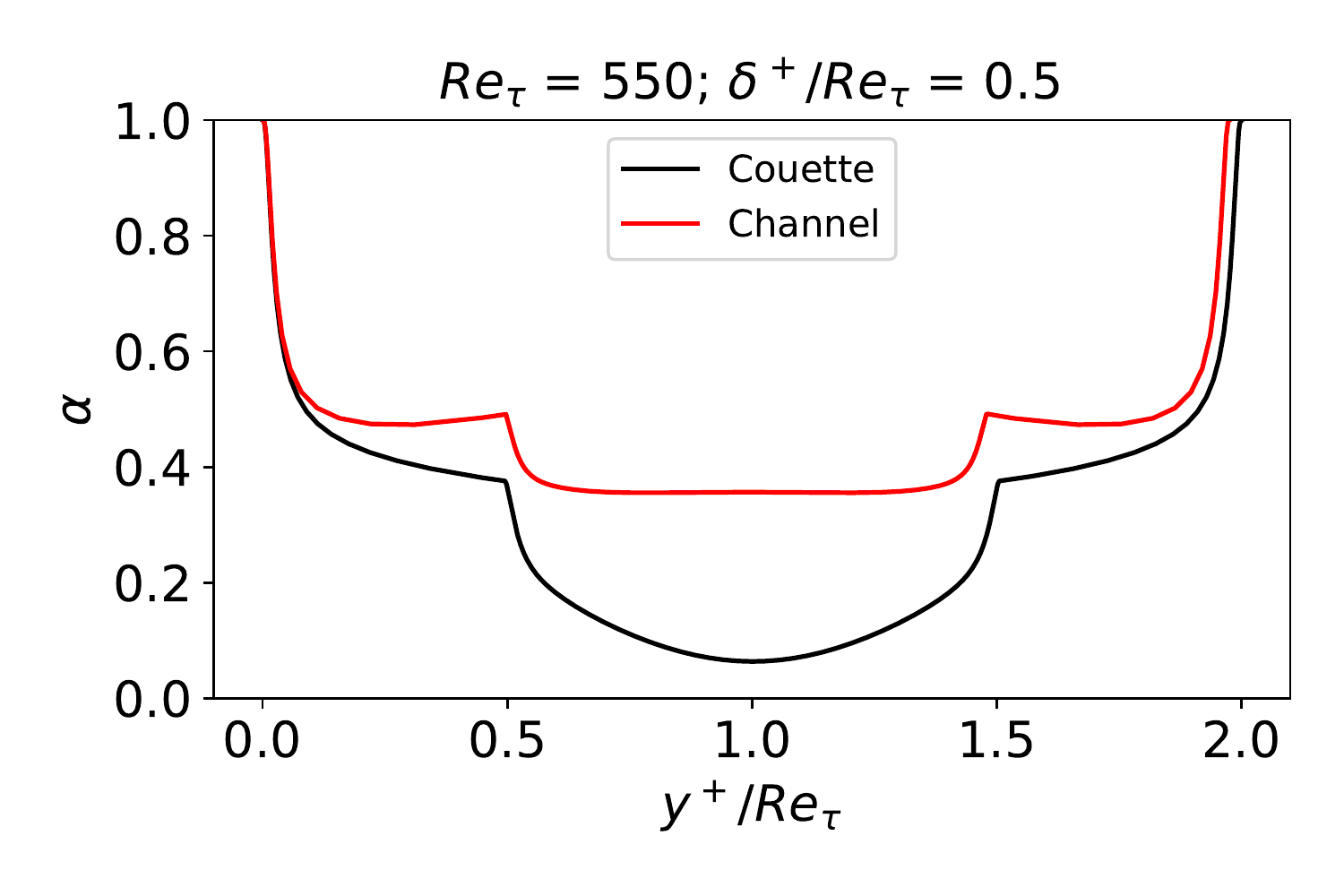}}} \\
     \subfloat[]{{\includegraphics[ width=0.5\textwidth]{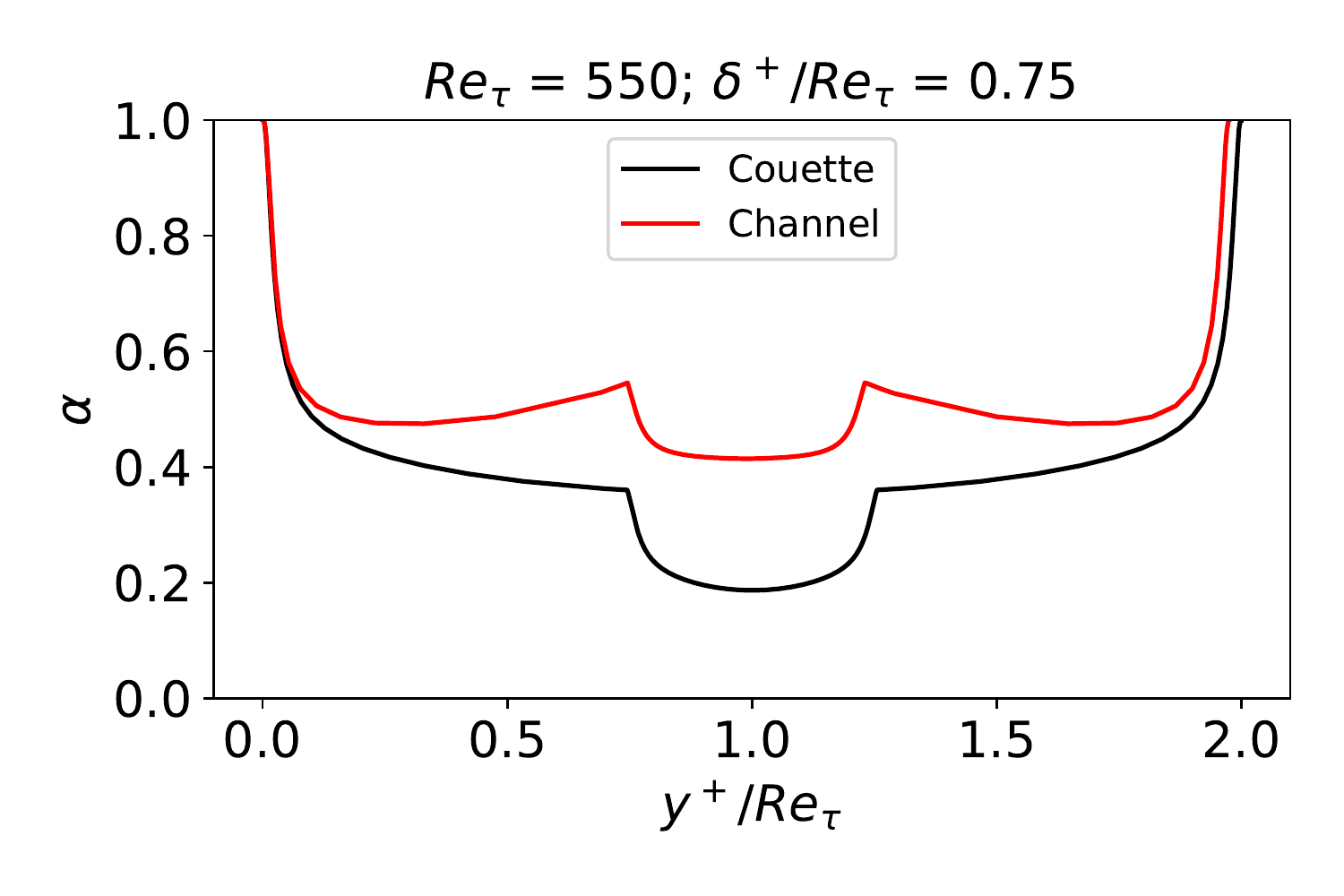}}} 
      \subfloat[]{{\includegraphics[ width=0.5\textwidth]{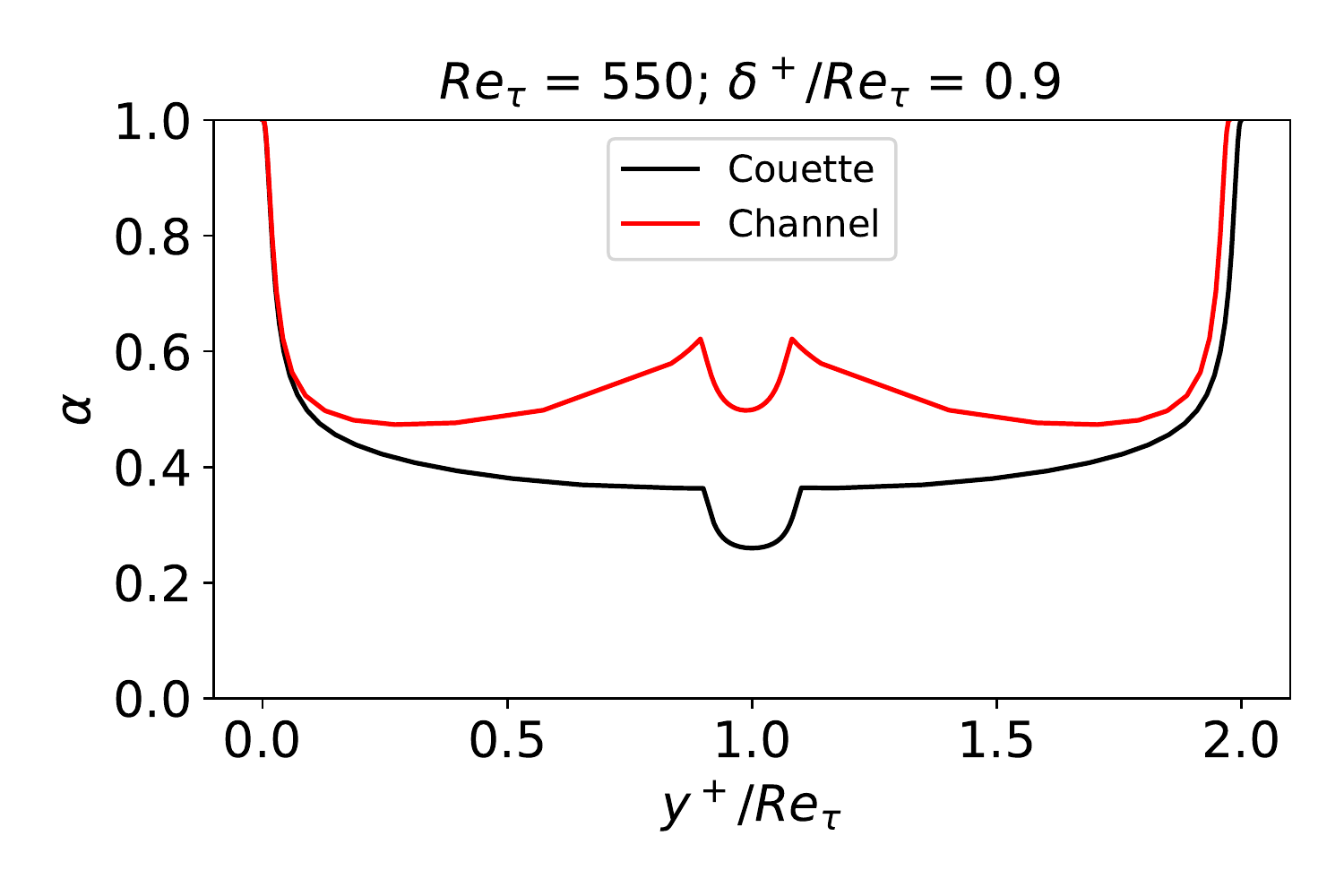}}} \\
       \subfloat[]{{\includegraphics[ width=0.5\textwidth]{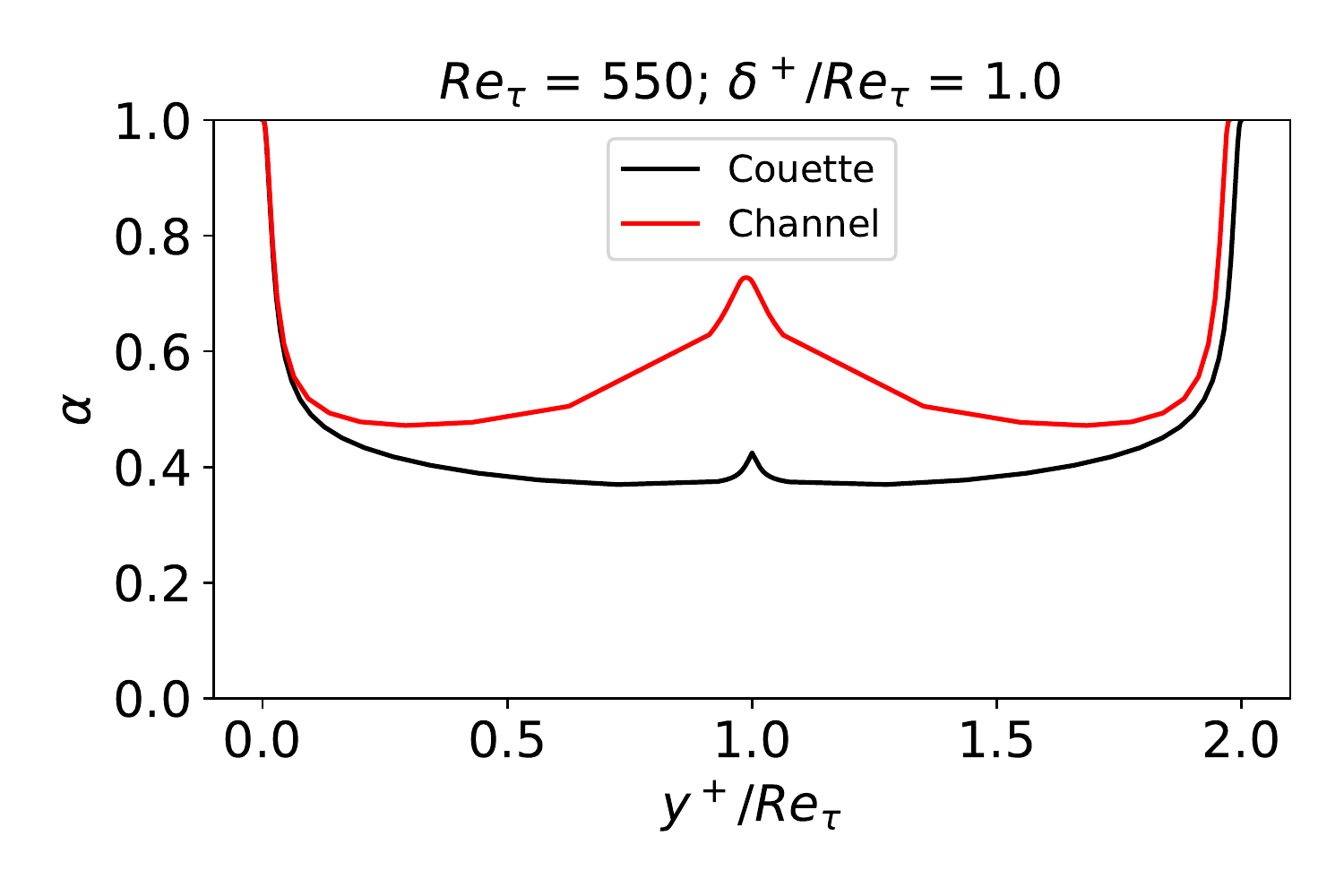}}}%
        \subfloat[]{{\includegraphics[ width=0.5\textwidth]{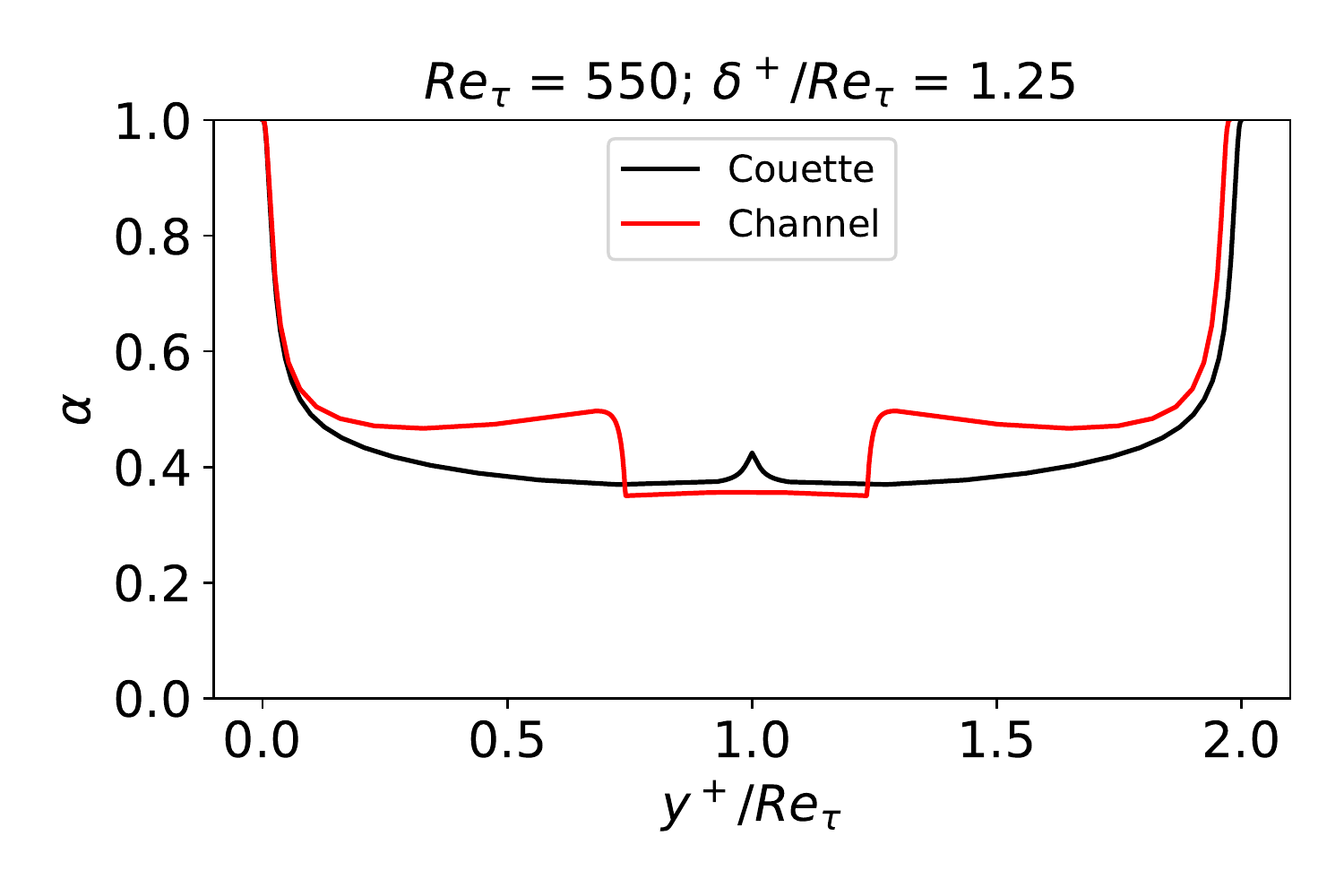}}}%
    \caption{Fractional order of truncated f-RANS model for Channel and Couette for a constant truncating parameter $\delta^+$. Recall that $\delta^+$ is a composite function for wall bounded flows, leading to a non-differentiable $\alpha$ curve.}
    \label{fig:frac_ydel_2sided_trunc}
\end{figure}

For our case of truncated f-RANS, $\delta^+ \in (0, 2Re_\tau)$, since the domain for this two-sided defination is $[0, 2 Re_\tau]$. Inferring from fig. \ref{fig:frac_ydel_2sided_trunc}, we observe,

\begin{itemize}
    \item The curve of fractional order is non-differentiable at the quoted $\delta^+ / Re_\tau$ values, since for wall bounded flows, $\delta^+$ is a composite function. Thus the domain of integration switches abruptly at the  quoted $\delta^+ / Re_\tau$ (or $y^+ / Re_\tau$). 
    
    \item Near the wall (within the viscous sub-layer), as in the previous cases the fractional order is unity, thus a local operator. This is physically consistent. Since, the flow is local in this region, the effect of truncation is not seen, since it only needs immediate neighbourhood.

    \item The couette flow still shows a notation of physical consistency, where the fractional order is lower in the regions of higher turbulence. While, the channel flow is a counter example for some values of $\delta^+$, we loosely attribute it to its strong symmetry. Thus, the $\delta^+$ needs to be chosen in a judicious manner to retain the physically interpretation of the fractional order.

    \item Similar to tempered f-RANS, we observe, the truncated f-RANS has multiple solutions too. Again, there is a minimum domain, upon further reducing the domain, no solutions can be found. 
\end{itemize}

\subsection{Equivalence between truncated and tempered f-RANS}

 By equivalence we imply that the fractional order of tempered f-RANS can be used for truncated f-RANS, and vice-versa. The fractional order is an important ingredient as a measure of non-locality, while the tempering parameter ($\lambda , \delta^+$) is associated with smooth or sharp cut-off. 
 
\begin{figure} [hbt!]
\centering
    \subfloat[Couette]{{\includegraphics[ width=0.5\textwidth]{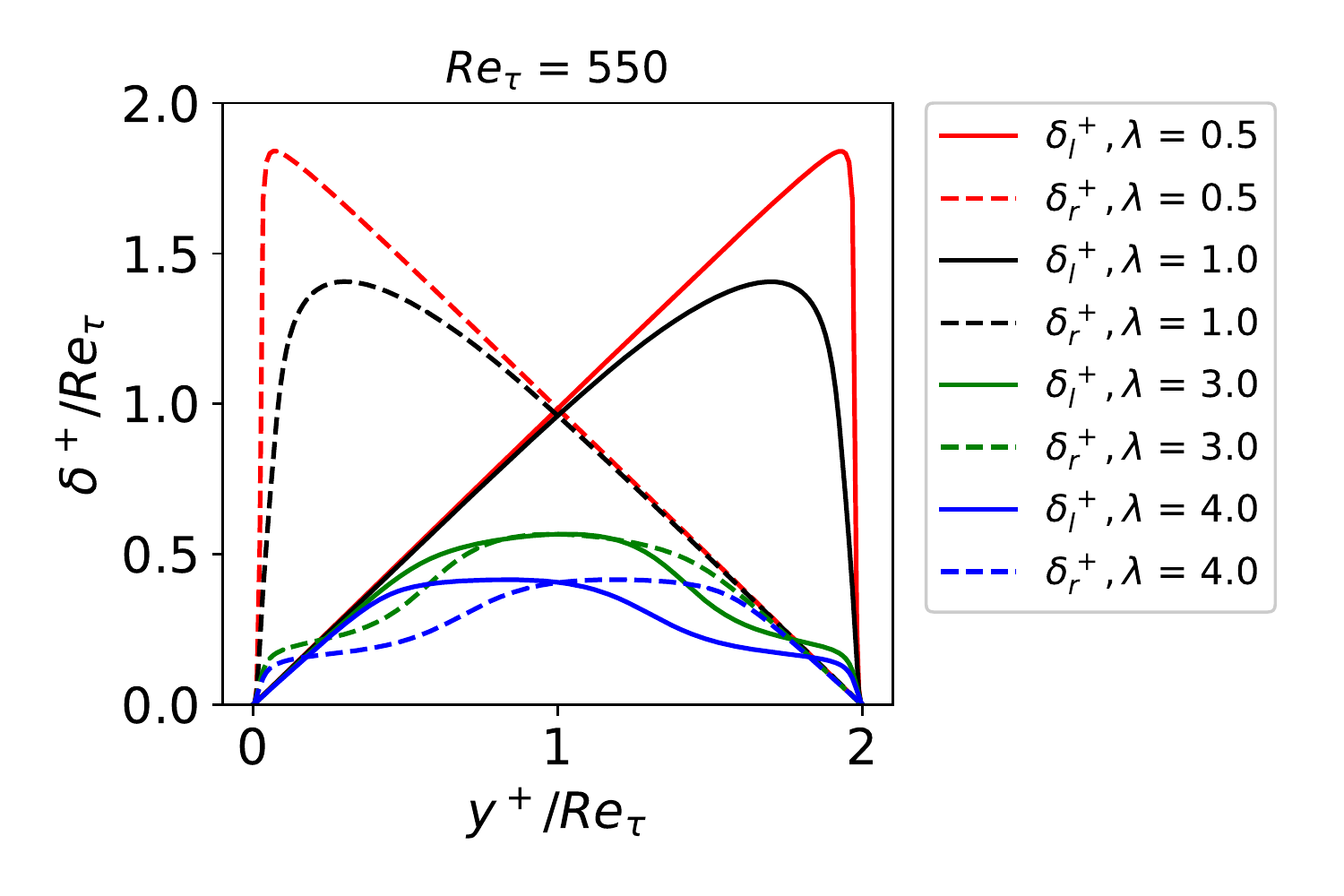}}}%
   \subfloat[Channel]{{\includegraphics[    width=0.5\textwidth]{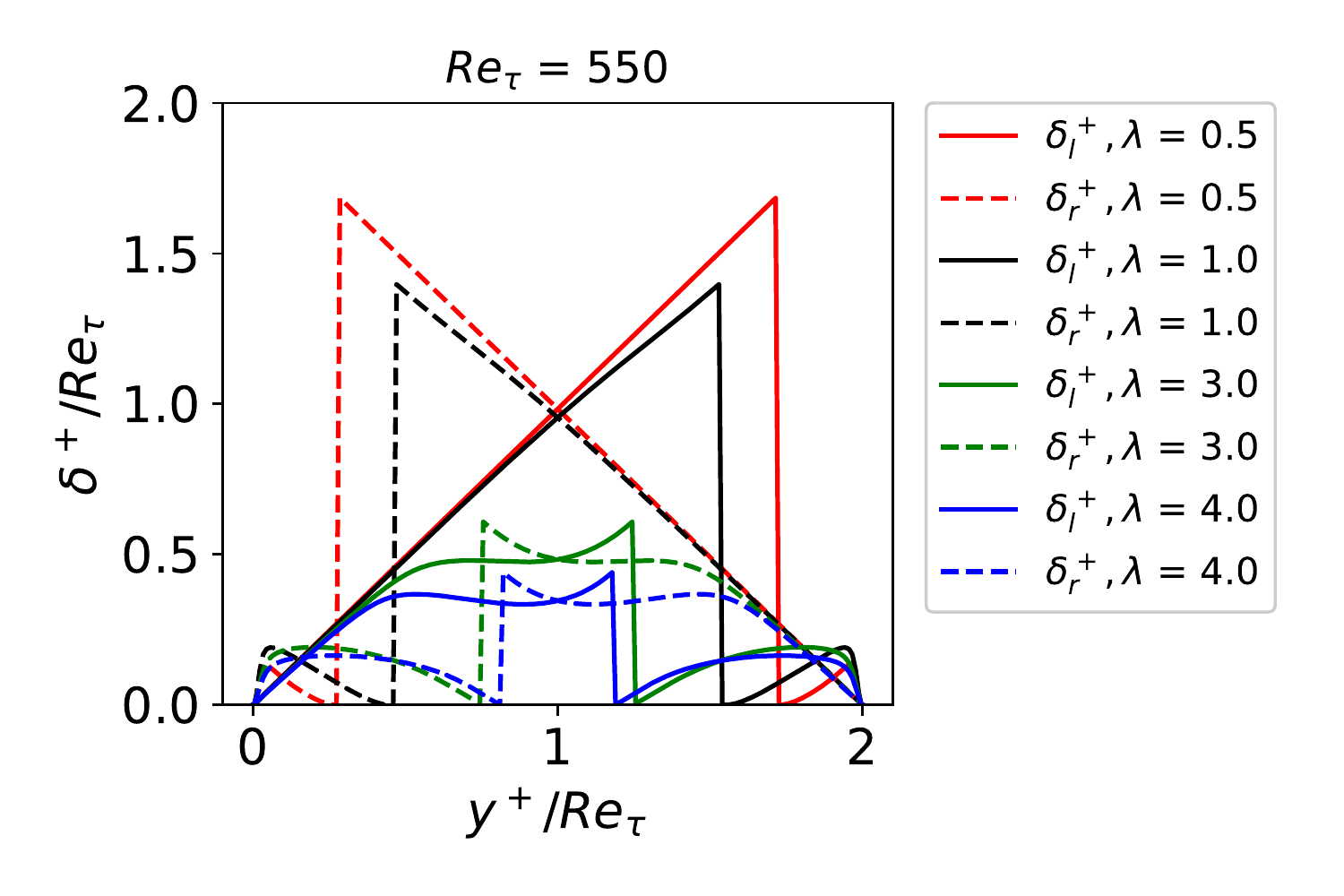}}}%
    \caption{Equivalence between tempered and truncated f-RANS model for Channel and Couette. In this experiment $\delta^+$ is found such that the truncated f-RANS model has the same fractional order as tempered f-RANS model given $\lambda$. Within the legend ($\delta^+_l, \lambda$) and ($\delta^+_r, \lambda$), the $\delta^+_l$ and $\delta^+_r$ are for the left- and right-sided truncated fractional derivative, respectively for given $\lambda$ of tempered f-RANS. We observe, that $\delta^+$ is a spatial varying function. Also, the left- and right-sided derivatives have different values of $\delta^+$.}
    \label{fig:frac_ydel_2sided_comp}
\end{figure}

 In this exercise, we investigate, the influence over $\delta^+$ associated with truncated f-RANS model, such that it produces the fractional order as that of tempered f-RANS with a given tempering parameter, $\lambda$. Inferring from fig.~\ref{fig:frac_ydel_2sided_comp}, we observe, 

\begin{itemize}
    \item The value for $\delta^+$ is no longer a constant, but a spatially varying function.  Recall from the previous section, that the fractional order of truncated f-RANS has a point of non-differentiable for a constant $\delta^+$ value, while the tempered f-RANS produce smooth curves for a fixed $\lambda$, thus inorder to produces a smooth curve for fractional order for truncated f-RANS,  $\delta^+$ has to be a spatially depended function, and not a constant.

    \item The left- and right-sided definition have different $\delta^+$, which are not equal to each other. 
    
    \item As the tempering parameter $\lambda$ increase, we need a smaller domain of integration as indicated by lowering of $\delta^+$.

    \item For the case of channel, a strong effect of symmetry is seen. 
\end{itemize}

In the literature (as per the authors knowledge) rigours results are obtained with exponential tempering fractional derivative (in our case, tempered f-RANS). By way of equivalence, these results can be translated to truncated f-RANS which are better suited for computations. 

Further, we remind the reader, $\delta^+$ is also the horizon of non-local interactions for the cases, when the fractional derivatives are defined over infinite or semi-infinite domains (refer section \ref{sec:horizon})

\section{Summary} \label{sec:summary}

The eddy viscosity hypothesis of stress-strain relation leads to a non-Fickain diffusion equation with a varying diffusivity ($\nu_t$) to address analomous diffusion process. As an alternative, in the paper, we proposed a fractional stress-strain relationship using variable-order Caputo fractional derivative in space for the closure problem. A fractional derivative is a non-local operator for non-integer order, while for integer order's it is a local operator (classical derivative). We use this fact, to address the amalgamation of local and non-local processes, by the viscous and turbulent action, respectively. Indeed, the fractional stress-strain relationship provides new opportunities for closure modeling, hopefully towards a generalised model applicable to all flow settings. 

It was not very clear at the beginning of our research, what kind of model formulations will lead to a physically meaningful results, thus we started with one-sided model, followed by two-sided model applied for channel, couette and pipe flow and tempered definition applied to channel and couette flow. To facilitate our investigation, algorithms were carefully designed, such as the pointwise fractional physics-informed neural networks, its extensions to tempered definitions and determining the equivalence of truncated and tempered f-RANS.

For the case of one-sided f-RANS model, the fractional order of couette flow was found to be universal. However, our later investigations of channel and pipe flow did not support our claim. This was due to the fact, non-locality was considered only from one wall. However, in reality non-locality at a point is the amalgamation of disturbances from all directions. This led us to formulate the two-sided f-RANS model. 

For the case of two-sided f-RANS model, we found that the fractional order can be decomposed into two distinct parts, the universal and wake regions, refereed to as the fractional law of wake. Such a wake-law also appears in velocity decomposition. We further find the expression for the fractional order in the wake region. Thereby subtracting the wake contribution, we receive an absolute universal curve. This universal curve (beyond the buffer region) shows a power-law scaling. A power-law in fractional order is analogous to log-law of the wall for velocity. We summarise this discussion in Table \ref{tab:summary}. If there was no wake contribution, both the log-law and fractional power-law would extend to infinity. We further note, a characteristic of the power-law is its asymptotic behaviour, thus the two-sided f-RANS model is valid at $Re_\tau \rightarrow \infty$. We verified it by both computational experiments and a asymptotic analysis of the expression of fractional order obtained by an elaborate fitting exercise. Indeed,  expression found for the fractional order in (\ref{eq:anal_couette}), (\ref{eq:anal_channel}) and (\ref{eq:anal_pipe}) solves the closure problem for couette, channel and pipe flow, respectively. In conclusion, the two-sided model leads to physically meaning results, where, 

\begin{itemize}
    \item In the viscous sub-layer the fractional order is unity (a local operator)
    \item The fractional order gradually decreases to address the rise in turbulence intensity (from onset of buffer layer to outer turbulent region) 
\end{itemize}

Thus the two-sided f-RANS model is valid for all regions of the flow. Motivated by the success of the two-sided model, we further formulated the truncated and tempered versions of the two-sided f-RANS model. Tempered fractional calculus ensures the second moment of a fractional derivative is finite. This is important for physical processes, since it does not allow infinite jump length of the L\'evy flight. For wall bounded flow (such as our case), the wall provides a natural truncation, and the second moment will exists. However, for unbounded case, truncated and tempered f-RANS model are important to mimic the physical reality. In a follow up paper, we will employ these definition for boundary layers. However, we were interested in its behaviour for wall bounded flow, thus we investigated for the case of channel and couette flow. 

The tempered f-RANS model provides physically meaningful results, similar to the case of two-sided f-RANS model. Besides, the fractional order, we also have a tempering parameter $\lambda$ as measure of non-locality. This can be a useful tool for future turbulence modeling exercise. Further, we found an equivalence between the tempered and truncated f-RANS models, such that it has the same fractional order. We further note that the truncated f-RANS model has computational advantage as it can be defined over a subset of the original domain. 

For the case of two-sided f-RANS model defined over the infinite or semi-infinite domain, its equivalent truncated f-RANS model can be defined with truncating parameter $\delta^+$, which in this case is the horizon of non-local interactions

All of our above investigations were for steady flows, however, our hypothesis won't change for unsteady flow, since the fractional stress-strain relation is considered in space, but we leave the study of unsteady flows as future work.

%%%%%%%%%%%%%%%%%%%%%%%%%%%%%%%%%%%%%%%%%%%%%%%%%%%%%%%%%%%%%%%%%%%%%%%%%%%%%%%%%%%%%%%%%%%%%%%%%%%%%%%%%%%%%%%%%%%%%%%%%%%%%%%%%%%%%%%%%%%%%%%

\section*{Acknowledgements}
This research was initiated and carried out at Brown University, USA. The author credits Prof. George Karniadakis (Brown University, USA) for proposing the use of fractional calculus for turbulence modeling following \cite{song2018universal}, and mentoring the author during his time at Brown University. 

Research was carried out using computational resources and services at the Center for Computation and Visualization, Brown University.

\bibliographystyle{tfnlm}
\bibliography{frac, turb, frac_review}
\appendix

%%%%%%%%%%%%%%%%%%%%%%%%%%%%%%%%%%%%%%%%%%%%%%%%%%%%%%%%%%%%%%%%%%%%%%%%%%%%%%%%%%%%%%%%%%%%%%%%%%%%%%%%%%%%%%%%%%%%%%%%%%%%%%%%%5
\section{Reynolds-averaged Navier-Stokes Equations} \label{sec:rans}

In this section we derive the Reynolds-averaged Navier-Stokes Equations by time-averaging the Incompressible Navier-Stokes Equations \cite{reynolds1895iv}. 

Consider the velocity-pressure form of Incompressible Navier-Stokes Equations equation in (\ref{eq:3.1}) and the continuity equation (or conservation of mass) in \ref{eq:3.0},

\begin{equation} \label{eq:3.0}
     {\partial U_j \over \partial x_j} = 0 ~~; j = 1, 2, 3.
\end{equation}

\begin{equation} \label{eq:3.1}
   {\partial U_i \over \partial t}  +  U_j {\partial U_i \over \partial x_j} =  - {1 \over \rho} {\partial P \over \partial x_i} +  \nu {\partial^2 U_i \over \partial x^2_j} ~~; ~i,j = 1, 2, 3
\end{equation}

where, $\nu$ is the molecular viscosity and $\rho$, density of fluid. Here, $U_i$ is the instantaneous velocity, which can be decomposed as $U_i = \overline{U_i} + u'_i$, where $\overline{U_i}$ is the time-averaged (or mean) velocity, such that $ \overline{u'_i} = 0$ for a sufficiently large, $T$. Similarly, the instantaneous pressure, $P$ can be decomposed as $P = \overline{P} + p'$, where $\overline{P}$ is the time-averaged (or mean) pressure, such that $ \overline{p'} = 0$ for a sufficiently large, $T$.

\subsection{Time averaging continuity equation}

Aplyting time-avergaing to (\ref{eq:3.0}), we have, 

\begin{equation} \label{eq:3.0.1}
    \overline{\partial  U_i \over \partial x_i} = {1 \over T} \int_{T-t/2}^{T+t/2} {\partial  U_i \over \partial x_i}  dt
\end{equation}

Since its continuous in space and time, (\ref{eq:3.0.1}) can be written as:

\begin{equation} \label{eq:3.0.2}
    \overline{\partial  U_i \over \partial x_i} = {1 \over T} {\partial \over \partial x_i} \int_{T-t/2}^{T+t/2} U_i  dt = {\partial  \overline{U_i} \over \partial x_i}
\end{equation}

Using (\ref{eq:3.0.2}) in (\ref{eq:3.0}), we have,

\begin{equation} \label{eq:3.0.3}
     {\partial \overline{U_j} \over \partial x_j} = 0 ~~; j = 1, 2, 3.
\end{equation}

\subsection{Deriving the RANS equations \cite{reynolds1895iv}}

Using the product rule, the non-linear term is $ U_j {\partial U_i \over \partial x_j} = {\partial U_i U_j \over \partial x_j} - U_i {\partial U_j \over \partial x_j}$, therefore, (\ref{eq:3.1}) is rewritten as (\ref{eq:3.2}),

\begin{equation} \label{eq:3.2}
   {\partial U_i \over \partial t}  + {\partial U_i U_j \over \partial x_j} - U_i {\partial U_j \over \partial x_j} =  - {1 \over \rho} {\partial P \over \partial x_i} +  \nu {\partial^2 U_i \over \partial x^2_j} ~~; ~i,j = 1, 2, 3
\end{equation}

From conservation of mass, it follows, ${\partial U_j \over \partial x_j} = 0$, thus, the incompressible Navier-Stokes Equation takes the form (\ref{eq:3.3}),

\begin{equation} \label{eq:3.3}
   {\partial U_i \over \partial t}  + {\partial U_i U_j \over \partial x_j}  =  - {1 \over \rho} {\partial P \over \partial x_i} +  \nu {\partial^2 U_i \over \partial x^2_j} ~~; ~i,j = 1, 2, 3
\end{equation}

Applying time average on both side of (\ref{eq:3.3}). Here, the primary assumption is the time-scale $T$ is large enough, we have,

\begin{equation} \label{eq:3.4}
   \overline{{\partial U_i \over \partial t}  +  {\partial U_i U_j \over \partial x_j}} =  - \overline{{1 \over \rho} {\partial P \over \partial x_i} +  \nu {\partial^2 U_i \over \partial x^2_j}} ~~; ~i,j = 1, 2, 3
\end{equation}

By linearity, we have,

\begin{equation} \label{eq:3.5}
   \overline{{\partial U_i \over \partial t}}  +  \overline{{\partial U_i U_j \over \partial x_j}} =  -{1 \over \rho} \overline{{\partial P \over \partial x_i}} +  \nu \overline{{\partial^2 U_i \over \partial x^2_j}} ~~; ~i,j = 1, 2, 3
\end{equation}

We begin with the terms on the right of eq. (\ref{eq:3.5})

\begin{equation} \label{eq:3.6}
    \overline{\partial^2  U_i \over \partial x_j^2} = {1 \over T} \int_{T-t/2}^{T+t/2} {\partial^2  U_i \over \partial x_j^2}  dt
\end{equation}

Since its continuous in space and time, (\ref{eq:3.6}) can be written as:

\begin{equation} \label{eq:3.7}
    \overline{\partial^2  U_i \over \partial x_j^2} = {1 \over T} {\partial^2 \over \partial x_j^2} \int_{T-t/2}^{T+t/2} U_i  dt = {\partial^2  \overline{U_i} \over \partial x_j^2}
\end{equation}

Similarly, the pressure term can be treated as, 

\begin{equation} \label{eq:3.8}
    \overline{\partial  P \over \partial x_i} = {1 \over T} \int_{T-t/2}^{T+t/2} {\partial  P \over \partial x_i}  dt = {1 \over T} {\partial \over \partial x_i} \int_{T-t/2}^{T+t/2} P  dt = {\partial  \overline{P} \over \partial x_i}
\end{equation}

Now, we look at the terms on the left of (\ref{eq:3.5}), for the temporal term of (\ref{eq:3.5}) apply the separation of scales:

\begin{equation} \label{eq:3.9}
    \overline{\partial U_i \over \partial t} = {1/T} \int_{T-t/2}^{T+t/2} {\partial \over \partial t} \left ( \overline{U_i} + u'_i \right) dt = \overline{\partial \overline{ U_i} \over \partial t} = {\partial  \overline{U_i} \over \partial t}
\end{equation}

This uses the fact the time-averaging of the fluctuations is zero. Now, we treat the second term of (\ref{eq:3.5}), it also relies on the fact, that the time average of fluctuations is zero.

\begin{equation}  \label{eq:3.10}
     \overline{{\partial \over \partial x_j} U_i U_j  } = {1 \over T}\int_{T-t/2}^{T+t/2} {\partial \over \partial x_j} \left [ (\overline{U_i} + u'_i) (\overline{U_j} +  u'_j) \right ] dt = {\partial \overline{U_i}\overline{U_j} \over \partial x_j} + {\partial \overline{u'_i u'_j}  \over \partial x_j} 
\end{equation}

Using (\ref{eq:3.7}, \ref{eq:3.8}, \ref{eq:3.9}, \ref{eq:3.10}), we write the time-averaged form of (\ref{eq:3.3}) as (\ref{eq:3.11}),

\begin{equation} \label{eq:3.11}
   {\partial \overline{U_i} \over \partial t}  + {\partial \overline{U_i U_j} \over \partial x_j}  =  - {1 \over \rho} {\partial \overline{P} \over \partial x_i} +  \nu {\partial^2 \overline{U_i} \over \partial x^2_j} - {\partial \overline{u'_i u'_j} \over \partial x_j} ~~; ~i,j = 1, 2, 3
\end{equation}

Finally, we use the product rule for the non-linear term as, ${\partial \overline{U_i  U_j} \over \partial x_j} = \overline{U_i} {\partial \overline{U_j} \over \partial x_j} + \overline{U_j} {\partial \overline{U_j} \over \partial x_j} $, we have, 

\begin{equation} \label{eq:3.12}
   {\partial \overline{U_i} \over \partial t}  +\overline{U_i} {\partial \overline{U_j} \over \partial x_j} + \overline{U_j} {\partial \overline{U_j} \over \partial x_j}  =  - {1 \over \rho} {\partial \overline{P} \over \partial x_i} +  \nu {\partial^2 \overline{U_i} \over \partial x^2_j} - {\partial \overline{u'_i u'_j} \over \partial x_j} ~~; ~i,j = 1, 2, 3
\end{equation}

From conservation of mass, it follows (\ref{eq:3.0.3}), ${\partial \overline{U_j} \over \partial x_j} = 0$, thus the final form of Reynolds-averaged Navier-Stokes equations \cite{reynolds1895iv} is given, by, 

\begin{equation} \label{eq:3.13}
   {\partial \overline{U_i} \over \partial t} + \overline{U_j} {\partial \overline{U_j} \over \partial x_j}  =  - {1 \over \rho} {\partial \overline{P} \over \partial x_i} +  {\partial \over \partial x_j} \left ( \nu {\partial \overline{U_i} \over \partial x_j} -  \overline{u'_i u'_j} \right ) ~~; ~i,j = 1, 2, 3
\end{equation}

The term $\overline{u'_i u'_j}$ are refereed as Reynolds stresses. The appearance of these new terms leads to a closure problem.

\subsection{Non-dimensional form  of RANS equations in wall units}

Substituting,
$$ \overline{{U^{+}_{i}}} = \frac{ \overline{U_{i}}} {U_{\tau}},~  {x^{+}_{i}} = \frac{x_{i} U_{\tau}}{\nu}, ~ t^+ = {t U^2_\tau \over \nu} , ~ \overline{u_{i}u_{j}}^+ = \frac{\overline{u_{i}u_{j}}}{U^{2}_{\tau}},~ \overline{{P^{+}}} = \frac{\overline{P}}{\rho U^{2}_{\tau}},  \mbox{ and } U_{\tau} = \sqrt{ \frac{\tau_{w}} {\rho}}
$$ in (\ref{eq:3.13}), yields,

\begin{equation}\label{eq:3.14} 
    {U^3_{\tau} \over \nu } {\partial \overline{U^+_i} \over \partial t^+} +  \frac{\overline{U^{+}_{j}} U^{3}_{\tau}} {\nu}\frac{\partial \overline{U^{+}_{i}}} {\partial x_{j}} = -\frac{\rho U^{3}_{\tau} } {\rho \nu} \frac{\partial \overline{P^{+}}} {\partial x^{+}_{i}} +  \frac{\nu U^{3}_{\tau}} {\nu^{2}}\frac{\partial^{2} \overline{U^{+}_{i}}} {\partial x^{+^{2}}_{j}} -  \frac{ U^{3}_{\tau}} {\nu}  \frac{\partial \overline{u_{i}u_{j}}^{+}} {\partial x_{j}}
\end{equation}

Removing the common terms and simplifying, thus we have the final non-dimensional form of RANS in wall units as (\ref{eq:3.15}), 

\begin{equation}\label{eq:3.15} 
 {\partial \overline{U^+_i} \over \partial t^+} + \overline{U^{+}_{j}} \frac{\partial \overline{U^{+}_{i}}} {\partial x^{+}_{j}} = -\frac{\partial \overline{P^{+}}} {\partial x^{+}_{i}} + \frac{\partial} {\partial x^{+}_{j}} \left( \frac{\partial \overline{U^{+}_{i}}} {\partial x^{+}_{j}} - \overline{u_{i}u_{j}}^{+} \right); ~ i, j = {1,2,3}.
\end{equation}

For flows which achieve statistical stantionrity, the temporal term vanishes.

\section{Turbulence Modeling (Eddy-viscosity)} \label{sec:turb}

%Von Neuman proposition of v\_t, followed by prandl

%cmpared to non-fickian equation of richordson

Reynolds stress ($\overline{u'_i u'_j}$) are the additional terms, which appear as a result of time-averaging. This leads to a closure problem, implying that there are more unknowns (or terms) than the number of equations. Indeed the closure problem is due to the fact there is a missing (long-range) information on account of not resolving all the scales of motion whilst directly solving for the mean fields. 

Inorder to close the RANS equations (\ref{eq:3.13}), a large class of eddy-viscosity turbulence models were proposed, which we shall review in this section (although not exhaustively).

\subsection{Linear Eddy Viscosity}

The first attempt to model the Reynolds stresses was the linear-eddy viscosity, which assumes a linear stress-strain relation, as follows, 

\begin{equation}  \label{eq:4.1}
    \overline{u'_i u'_j} = 2/3 k \delta_{ij} - \nu_t \left (  { \partial \overline{U_i} \over \partial x_j} +  { \partial \overline{U_j} \over \partial x_i} \right )
\end{equation}

Here, ($S_{ij} =   { \partial \overline{U_i} / \partial x_j} +  { \partial \overline{U_j} / \partial x_i} $ ) is the strain-rate; and $\nu_t$ is the turbulent viscosity (also, know as eddy or artificial viscosity). The proposition to use eddy viscosity was in \cite{vonneumann1950method}. Further, by plugging (\ref{eq:4.1}) in (\ref{eq:3.13}) and simplifying the diffusion terms are given as (\ref{eq:4.2}),

\begin{equation}  \label{eq:4.2}
    \frac{\partial} {\partial x_{j}} \left( { (\nu + \nu_t) \frac{\partial \overline{U_{i}}} {\partial x_{j}}} \right) + \frac{\partial} {\partial x_{j}} \left( { \nu_t \frac{\partial \overline{U_{j}}} {\partial x_{i}}} \right) ~; ~ i,j = {1,2,3}.
\end{equation}

Here, the effective viscosity,  ($\nu_{eff}$) is given as $\nu_{eff} = \nu  + \nu_t$, where, $\nu_t$ is a spatially varying function. Indeed, (\ref{eq:4.2}) is a more complicated version on non-fickian diffusion equation, which was proposed for turbulence in \cite{richardson1926atmospheric}, where the diffusivity is no longer a constant. This is an important remark, as it bridges the two literature namely, anomalous diffusion using non-Fickian diffusion equation \cite{richardson1926atmospheric} and turbulence modeling using eddy viscosity, where both have non-constant diffusivity. 

\subsubsection{Simple Mixing length model}

In-order to determine, the eddy viscosity ($\nu_t$), the earliest attempt was made by Prandtl \cite{Prandtl}, where ($\nu_t$) is a function of the mixing-length ($l$), 

\begin{equation} \label{eq:4.3}
    \nu_t = l^2 \sqrt{S^2_{ij}}
\end{equation}

Here, the length-scale ($l$) is spatial varying, almost linearly with the distance from the wall. It's evident for complex flows it's impossible to determine a suitable length-scale ($l$). 

Inorder to mitigate the manual prescription of length-scale ($l$), a more complicated one- and two-equation linear eddy viscosity models were proposed. We skip the review of one-equation model and review a more important class of two-equation linear eddy viscosity. 

\subsubsection{$k-\epsilon$ model}

One of the most important and the first practical model is the $k-\epsilon$ model \cite{jones1972prediction} \cite{launder1974application} \cite{LAUNDER1974269}. Indeed, it gave rise to an overwhelming amount of literature and still continues to be in use. In this model, the eddy viscosity is given by (\ref{eq:4.4}),

\begin{equation} \label{eq:4.4}
    \nu_t = {c_\mu k^2 \over \epsilon}
\end{equation}

Here, ($k$) is the turbulent kinetic energy obtained by the its transport equation (\ref{eq:4.5}) and ($\epsilon$) is the dissipation rate, obtained by its transport equation (\ref{eq:4.6}),

\begin{equation}  \label{eq:4.5}
    {D k \over D t} = P_k - \epsilon + {\partial \over \partial x_j} \left [  (\nu + {\nu_t \over \sigma_k}) {\partial k \over \partial x_j}  \right]  
\end{equation}

\begin{equation} \label{eq:4.6}
    {D \epsilon \over D t} = c_{\epsilon_1} { \epsilon \over k } P_k -  c_{\epsilon_2} { \epsilon^2 \over k } + {\partial \over \partial x_j} \left [  (\nu + {\nu_t \over \sigma_{\epsilon}}) {\partial \epsilon \over \partial x_j}  \right]  
\end{equation}

The terms appearing in (\ref{eq:4.4}, \ref{eq:4.5}, \ref{eq:4.6}) are as follows, 

\begin{itemize}
    \item $P_k$ is the generation of turbulent kinetic energy, given as $P_k = - \overline{u'_i u'_j} { \partial U_i / \partial x_j}$. 
    \item  $c_{\epsilon_1}$, $c_{\epsilon_2}$, $c_{\mu}$, $\sigma_k$ and $\sigma_{\epsilon}$ are constants tuned for a given flow. 
\end{itemize}

\subsubsection{$k-\omega$ model}

In order to mitigate the poor performance of $k-\epsilon$ model, a $k-\omega$ model was proposed in \cite{wilcox1988reassessment}, where instead of the ($\epsilon$) transport equation (\ref{eq:4.6}), a vorticity ($\omega$) transport equation is solved (\ref{eq:4.8}) along with transport equation for turbulent kinetic energy ($k$) given as (\ref{eq:4.5}), then the eddy viscosity is obtained as (\ref{eq:4.7}), 

\begin{equation} \label{eq:4.7}
    \nu_t = {c_\mu k \over \omega}
\end{equation}

\begin{equation} \label{eq:4.8}
    {D \omega \over D t} = c_{\omega_1} { \omega \over k } P_k -  c_{\omega_2} \omega^2 + {\partial \over \partial x_j} \left [  (\nu + {\nu_t \over \sigma_{\omega}}) {\partial \omega \over \partial x_j}  \right]  
\end{equation}

Here, $c_{\omega_1}$, $c_{\omega_2}$, $c_{\mu}$ and $\sigma_{\omega}$ are constants tuned for a given flow.

\subsubsection{$k-\omega$ SST model}

A blending function ($F_1$) is introduced in (\ref{eq:4.9}), thus taking advantages of both ($\epsilon$) and ($\omega$) equation given (\ref{eq:4.6}) and (\ref{eq:4.8}) respectively. This new model is refereed as $k-\omega$ SST introduced in \cite{menter1993zonal}. This model solves for ($\omega$) in the near wall region, thereby retaining performance, while mitigates the sensitivity of ($\omega$) away from the wall by solving ($\epsilon$).

\begin{equation} \label{eq:4.9}
    {D \omega \over D t} = c_{\omega_1} { \omega \over k } P_k -  c_{\omega_2} \omega^2 + {\partial \over \partial x_j} \left [  (\nu + {\nu_t \over \sigma_{\omega}}) {\partial \omega \over \partial x_j}  \right]  + 2(1 - F_1) {\sigma_{\omega_2} \over \omega} { \partial k \over \partial x_j } { \partial \omega \over \partial x_j}
\end{equation}

\subsection{Non-linear Eddy Viscosity}

In the previous cases, there are many notable deficiency's, for examples, it poorly preforms for jet impingement and streamline curvature (see \cite{craft1993impinging, suga_cubic});  separated flows \cite{speziale_quad}; and amongst many others. Further they cannot be readily applied for non-interial reference frame and in-variance \cite{speziale_quad}. To mitigate, the limitations of linear-eddy viscosity models, a class of non-linear eddy viscosity models, following a non-linear stress-strain relationship  was proposed (\ref{eq:4.10}), 

\begin{equation}  \label{eq:4.10}
    \overline{u'_i u'_j} = 2/3 k \delta_{ij} - \nu_t S_{ij} + f(S_{ij}, \Omega_{ij}) 
\end{equation}

where, $f(.)$ is a non-linear function, $ S_{ij} = ({ \partial \overline{U_i} / \partial x_j} +  { \partial \overline{U_j} / \partial x_i})$ and $\Omega_{ij} = ({ \partial \overline{U_i} / \partial x_j} -  { \partial \overline{U_j} / \partial x_i}) $ are the strain rate and rotation tensors respectively.  

In the forthcoming section, we limit ourselves to giving the non-linear stress-strain relationship, which is only ingredient required to motivate how elaborate forms of non-Fickian diffusion can be constructed. Indeed, specific details of the models can be found in the cited articles.

\subsubsection{Quadratic : Speziale model \cite{speziale_quad} }

The quadratic stress-strain relationship for Speziale's non-linear $k-\epsilon$ model \cite{speziale_quad} takes the form, 

\begin{align}
\begin{split}
         \overline{u'_i u'_j} = 2/3 k \delta_{ij} - k^{1/2} l ~ S_{ij} &- C_D l^2 \left ( S_{im} S_{mj} - {1 \over 3} S_{mn} S_{mn} \delta_{ij}  \right ) \\ &- C_E l^2 \left ( D_{ij} - {1 \over 3} D_{mm} \right )
\end{split}
\end{align}

where, 

\begin{equation}
    l = C {k^{3/2} \over \epsilon}
\end{equation}

and, 

\begin{equation}
    D_{ij} = {\partial S_{ij} \over \partial t} + \bold{\overline{U}} . \nabla S_{ij} - { \partial \overline{U_i} \over \partial x_k} S_{kj}  - { \partial \overline{U_j} \over \partial x_k} S_{ki}    
\end{equation}

where, $ \bold{\overline{U}} = \overline{U}_1 \bold{i} +  \overline{U}_2 \bold{j} + \overline{U}_3 \bold{k}$ and $D_{ij}$ is the frame-indifferent Oldroyd derivative \cite{Oldroyd} of $S_{ij}$. Here, $C_D$ and $C_E$ are dimensionless constants. This model meets the in-variance requirements. Further, in \cite{speziale_quad} the model is shown to predict the reattachment point of the separated flow following a backward facing step.

\subsubsection{Cubic model: Craft, Launder and Suga \cite{suga_cubic}}

Although, the quadratic stress-strain relationship, was an improvement over linear eddy viscosity models; the authors in \cite{suga_cubic} argue little generalisibility, thus a cubic model is proposed. 

The stress-strain relationship for a general cubic non-linear eddy-viscosity takes the form \cite{suga_cubic}, 

\begin{align}
    \begin{split}
          {\overline{u'_i u'_j} - 2/3 k \delta_{ij} \over k } = & -{\nu_t \over k} S_{ij} \\ &+ c_1 {\nu_t \over \epsilon} \left (S_{ik} S_{jk}  - 1/3 S_{kl} S_{kl} \delta_{ij}  \right) \\ &+ c_2 {\nu_t \over \epsilon } \left ( \Omega_{ik} S_{jk} + \Omega_{jk} S_{ik}   \right) \\
         &+ c_3 {\nu_t \over \epsilon } \left ( \Omega_{ik} S_{jk} - 1/3  \Omega_{kl} \Omega_{kl} \delta_{ij}   \right) \\ 
          &+ c_4 {\nu_t k \over {\epsilon}^2 } \left ( \Omega_{lj} S_{ki} + \Omega_{ii} S_{kj}   \right) S_{kl} \\
          &+ c_5 {\nu_t k \over {\epsilon}^2 } \left ( \Omega_{il} \Omega_{lm} S_{mj} + S_{il} \Omega_{lm} \Omega_{mj}  - 2/3 S_{lm} \Omega_{mn} \Omega_{nl} \delta_{ij}  \right) \\
          &+ c_6 {\nu_t k \over {\epsilon}^2 } S_{ij} S_{kl} S_{kl} \\ &+ c_7 {\nu_t k \over {\epsilon}^2 } S_{ij} \Omega_{kl} \Omega_{kl}
    \end{split}
\end{align}

where, $c_1 - c_7$ are the coefficients for the model, tuned appropriately. Such, a cubic model has been shown to preform well for the challenging cases, such as jet impingement and strong stream-line curvature \cite{suga_cubic}.

\end{document}